\documentclass[10pt,preprint]{aastex} 
\usepackage{longtable}
\usepackage{multirow}
\usepackage{graphicx}
\usepackage{epstopdf}
\usepackage{hyperref}
\usepackage{booktabs}
\newsavebox{\myimage}

\begin{document}
\newcommand{\kms}{km\,s$^{-1}$}
\newcommand{\tbm}{\tablenotemark}

\title{An Interferometric Spectral-Line and Imaging Survey\\ 
       of VY~Canis Majoris in the 345~GHz Band}

\author{T. Kami\'{n}ski\altaffilmark{1}, C. A. Gottlieb\altaffilmark{2},
        K. H. Young\altaffilmark{2}, K. M. Menten\altaffilmark{1}, N. A. Patel\altaffilmark{2}
        } 
 
\altaffiltext{1}{Max-Planck Institut f\"ur Radioastronomie, Auf dem H\"ugel 69, 53121 Bonn, Germany}
\altaffiltext{2}{Harvard-Smithsonian Center for Astrophysics, 60 Garden Street, Cambridge, MA 02138, USA}
\email{kaminski@mpifr.de}
\shorttitle{SMA survey of VY\,CMa at 345\,GHz}
\shortauthors{Kami\'{n}ski et al.}

\begin{abstract} 
A spectral line survey of the oxygen-rich red supergiant VY\,Canis Majoris was made between 279 and 355\,GHz with the Submillimeter Array. Two hundred twenty three spectral features from 19 molecules (not counting isotopic species of some of them) were observed, including the rotational spectra of TiO, TiO$_2$, and AlCl for the first time in this source.  The parameters and an atlas of all spectral features is presented. Observations of each line with a synthesized beam of $\sim$0\farcs9, reveal the complex kinematics and morphology of the nebula surrounding VY\,CMa. Many of the molecules are observed in high lying rotational levels or in excited vibrational levels. From these, it was established that the main source of the submillimeter-wave continuum (dust) and the high excitation molecular gas (the star) are separated by about 0\farcs15. Apparent coincidences between the molecular gas observed with the SMA, and some of the arcs and knots observed at infrared wavelengths and in the optical scattered light by the Hubble Space Telescope are identified.   
The observations presented here provide important constraints on the molecular chemistry in oxygen-dominated circumstellar environments and a deeper picture of the complex circumstellar environment of VY\,CMa. 
 
\end{abstract}

\keywords{astrochemistry -- circumstellar matter -- line: identification -- stars: individual (VY CMa) -- supergiants -- surveys}

\section{Introduction \label{intro}}

Most recent studies of circumstellar envelopes with radio telescopes have focused on IRC$+10\degr216$ 
(CW\,Leo), a nearby carbon star on the asymptotic giant branch \citep[AGB;][and references therein]{patel}. However, studies of the rich and complex chemistry in carbon-rich sources do not advance an understanding of oxygen-rich environments, like those of M-type AGB stars or red supergiants. 
Because the abundances of the most chemically active species in oxygen-rich circumstellar environments are closer to the average Galactic (``cosmic'') abundances (such as those of star forming regions), oxygen-rich envelopes provide laboratories of chemical processes that have a much broader application than those of carbon-rich envelopes. Recently, it has been shown that the millimeter-wave spectrum of the oxygen-rich supergiant VY\,Canis Majoris (VY\,CMa) is very rich, owing to approximately 20 molecules in the circumstellar material that have now been identified in deep observations of its envelope
 \citep{ziurys_nat,tenen_survey}. An observational effort to explore the spectrum of VY\,CMa in a broad range of wavelengths may bring us closer to an understanding of the chemical processes that occur in oxygen-rich environments, but this can be only achieved if the physical and dynamical structure of the envelope is well characterized. This is not a trivial problem, especially in the case of a supergiant such as VY\,CMa.  Despite extensive studies over many decades, there still remain many questions about the physics, chemistry, and dynamics of this object.

VY\,CMa, predicted to become a supernova \citep{smith_CO}, is one of the most intrinsically luminous red stars in the Galaxy \citep[$\sim 3 \times 10^5 L_\sun$;][]{yoon}. It is a large ($R_{\star} = 1500 - 3000$\,R$_\sun$, or $\sim$10\,mas at 1.2\,kpc), massive (presently $\sim 17$\,M$_\sun$, but originally $\sim 25$\,M$_\sun$; \citealp{wittk_2012}) star, with a very high mass-loss rate \citep[$4 \times 10^{-4}$\,M$_\sun$\,yr$^{-1}$,][]{monnier}. Although it is certain it is a cool object, there is still some controversy about its actual spectral type and effective temperature:  for example, \cite{sidaner} refer to it as an M4--M5 star with $T_{\rm eff}\approx$3000\,K, whereas  \cite{massey} as an  M1 star with $T_{\rm eff}$=3650\,K. The distance to VY\,CMa is well known ($d=1.2\pm0.1$\,kpc), owing to recent precise measurements of its parallax at radio wavelengths \citep{bo,yoon}.

\begin{figure}[ht!]\centering
\includegraphics[angle=270, width=0.7\textwidth]{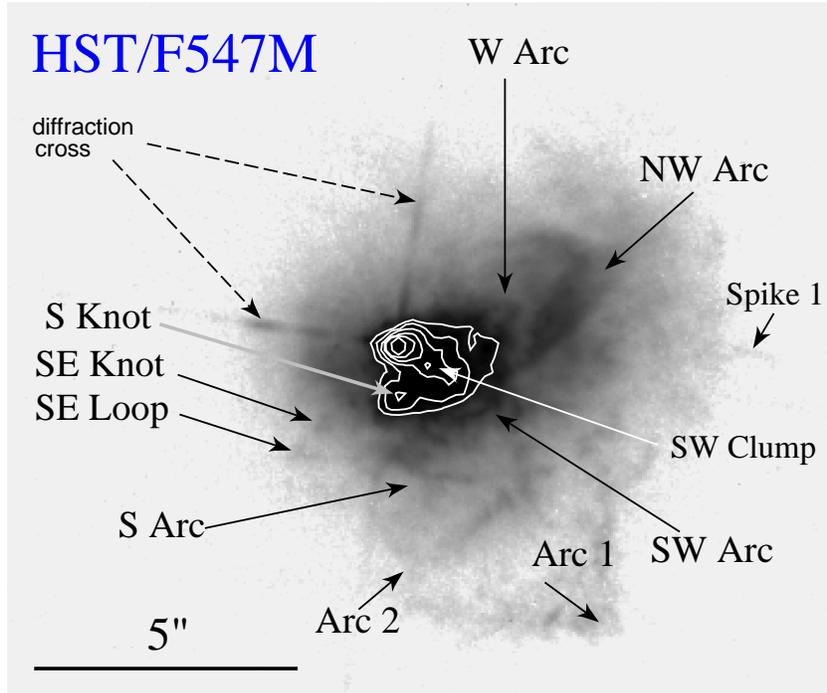}
\caption{Scattered light image of the nebula surrounding VY\,CMA obtained with the HST (F547M filter). Indicated are some of the main spatial features identified in \citet{hump-hst}. A diffraction pattern of the strongest central source was not removed from the image. The white contours trace the structure of the bright inner part of the nebula.
\label{fig-HST}}
\end{figure}

The low temperature and high mass-loss rate has produced a dusty envelope around VY\,CMa with a very complex morphology which manifests itself as an emission and reflection nebula. A composite picture of the nebula derived from near infrared interferometry \citep{monnier} and images obtained with the Hubble Space Telescope \citep[HST;][]{smith-HST,hump-hst}, reveal complex circumstellar ejecta with arcs, filaments, ``spikes'', and knots. An optical HST image of the nebula in which some of the most characteristic features are labeled is shown in Fig.\,\ref{fig-HST}. The most sensitive optical observations indicate that the dimensions of the entire nebula are up to about 10\arcsec\ (0.06\,pc). In the current picture, the entangled nebula was formed by multiple discrete ejections, some of which happened up to $\sim$1700 years ago \citep{hump-hst}, and were possibly accompanied by a more steady spherical wind \citep{smith_CO}. The random orientations of the arcs suggests that they might have been produced by localized eruptions on the stellar surface, but the high extinction of the dust shell poses an inherent challenge to an unambiguous interpretation of the innermost circumstellar structure on the basis of the continuum emission at infrared and visible wavelengths. Extinction and, most importantly, scattering are negligible at far-infrared and longer wavelengths, thereby providing an opportunity to disentangle the inner regions when sufficient angular resolution in the submillimeter band is attained.

Although several diatomic molecules have been observed in emission in the optical spectrum of VY\,CMa, including TiO, VO, ScO, YO, and most recently AlO \citep[][and references therein]{kami_alo}, sensitive observations in the radio and (sub-)millimeter domain have allowed to discover and explore many more molecular species in the material around VY\,CMa. Early in the development of molecular radio astronomy, observations with single antennas of OH, H$_2$O, and SiO in the circumstellar envelope of VY\,CMa  revealed complex line profiles.  Soon afterwards, VY\,CMa became the focus of interferometric VLBI\footnote{Very Large Baseline Interferometry} measurements of intense maser emission of SiO and H$_2$O close to the star \citep[see,][and references therein]{bo}, and of OH at 1612\,MHz at a couple of arcsec extent \citep{OH}. Following these initial observations more than 35~years ago, a number of diatomic and a few small polyatomic molecules were observed in the intervening years, including CO, CS, SiS, SO, SO$_2$, H$_2$S, NS, CN, HCN, HNC, HCO$^+$, and NH$_3$; and the metal-bearing species NaCl, AlO, AlOH, PO, and PN  in the past six years. Except for CO and SO \citep{muller}; and SO$_2$, SiS, and PN \citep{fu}. Most were observed with single antennas in the millimeter band at an angular resolution of typically $\ge 25\arcsec$ (\citealp{menten_water}; \citealp{tenen_survey} and references therein). 

Because the angular size of some of the arcs and knots of VY\,CMa's nebula is comparable to the angular resolution of modern interferometers, interferometric observations of molecular lines may yield fundamental information on the chemical composition, temperature, density, and kinematics of these features. Owing to the low angular resolution in most prior observations, and to the complex kinematic structure of the nebula, the apparent association of most molecules  with specific morphological features observed in optical and infrared images (except those of OH, H$_2$O, and SiO), could only be inferred from tentative interpretations of spatially unresolved velocity features in the broad asymmetric line profiles. The high angular resolution and sensitivity of current submillimeter-wave interferometers, as well as the lack of extinction effects, makes this wavelength band especially promising for such studies.

Motivated by the desire to explore the (sub-)millimeter spectrum of VY\,CMa and to investigate the spatio-kinematical structure of its nebula, we undertook an interferometric spectral imaging survey of VY\,CMa between 279 and 355\,GHz with the Submillimeter Array (SMA)\footnote{The Submillimeter Array is a joint project between the Smithsonian Astrophysical Observatory and the Academia Sinica Institute of Astronomy and Astrophysics and is funded by the Smithsonian Institution and the Academia Sinica.} at an angular resolution of 0\farcs9. This paper presents the details of the observations, calibration of the interferometric measurements, and generation of the maps and spectra (Sect.\,\ref{obs}); describes the line identifications (Sect.\,\ref{results}); provides a detailed description of the line profiles and spatial distributions of each molecule (Sect.\,\ref{Sect-species}); 
classifies each species according to morphological type (Sect.\,\ref{morph}); and discusses the continuum and location of the centers of molecular emission with respect to the peak of the continuum (Sect.\,\ref{Sect-conti}). From an analysis of population diagrams, approximate rotational temperatures and column densities of eleven molecules were derived (Appendix\,\ref{sect-RDintro}). A detailed discussion of the chemistry in the circumstellar gas and the structure of the outflow of VY\,CMa will be presented elsewhere.


\section{Observations and data reduction \label{obs}}

The survey of VY\,CMa was obtained together with an AGB star IK\,Tau, whose observations will be presented elsewhere. The observations  of VY\,CMa were made with the SMA in its extended configuration over ten nights during January and February of 2010 (see Table\,\ref{Tab-log}). In that configuration, unprojected baselines range in length from 44 to 226\,m producing a synthesized beam with a full width at half maximum (FWHM) of typically $0\farcs$9. On each night, observing began around 03:30 UT; IK\,Tau was observed for about two hours until VY\,CMa rose, at which time observations of IK\,Tau and VY\,CMa were interleaved for five hours until IK\,Tau set. After that, VY\,CMa was observed alone for two more hours. The phase center for IK\,Tau was 
$\alpha (2000) = 03^{\rm{h}}53^{\rm{m}}28.866^{\rm{s}}, 
\delta (2000) = 11^{\circ}24^{\prime}22.29\farcs2 $;
for VY\,CMa
$\alpha (2000) = 07^{\rm{h}}22^{\rm{m}}58.332^{\rm{s}},
\delta (2000) = -25^{\circ}46^{\prime}03.17\farcs2 $
was used. The online software Doppler-tracked VY\,CMa during the entire observation, the frequency scale for the IK\,Tau data was corrected offline to compensate for that. Nearby quasars were observed every 20\,min for gain calibration. The quasar 0423-013 was used to calibrate IK\,Tau data, and 0730-116 was used for VY\,CMa. Titan was observed for flux calibration on the nights when it was separated from Saturn by several arcmin. On other nights Uranus or Vesta was used as the primary flux calibrator. The standard source 3C\,273 was observed at the end of each track to provide bandpass calibration.

The correlator was configured with a uniform resolution of 812.5\,kHz per channel, providing a velocity resolution of 0.69--0.87\,\kms. For each tuning, there is a gap of 32\,MHz at the center of the 4\,GHz total frequency coverage, which was required to avoid interference from a downconverter local oscillator (the frequency coverage is given in Table\,\ref{Tab-log}).  On some nights, there was a phase drift between the upper and lower 2\,GHz halves of the intermediate frequency (IF), which required the two halves to be calibrated separately (see Table\,\ref{Tab-log}).

\begin{deluxetable}{l cc cc c@{}r r c@{}c cc}
\rotate
\tabletypesize{\scriptsize}
\tablecaption{Technical details of the survey observations with SMA\label{Tab-log}}
\tablewidth{0pt}
\tablehead{
& 
\multicolumn{4}{c}{Coverage in rest frequency (GHz)}& 
\multicolumn{2}{c}{Synthesized}&
& 
&& 
&\\ 
\cmidrule{2-5}
& 
\multicolumn{2}{c}{LSB}&
\multicolumn{2}{c}{USB}& 
\multicolumn{2}{c}{beam}&
\colhead{}&   
\multicolumn{2}{c}{$T_{\rm{sys}}$\,(K)\tablenotemark{b}}&
\multicolumn{2}{c}{rms\,(Jy)\tablenotemark{c}}\\
\cmidrule(r){2-3}
\cmidrule(l){4-5}
\cmidrule(l){6-7}
\cmidrule(r){9-10}
\cmidrule(l){11-12}
\multicolumn{1}{c}{UT date}&
\multicolumn{1}{c}{High IF}&
\multicolumn{1}{c}{Low IF}&
\multicolumn{1}{c}{Low IF}&
\multicolumn{1}{c}{High IF}&
\colhead{FWHM\,(\arcsec)}&
\multicolumn{1}{c}{PA\,(\degr)}&
\colhead{$\tau$\tablenotemark{a}} & 
\colhead{IK\,Tau}& 
\colhead{VY\,CMa}&
\colhead{LSB}&
\colhead{USB}
}
\startdata
2010 Jan 17                  & $339.071-341.039$ & $341.071-343.040$ &  $351.073-353.041$ & $353.073-355.041$  & $0.84\times0.81$& $ 34.2$ & 0.065 & 245 & 333 & 0.120  &  0.138\\
2010 Jan 20                  & $335.071-337.039$ & $337.071-339.039$ &  $347.072-349.041$ & $349.073-351.041$  & $0.84\times0.80$& $-79.5$ & 0.035 & 158 & 180 & 0.084  &  0.082\\
2010 Jan 21                  & $331.071-333.039$ & $333.071-335.039$ &  $343.072-345.041$ & $345.073-347.041$  & $0.94\times0.84$& $ 58.6$ & 0.040 & 161 & 180 & 0.057  &  0.066\\
2010 Jan 22\tablenotemark{d} & $315.071-317.039$ & $317.071-319.040$ &  $327.073-329.041$ & $329.073-331.041$  & $1.03\times0.73$& $  2.7$ & 0.035 & 199 & 264 & 0.081  &  0.102\\
2010 Jan 23                  & $311.071-313.039$ & $313.071-315.039$ &  $323.072-325.041$ & $325.073-327.041$  & $0.96\times0.83$& $-39.2$ & 0.085 & 464 & 761 & 0.157  &  2.397\\
2010 Jan 25                  & $307.071-309.039$ & $309.071-311.040$ &  $319.073-321.041$ & $321.073-323.041$  & $0.93\times0.87$& $ 13.2$ & 0.080 & 305 & 371 & 0.166  &  0.379\\
2010 Jan 26\tablenotemark{d} & $303.071-305.039$ & $305.071-307.039$ &  $315.073-317.041$ & $317.073-319.041$  & $0.97\times0.86$& $-20.6$ & 0.055 & 181 & 233 & 0.069  &  0.080\\
2010 Jan 27                  & $287.071-289.039$ & $289.071-291.040$ &  $299.073-301.041$ & $301.073-303.041$  & $0.95\times0.93$& $  8.8$ & 0.060 & 153 & 192 & 0.064  &  0.084\\
2010 Feb 01\tablenotemark{d} & $283.071-285.039$ & $285.071-287.040$ &  $295.073-297.041$ & $297.073-299.041$  & $1.01\times0.95$& $-13.1$ & 0.035 & 134 & 162 & 0.056  &  0.074\\
2010 Feb 02\tablenotemark{d} & $279.071-281.039$ & $281.071-283.039$ &  $291.073-293.041$ & $293.073-295.041$  & $1.02\times0.94$& $-22.0$ & 0.040 & 134 & 159 & 0.059  &  0.067\\

\enddata
\tablenotetext{a}{Zenith optical depth at 225\,GHz measured with the radiometer of Caltech Submillimeter Observatory.}
\tablenotetext{b}{Average of $T_{\rm{sys}}$ (for double side band) over all antennas and elevations during the observations.}
\tablenotetext{c}{Typical rms noise level in the 4-GHz band per 0.8\,MHz (for VY\,CMa only).}
\tablenotetext{d}{For this data set, the upper and lower 2\,GHz halves of the IF had to be
calibrated separately, because of a local oscillator frequency drift.}
\end{deluxetable}

The data were calibrated using the MIR-IDL\footnote{\url{https://www.cfa.harvard.edu/~cqi/mircook.html}} package. A bandpass calibration was applied first.  The system temperature, $T_{\rm{sys}}$, for each
baseline (the geometric mean of the $T_{\rm{sys}}$ values of all antennas) was examined for data outliers, and then used to convert the correlator's output to approximate Jy units. Weights for the
visibility data were set equal to the scan integration time divided by the square of $T_{\rm{sys}}$.   The visibility amplitudes of the complex gain calibrators were compared with the amplitudes of each track's primary flux calibrator (usually Titan), at a time when the gain and flux calibrators were at similar elevations. This produced a more accurate flux for the gain calibrators than did the $T_{\rm{sys}}$ correction. After all ten tracks had been calibrated, it was found that the tracks for which the primary flux calibrator had been Titan gave much more consistent flux values for the gain calibrators than did the tracks using Uranus or Vesta. Because of this, the Titan-calibrated flux values were used to construct linear models for the flux of each gain calibrator as a function of frequency, and the tracks for which no Titan data was available were recalibrated using these model flux values, rather than obtaining the calibration from Uranus or Vesta scans. After the flux of the gain calibrators had been established, they were used to correct the amplitude and phase of the VY\,CMa and IK\,Tau data as a function of time.  This gain calibration step also modified the data weights, to compensate for the fact that the receiver gain and atmospheric noise differed in the receivers' two sidebands. After a careful inspection of the absolute flux calibration, we estimate that its typical uncertainty is about 15\%.

After the visibility data points were calibrated, maps of each channel were produced using the Miriad package \citep{miriad}. (The pixel size in the maps was usually 0\farcs1.) The continuum of VY\,CMa was very clearly detected (0.5--0.9\,Jy), and many of the spectral features were offset from the continuum position. For each sideband of each track, a continuum map was produced from the channels which appeared to be free of strong line features. The continuum map was then used to self-calibrate the data (in phase only) to correct for some of the phase variations which occurred on timescales too short to be corrected by the twenty-minute gain calibration cycle. Continuum and line map cubes were then produced from the recalibrated visibilities. All of the spectra presented and analyzed in this study have the continuum subtracted. Because the data were self-calibrated, the \emph{absolute} astrometric information was not available in the data used in this study and offsets scales used here are given with respect to the continuum peak. The errors in relative positioning between different tracks are related to the uncertainty in the position of the continuum peak which is less than 11\,mas (3$\sigma$).

The noise is fairly homogeneous throughout most of the survey. The typical rms noise level is of 0.1\,Jy per 0.8\,MHz, except for the region between about 320 and 327\,GHz, where it is as high as 2.4\,Jy owing to a telluric water band. In Fig.\,\ref{Fig-Full_survey}, the entire spectrum is compared with the transmission of the earth's atmosphere. The rms noise in each 4-GHz band is given in Table\,\ref{Tab-log}.

In order to assess the fraction of the flux filtered out in the SMA observations, we compared several strong lines observed with the interferometer and with single antennas. We mainly referred to observations in the archives of the Atacama Pathfinder Experiment (APEX), which we converted to the flux scale in the Rayleigh-Jeans limit. To compare the two, the SMA spectra were averaged over a  solid angle that corresponds to the diameter of the APEX antenna beam. We found that at angular scales of a few arcsec, as much as half of the flux measured with the SMA is missing for the most extended emission such as that observed in CO, and the line shapes in the two spectra look different. Although this places constraints on the analysis of the most extended features in the nebula, the small scale structures of extended emission is enhanced. The flux is fully recovered for sources smaller than about 2\arcsec.  

\section{Results\label{results}}

The SMA observations reveal a very rich molecular spectrum of the  material around the star.
Because the emission nebula of VY\,CMa is extended in many of the molecular transitions, our interferometric survey allows us to study the complex spatial distribution and kinematics of the molecular gas. The data were explored by extracting spectra for a chosen region (``aperture'') and obtaining maps of integrated emission within a chosen frequency range. Most of the spectral analysis reported here was obtained for an aperture covering the central, unresolved source, namely in a 1\arcsec$\times$1\arcsec\ box which exhibits the greatest variety of molecular species. A full spectrum extracted for this aperture is shown in Fig.\,\ref{Fig-Full_survey} and in the spectral atlas in Appendix\,\ref{atlas} (Fig.\,\ref{Fig-Taco}).   

\begin{figure*}[ht!]\centering
\includegraphics[angle=0, width=0.7\textwidth]{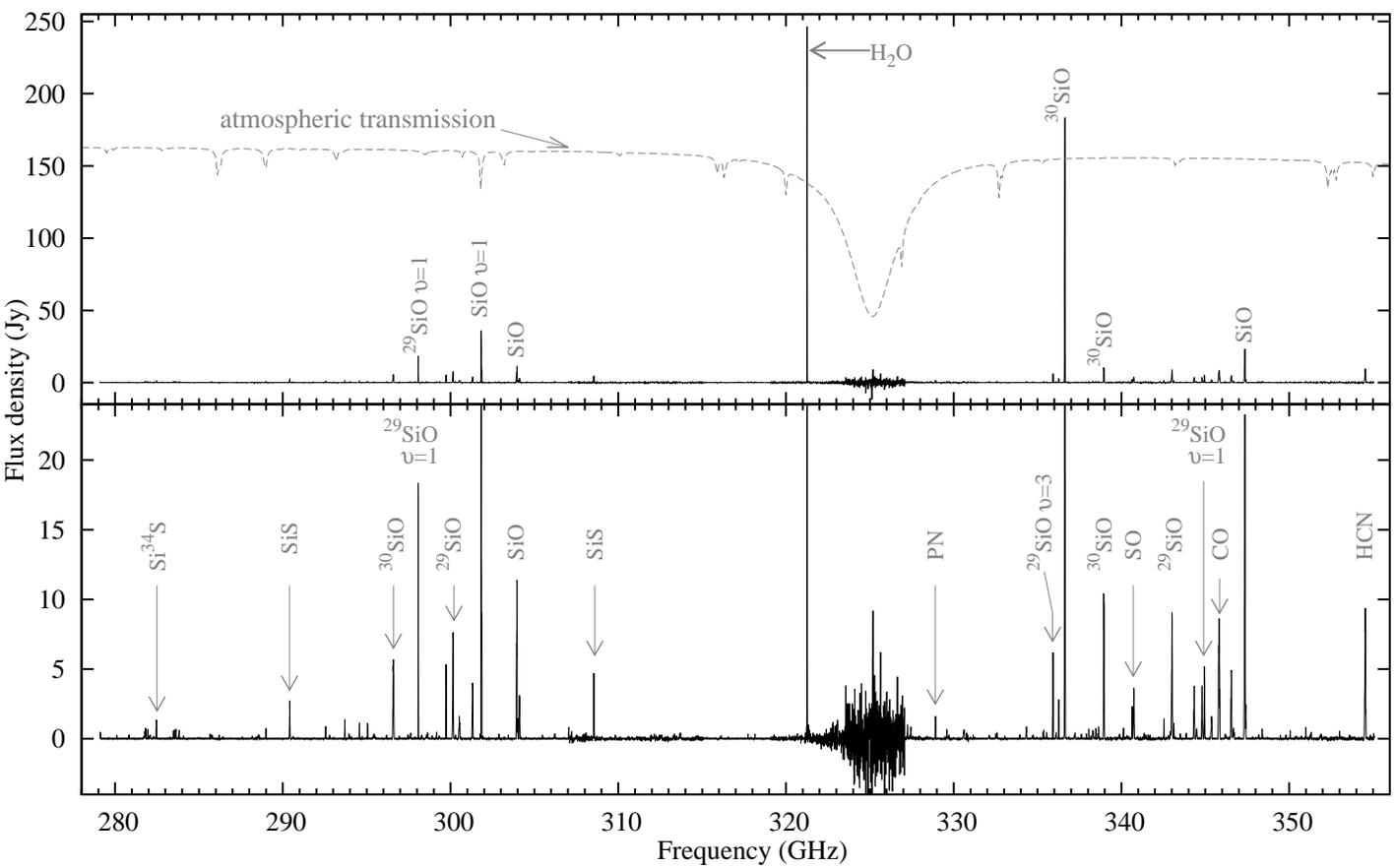}
\caption{Full spectrum of VY\,CMa obtained in the SMA survey. The two panels show the most intense lines with two different intensity scales. Plotted in the upper panel (dashed line) is the atmospheric transmission \citep{ATM}.
\label{Fig-Full_survey}}
\end{figure*}

The identification of spectroscopic features proceeded in several steps. At first, tentative identifications were made on the basis of the close coincidence between the astronomical frequency (on the assumption that the $V_{\rm{LSR}}$ of the central compact component is $\sim 22$\,km\,s$^{-1}$; \citealp{tenen_survey}), and calculated frequencies tabulated in the standard catalogs of spectral lines, i.e., that of Jet Propulsion Laboratory \citep[JPL;][]{jpl} and Cologne Database for Molecular Spectroscopy \citep[CDMS;][]{cdms1,cdms2}. The assignments were subsequently confirmed by manual inspection of the spectra, and by referring to Splatalogue\footnote{ \url{http://splatalogue.net/}} and papers describing the primary laboratory measurements. Nearly all the molecules observed in this survey have more than one transition at comparable predicted intensity. A detection of a particular molecule was considered certain if all transitions were observed with the correct predicted intensity as determined either from (1) an analysis of a population diagram (see Appendix\,\ref{sect-RDintro}), or (2) for an assumed excitation temperature, if there was an insufficient number of lines for a full analysis of population diagrams. 
Otherwise, in the absence of confirming transitions the carrier was designated either as unidentified (U), or by a question mark appearing beside the tentatively identified species. 
The first identifications were made on the spectrum of high signal-to-noise (S/N) extracted for the central source of 1\arcsec, and the same procedure was subsequently applied to spectra averaged over larger apertures, allowing for identification of features arising in outer parts of the nebula (e.g., emission of CN). The spectral identification of each feature is indicated in the atlas in Fig.\,\ref{Fig-Taco}, where the lines are labelled by the sequential index numbers in Table\,\ref{Tab-main}. Further details on the identifications are provided in Sect.\,\ref{Sect-species}.

Table\,\ref{Tab-summary} summarizes the identified molecules and their isotopic species, with the number of transitions detected in each, and the corresponding reference numbers in Table\,\ref{Tab-main}. In all, 223 lines were detected in VY\,CMa in the 77\,GHz-wide band, with all but twelve of these assigned to known carriers. We identified 33 molecules (main plus rare isotopic species), including the first detection in this source of the rotational spectrum of TiO and the first astronomical observation of  TiO$_2$ \citep[see][]{kami_tio}, and AlCl for the first time in VY\,CMa. Many of the species (e.g., PO, PN, AlO, AlOH, and NaCl) were observed by \citet{tenen_survey} at a very modest S/N, but our interferometric survey allows a more detailed look at these exotic species. The molecules with the highest number of transitions are SO$_2$ (50 assigned), TiO$_2$ (33), NaCl (21), and SO (17). Most are in the ground vibrational state, but rotational transitions in vibrationally excited levels of NaCl, SiS, SO,  and in the well-studied SiO and H$_2$O are all observed with good S/N. In addition to the species with the most abundant isotopes, several rare isotopologues were observed (e.g., $^{34}$S- and  $^{13}$C-bearing molecules; Na$^{37}$Cl; $^{29}$Si- and $^{30}$Si-bearing molecules; and Si$^{18}$O). 

Table\,\ref{Tab-main} gives the measured properties of each feature including: the centroid velocity (signal-weighted mean velocity of the feature) with respect to the Local Standard of Rest ($V_{\rm LSR}$), the approximate velocity range of the feature, the flux within the specified velocity interval, maximum flux density within that interval (given for a bin of 0.8\,\kms), and the full-width at half-maximum (FWHM) obtained from a least-squares fit of a Gaussian profile to those features consisting of a single peak with sufficiently high S/N. The measurements  were performed on a spectrum extracted with a 1\arcsec$\times$1\arcsec\ box centered on the continuum peak. A detailed description of the line profiles, the central velocities, line widths, and spatial distributions of each species is given in Sect.\,\ref{Sect-species}. Summarized in Sect.\,\ref{morph}, is the classification of each species into the different morphological types. 


In addition to the variety of molecular features, an important result of the survey are sensitive measurements of the (sub-)millimeter-wave continuum in VY\,CMa. These are presented in Sect.\,\ref{Sect-conti}, where the positions of the centers of molecular emission and the location of the stellar photosphere with respect to the peak of the continuum emission are discussed. 

\clearpage
\begin{deluxetable}{c@{}r@{ } lcl @{ }c@{} c rr ccc c}
\tabletypesize{\tiny}
\tablewidth{0pt}
\tablecaption{Spectral features observed in VY\,CMa with the SMA\label{Tab-main}}
\tablehead{
\multicolumn{1}{c}{(1)}& 
\colhead{}  
&\multicolumn{1}{c}{(2)} & \colhead{(3)} &\colhead{(4)} &\colhead{} &\colhead{(5)} 
&\multicolumn{2}{c}{(6)}
&\colhead{(7)} &\colhead{(8)} &\colhead{(9)}  &\colhead{(10)} \\
\multicolumn{1}{c}{No.} &
\colhead{} & 
\multicolumn{1}{c}{Molecule} & 
\colhead{Quantum} & 
\colhead{$\nu_{\textit{ref}}$} & 
\colhead{} & 
\colhead{Centroid} & 
\multicolumn{2}{c}{Velocity\,range} & 
\colhead{Flux} & 
\colhead{Max}  & 
\colhead{FWHM} &
\colhead{Comments}\\
\cline{8-9}
\colhead{} &   \colhead{} & \colhead{} &
\colhead{numbers} & 
\colhead{(MHz)} &  & 
\colhead{(km s$^{-1}$)} & 
\multicolumn{2}{c}{(km s$^{-1}$)} & 
\colhead{(Jy\,km\,s$^{-1}$)} & 
\colhead{(Jy)} &
\colhead{(km\,s$^{-1}$)} &
\colhead{} } 
\startdata

\nodata  & &{\it Instr.279096}& \nodata & \nodata &  & \nodata & \nodata & \nodata & \nodata & \nodata & \nodata & a \\
1& &NaCl $v$=3 & 22--21 &279830.9588&  &24.0&1.6&44.3&1.7&0.17&18&  \\
\multirow{2}*{2} &\multirow{2}*{$\{$} &Na$^{37}$Cl & 22--21 &280063.6927& \multirow{2}*{$\}$} & \multirow{2}*{14.1} & \multirow{2}*{--32.6} & \multirow{2}*{59.1} & \multirow{2}*{9.5} & \multirow{2}*{0.40} & \multirow{2}*{32} &  \\
                         &   & $^{30}$SiS & 16--15 &280079.6845&  &  &  &  &  &  &  &  \\
3&&Si$^{18}$O $v$=1? & 7--6 &280542.66&  &25.8&--0.1&47.9&0.3&0.19&5&  \\
4&&SO$_2$ & 26(4,22)--26(3,23) &280807.2458&  &28.8&--38.7&91.1&9.6&0.41&46&  \\
\multicolumn{3}{l}{\it gap 281039--281071\,MHz} &  &  &  &  &  &  &  &  &  \\
5&&SO$_2$ & 36(4,32)--36(3,33) &281688.9306&  &24.9&--11.5&89.7&9.1&0.28&74&  \\
6&&SO$_2$ & 15(1,15)--14(0,14) &281762.6003&  &27.8&--30.9&68.1&27.9&0.77&135&  \\
\multirow{2}*{7}&\multirow{2}*{$\{$}& PN & 6--5 &281914.2049& \multirow{2}*{$\}$} & \multirow{2}*{18.5} & \multirow{2}*{--27.3} & \multirow{2}*{63.2} & \multirow{2}*{20.5} & \multirow{2}*{0.91} & \multirow{2}*{27} &  \\
                         &&NaCl $v$=2 & 22--21 &281932.3655&  &  &  &  &  &  &  &  \\
8&&SO$_2$ & 6(2,4)--5(1,5) &282036.5663&  &35.6&--18.1&86.0&10.0&0.44&70&  \\
9&&SO$_2$ & 20(1,19)--20(0,20) &282292.8063&  &28.8&--16.6&79.5&8.4&0.28&50&  \\
\multirow{2}*{10}&\multirow{2}*{$\{$} &Si$^{34}$S & 16--15 &282431.1631& \multirow{2}*{$\}$} & \multirow{2}*{18.0} & \multirow{2}*{--18.7} & \multirow{2}*{62.0} & \multirow{2}*{38.1} & \multirow{2}*{1.48} & \multirow{2}*{20} &  \\
                               &&Si$^{18}$O & 7--6   &282434.7338&  &  &  &  &  &  &  &  \\
11&&SO$_2$ $\nu_2$=1 & 16(0,16)--15(1,15) &282803.4748&  &17.3&--22.9&56.3&2.2&0.20& \nodata &  \\
\multicolumn{3}{l}{\it gap 283039--283071\,MHz} &  &  &  &  &  &  &  &  &  \\
12&&AlOH & 9--8 &283253.884&  &17.2&--21.3&53.1&2.4&0.20&19& b \\
13&&SO$_2$ & 16(0,16)--15(1,15) &283464.7685&  &25.0&--22.9&79.4&30.2&0.75&74&  \\
\multirow{2}*{14}&\multirow{2}*{$\{$} &TiO$_2$ & 20(1,19)--19(2,18) &283567.7163& \multirow{2}*{$\}$} & \multirow{2}*{1.7} & \multirow{2}*{--28.7} & \multirow{2}*{51.1} & \multirow{2}*{23.3} & \multirow{2}*{0.79} & \multirow{2}*{28} & \multirow{2}*{b} \\
                                 &&PO & $^2 \Pi_{1/2}\,J$=13/2--11/2, $e$ &283589.76&  &  &  &  &  &  &  &  \\
15&&$^{29}$SiS $v$=1? & 16--15 &283672.3388&  &36.4&6.5&63.4&1.6&0.16& \nodata &  \\
16&&PO & $^2\Pi_{1/2}\,J$=13/2--11/2, $f$ &283781.21&  &22.9&--13.1&73.6&22.1&0.78&25& b\\
17&&U283839& \nodata & \nodata &  & \nodata & \nodata & \nodata &1.3&0.15& \nodata &  \\
18&&TiO$_2$ & 29(7,23)--29(6,24) &283942.5446&  &51.1&8.8&74.3&2.5&0.18& \nodata &  \\
19&&NaCl $v$=1 & 22--21 &284047.6265&  &22.5&--1.3&45.4&4.2&0.34&15&  \\
20&&TiO$_2$ & 27(7,21)--27(6,22) &284094.3206&  &18.5&--0.2&44.8&0.7&0.15& \nodata &  \\
21&&TiO$_2$ & 21(1,21)--20(0,20) &284371.901&  &22.0&--9.2&73.8&4.1&0.22&\nodata&  \\
22&&TiO & $^3\Delta_1\,J$=9--8  &284875.3115&  &22.0&--8.2&74.0&4.2&0.22&23&  \\
23&&TiO$_2$ & 20(2,18)--19(3,17) &284951.3104&  &22.1&--9.6&74.3&4.2&0.22& \nodata &  \\
24&&$^{29}$SiS & 16--15  &285046.9528&  &\nodata&\nodata&56.4&\nodata&\nodata& \nodata & c \\
\multicolumn{3}{l}{\it gap 285039--285071\,MHz} &  &  &  &  &  &  &  &  &  \\
25&&PO & $^2 \Pi_{3/2}\,J$=13/2--11/2 &285634.28&  &21.1&--27.7&73.1&14.7&0.48&34& b \\
26&&SO$_2$ & 17(3,15)--17(2,16) &285743.5877&  &32.9&--20.7&78.8&13.6&0.35&57&  \\
27&&TiO$_2$ & 25(7,19)--25(6,20) &285859.6779&  &32.9&2.2&64.1&2.5&0.13&54&  \\
28&&NaCl & 22--21 &286176.3466&  &24.1&--0.1&47.1&7.2&0.44&20&  \\
29&&SO? & 1(1)--1(0) &286340.152&  &78.9&62.0&97.3&1.1&0.17& \nodata &  \\
30&&SO$_2$ & 22(2,20)--21(3,19) &286416.2712&  &21.4&--15.2&63.8&8.0&0.32&\nodata&  \\
\multicolumn{3}{l}{\it gap 287040--287071\,MHz} &  &  &  &  &  &  &  &  &  \\
31&&TiO$_2$ & 22(7,15)--22(6,16) &287253.6127&  &27.7&--0.2&69.4&3.0&0.20& \nodata &  \\
32&&SiS $v$=2 & 16--15 &287553.7256&  &18.2&--0.7&39.5&1.9&0.20&9&  \\
33&&TiO & $^3\Delta_2\,J$=9--8  &288156.4323&  &19.9&4.2&34.6&2.5&0.27&7&  \\
&&SO$_2$ & 18(1,17)--17(2,16) &288519.9964&  &36.2&--17.8&95.1&13.8&0.38&\nodata&  \\
35&&SiS $v$=1 & 16--15 &288967.3023&  &20.3&0.3&43.9&13.8&0.96&14&  \\
\multicolumn{3}{l}{\it gap 289039--289071\,MHz} &  &  &  &  &  &  &  &  &  \\
36&&SiS & 16--15 &290380.757&  &21.0&--24.4&68.1&82.4&2.86&25&  \\
\multirow{2}*{37}&\multirow{2}*{$\{$}&$^{34}$SO & 6(7)--5(6) &290562.238& \multirow{2}*{$\}$} & \multirow{2}*{15.6} & \multirow{2}*{--49.9} & \multirow{2}*{62.8} & \multirow{2}*{9.6} & \multirow{2}*{0.25} & \multirow{2}*{\nodata} &  \\
                               && Na$^{37}$Cl $v$=1 & 23--22 &290602.3735&  &  &  &  &  &  &  &  \\
38&&TiO$_2$ & 21(7,15)--21(6,16) &290830.5183&  &20.0&3.8&48.7&2.1&0.23& \nodata &  \\
39&&TiO & $^3\Delta_3\,J$=9--8 &290998.0368&  &16.3&2.6&35.9&0.8&0.24& \nodata &  \\
\multicolumn{3}{l}{\it gap 291040--291073\,MHz} &  &  &&  &  &  &  &  &  \\
40&&TiO$_2$ & 20(7,13)--20(6,14) &291163.3404&  &25.3&3.7&66.5&1.5&0.16& \nodata &  \\
41&&AlCl & 20--19 &291309.36&  &25.6&5.7&54.5&2.4&0.17& \nodata & b \\
42&&SO$_2$ $\nu_2$=1 & 17(3,15)--17(2,16) &291752.929&  &8.2&--20.5&35.1&1.7&0.20& \nodata &  \\
\multirow{2}*{43}&\multirow{2}*{$\{$} &$^{30}$SiO $v$=2 & 7--6 &292505.3264& \multirow{2}*{$\}$} & \multirow{2}*{3.4} & \multirow{2}*{--33.1} & \multirow{2}*{52.6} & \multirow{2}*{28.4} & \multirow{2}*{0.98} & \multirow{2}*{26} &  \\
                              &&NaCl $v$=3 & 23--22 &292512.2299&  &  &  &  &  &  &  &  \\
44&&Na$^{37}$Cl & 23--22 &292756.8289&  &21.5&0.2&41.9&5.7&0.51&14&  \\
45&&SO$_2$ & 13(6,8)--14(5,9) &292882.697&  &26.5&1.1&54.4&2.2&0.18& \nodata &  \\
46&&TiO$_2$ & 19(7,13)--19(6,14) &293001.563&  &16.4&--1.2&37.2&2.7&0.24& \nodata &  \\
\multicolumn{3}{l}{\it gap 293041--293073\,MHz} &  &  &  &  &  &  &  &  &  \\
47&&H$_2$O $\nu_2$=1   & 6(6,1)--7(5,2) &293664.442&  &21.4&--6.1&52.6&31.0&1.53&18& d  \\
48&&CS  & 6--5 &293912.244&  &16.7&--22.7&59.4&16.3&0.43&51& e  \\
49&&SO $v$=1 & 6(7)--5(6) &294047.815&  &21.0&7.5&31.7&2.1&0.26& \nodata &  \\
50&&$^{30}$SiO $v$=1 & 7--6 &294539.5615&  &20.5&--1.8&43.0&31.9&1.39&25&  \\
\multirow{2}*{51}&\multirow{2}*{$\{$} &TiO$_2$ & 17(7,11)--17(6,12) &294696.6722& \multirow{2}*{$\}$} & \multirow{2}*{9.0} & \multirow{2}*{--6.6} & \multirow{2}*{26.1} & \multirow{2}*{4.0} & \multirow{2}*{0.27} & \multirow{2}*{17} &  \\
                              &&NaCl $v$=2 & 23--22 &294709.2058&  &  &  &  &  &  &  &  \\
\nodata &&{\it Instr.294986}&   \nodata &\nodata&  & \nodata & \nodata & \nodata & \nodata & \nodata & \nodata & a \\
\nodata &&{\it Instr.295023}&   \nodata &\nodata&  & \nodata & \nodata & \nodata & \nodata & \nodata & \nodata & a \\
\multicolumn{3}{l}{\it gap 295041--295073\,MHz} &  &  &  &  &  &  &  &  &  \\
52&&TiO$_2$ & 16(7,9)--16(6,10) &295303.3501&  &24.1&2.9&45.8&2.4&0.15&30&  \\
53&&$^{34}$SO & 7(7)--6(6) &295396.334&  &9.5&--50.0&96.5&25.1&0.54&66& f \\
54&&SiO $v$=4? & 7--6 &295488.952&  &14.0&--10.1&38.2&3.2&0.19&\nodata&  \\
55&&$^{29}$SiO $v$=2 & 7--6 &295976.7626&  &11.9&--3.5&27.6&1.6&0.19& \nodata &  \\
56&&SO$_2$ & 26(2,24)--26(1,25) &296168.6746&  &29.4&--11.0&62.1&11.0&0.45&44&  \\
\multirow{2}*{57}&\multirow{2}*{$\{$} &SO & 6(7)--5(6) &296550.064& \multirow{2}*{$\}$} & \multirow{2}*{7.5} & \multirow{2}*{--55.7} & \multirow{2}*{85.8} & \multirow{2}*{335.9} & \multirow{2}*{6.13} & \multirow{2}*{58} & \multirow{2}*{g} \\
                               &&$^{30}$SiO & 7--6 &296575.7401&  &  &  &  &  &  &  &  \\
58&&NaCl $v$=1 & 23--22 &296920.3737&  &21.5&6.7&36.3&5.1&0.32&18&  \\
\multicolumn{3}{l}{\it gap 297041--297073\,MHz} &  &  &  &  &  &  &  &  &  \\
59&&H$_2$O $\nu_2$=1& 6(6,0)--7(5,3) &297439.107&  &20.7&4.4&46.3&5.3&0.38&9&  \\
\multirow{2}*{60}&\multirow{2}*{$\{$} &$^{30}$SiS & 17--16 &297572.2031& \multirow{2}*{$\}$} & \multirow{2}*{12.4} & \multirow{2}*{--44.6} & \multirow{2}*{87.5} & \multirow{2}*{22.3} & \multirow{2}*{0.62} & \multirow{2}*{52} & \multirow{2}*{d} \\
                               &&SiO $v$=3 & 7--6 &297595.4518&  &  &  &  &  &  &  &  \\
61&&$^{29}$SiO $v$=1 & 7--6 &298047.6249&  &22.0&--7.1&47.1&315.7&30.06&13&  \\
62&&$^{34}$SO & 8(7)--7(6) &298257.982&  &25.3&--13.7&79.9&10.8&0.44&33&  \\
\multirow{2}*{63}&\multirow{2}*{$\{$} &TiO$_2$ & 21(2,20)--20(1,19) &298517.2541& \multirow{2}*{$\}$} & \multirow{2}*{--14.2} & \multirow{2}*{--81.5} & \multirow{2}*{57.5} & \multirow{2}*{18.1} & \multirow{2}*{0.75} & \multirow{2}*{80} &  \\
                               &&SO$_2$ & 9(2,8)--8(1,7) &298576.3068&  &  &  &  &  &  &  &  \\
64&&SO $v$=1 & 7(7)--6(6) &298878.699&  &20.1&--1.6&44.7&5.5&0.34&18&  \\
\multicolumn{3}{l}{\it gap 299041--299073\,MHz} &  &  &  &  &  &  &  &  &  \\
65&&NaCl & 23--22 &299145.7002&  &23.6&1.7&52.7&11.4&0.62&19&  \\
66&&$^{34}$SO? & 1(2)--1(0) &299181.8161&  &9.5&--0.1&23.5&0.4&0.11&\nodata&  \\
67&&SO$_2$ & 19(3,17)--19(2,18) &299316.8185&  &35.1&--13.7&81.2&12.2&0.36&55&  \\
68&&SiO $v$=2 & 7--6 &299703.909&  &21.2&--11.3&60.7&88.1&6.40&10& h \\
\multirow{2}*{69}&\multirow{2}*{$\{$} &Si$^{34}$S & 17--16 &300070.4386& \multirow{2}*{$\}$} & \multirow{2}*{--25.7} & \multirow{2}*{--72.3} & \multirow{2}*{40.5} & \multirow{2}*{318.0} & \multirow{2}*{7.90} & \multirow{2}*{40} &  \\
                               &&$^{29}$SiO & 7--6 &300120.477&  &  &  &  &  &  &  &  \\
70&&SO$_2$ & 32(3,29)--32(2,30) &300273.4184&  &25.7&--9.9&67.2&8.3&0.51&33&  \\
71&&H$_2$S & 3(3,0)--3(2,1) &300505.56&  &29.1&--15.1&83.0&83.3&1.89&55&  \\
\multirow{2}*{72}&\multirow{2}*{$\{$} &TiO$_2$? & 36(5,31)--36(4,32) &300574.1727& \multirow{2}*{$\}$} & \multirow{2}*{--19.1} & \multirow{2}*{--57.0} & \multirow{2}*{16.3} & \multirow{2}*{4.7} & \multirow{2}*{0.28} & \multirow{2}*{55} &  \\
                              &&TiO$_2$? & 10(2,8)--9(1,9) &300625.113&  &  &  &  &  &  &  &  \\
\multicolumn{3}{l}{\it gap 301041--301073\,MHz} &  &  &  &  &  &  &  &  &  \\
73&&SO & 7(7)--6(6) &301286.124&  &24.1&--19.5&79.0&147.7&4.42&44&  \\
74&&SO $v$=1 & 8(7)--7(6) &301733.518&  &17.8&--7.5&38.1&7.4&0.58&14&  \\
75&&SiO $v$=1 & 7--6 &301814.332&  &37.6&--12.8&64.3&869.1&45.91&12&  \\
76&&SO$_2$ & 19(2,18)--19(1,19) &301896.6287&  &20.2&--14.1&59.4&5.4&0.48&22&  \\
77&&$^{29}$SiS & 17--16 &302849.4789&  &19.9&--7.0&46.0&9.9&0.52&20&  \\
\multicolumn{3}{l}{\it gap 303041--303071\,MHz} &  &  &  &  &  &  &  &  &  \\
78&&Na$^{37}$Cl $v$=1 & 24--23 &303196.9895&  &18.5&--7.9&41.0&2.1&0.28&21&  \\
79&&U303413& \nodata & \nodata &  &   \nodata & \nodata  & \nodata  &4.3&0.37&12& i \\
80&&SiO & 7--6 &303926.8092&  &24.4&--37.9&80.8&525.6&11.57&48&  \\
81&&SO & 8(7)--7(6) &304077.844&  &24.7&--16.1&75.4&122.8&3.47&46&  \\
82&&TiO$_2$ & 19(3,17)--18(2,16) &304189.864&  &39.3&--0.8&62.1&3.3&0.37& \nodata &  \\
83&&SO$_2$ $\nu_2$=1 & 24(4,20)--24(3,21) &304474.821&  &19.7&--1.4&53.3&2.7&0.23&29&  \\
\multicolumn{3}{l}{\it gap 305039--305071\,MHz} &  &  &  &  &  &  &  &  &  \\
84&&NaCl $v$=3 & 24--23 &305188.2103&  &34.5&8.3&60.1&2.2&0.34&\nodata&  \\
85&&Na$^{37}$Cl  & 24--23 &305445.0269&  &23.2&--14.1&55.3&7.4&0.42&13&  \\
86&&SiS $v$=2 & 17--16 &305512.4102&  &18.7&--2.7&34.1&2.3&0.20&17&  \\
87&&AlCl & 21--20 &305849.02&  &19.9&4.0&40.9&2.2&0.24&23& b \\
88&&AlO & $N$=8--7 &306200.29&  &29.5& --10.5 &100.1&22.1&0.50&53& b \\
89&&SiS $v$=1 & 17--16 &307014.3489&  &20.2&--2.0&41.4&15.9&0.97&15&  \\
\multicolumn{3}{l}{\it gap 307039--307071\,MHz} &  &  &  &  &  &  &  &  &  \\
90&&SO$_2$ & 38(4,34)--38(3,35) &307185.305&  &26.1&--4.2&78.9&10.8&0.86&47&  \\
91&&U308161& \nodata & \nodata &  & \nodata & \nodata & \nodata &4.2&0.86&20&  \\
92&&SiS & 17--16 &308516.1439&  &21.0&--16.5&63.1&133.2&5.29&20&  \\
\multicolumn{3}{l}{\it gap 309039--309071\,MHz} &  &  &  &  &  &  &  &  &  \\
93&&NaCl $v$=1 & 24--23 &309787.8261&  &22.9&1.3&40.3&5.0&0.44&22&  \\
94&&TiO$_2$ & 22(1,21)--21(2,20) &310554.7347&  &27.5&--3.5&70.6&3.6&0.25&\nodata&  \\
95&&U310641& \nodata & \nodata &  & \nodata & \nodata & \nodata &1.6&0.21&10&  \\
96&&TiO$_2$ & 23(1,23)--22(0,22) &310782.7126&  &29.5&3.9&63.2&4.5&0.34& \nodata &  \\
\multicolumn{3}{l}{\it gap 311040--311071\,MHz} &  &  &  &  &  &  &  &  &  \\
97&&NaCl & 24--23 &312109.9004&  &20.9&--2.5&52.4&10.0&0.68&21&  \\
98&&SO$_2$ & 22(4,18)--22(3,19) &312542.5195&  &39.7&--13.6&111.5&13.9&0.50&79&  \\
\multicolumn{3}{l}{\it gap 313039--313071\,MHz} &  &  &  &  &  &  &  &  &  \\
99&&SO$_2$ & 3(3,1)--2(2,0) &313279.7176&  &27.1&--14.4&77.2&9.4&0.49&\nodata&  \\
100&&SO$_2$ & 17(1,17)--16(0,16) &313660.8516&  &25.3&--42.1&70.6&17.3&0.63&\nodata&  \\
\multicolumn{3}{l}{\it gap 315039--315071\,MHz} &  &  &  &  &  &  &  &  &  \\
101&&U315138& \nodata & \nodata &  & \nodata & \nodata & \nodata &0.8&0.22&\nodata&  \\
\nodata &&{\it Instr.315266}& \nodata & \nodata &  & \nodata & \nodata & \nodata & \nodata & \nodata & \nodata & a \\
102&&Na$^{37}$Cl $v$=1 & 25--24 &315786.4642&  &18.8&--6.3&34.0&1.8&0.20&8&  \\
103&&SO$_2$ & 21(3,19)--21(2,20) &316098.8737&  &32.9&--13.9&80.2&13.7&0.41&57&  \\
104&&U316184& \nodata & \nodata &  & \nodata & \nodata & \nodata &1.2&0.13&17&  \\
105&&TiO & $^3\Delta_1$ $J$=10--9 &316518.994&  &23.2&2.1&55.6&4.6&0.30&11&  \\
\multicolumn{3}{l}{\it gap 317041--317071\,MHz} &  &  &  &  &  &  &  &  &  \\
106&&TiO$_2$ & 30(8,22)--30(7,23) &317120.5331&  &27.9&12.4&45.9&2.0&0.26&\nodata&  \\
107&&Si$^{34}$S & 18--17 &317707.4101&  &20.6&--5.8&49.1&7.8&0.45&15&  \\
\multirow{2}*{108}&\multirow{2}*{$\{$} &NaCl $v$=3 & 25--24 &317858.9303& \multirow{2}*{$\}$} & \multirow{2}*{0.5} & \multirow{2}*{--36.4} & \multirow{2}*{33.3} & \multirow{2}*{4.6} & \multirow{2}*{0.27} & \multirow{2}*{\nodata} &  \\
                                 &&TiO$_2$ & 22(2,20)--21(3,19) &317898.713&  &  &  &  &  &  &  &  \\
109&&Na$^{37}$Cl & 25--24 &318128.072&  &22.5&--1.4&43.1&6.0&0.41&17&  \\
\multicolumn{3}{l}{\it gap 319041--319073\,MHz} &  &  &  &  &  &  &  &  &  \\
110&&SO$_2$ & 40(5,35)--40(4,36) &319277.0535&  &27.6&2.4&60.9&8.3&0.60&27&  \\
111&&U319688& \nodata & \nodata &  & \nodata & \nodata & \nodata &2.8&0.42&8&  \\
112&&TiO & $^3\Delta_2$ $J$=10--9 &320159.347&  &20.8&1.1&36.8&3.5&0.50&14&  \\
113&&NaCl $v$=2 & 25--24 &320246.7249&  &23.0&7.9&42.4&2.3&0.48&13&  \\
114&&AlCl & 22--21 &320384.6038&  &35.5&8.2&51.9&3.8&0.55&18& b \\
\multirow{2}*{115}&\multirow{2}*{$\{$} &Si$^{18}$O $v$=1? & 18--17 &320607.7961& \multirow{2}*{$\}$} & \multirow{2}*{--10.2} & \multirow{2}*{--51.1} & \multirow{2}*{53.8} & \multirow{2}*{12.0} & \multirow{2}*{0.58} & \multirow{2}*{25} & \multirow{2}*{j} \\
                                &&$^{29}$SiS? & 8--7 &320649.6581&  &  &  &  &  &  &  &  \\
\multicolumn{3}{l}{\it gap 321041--321073\,MHz} &  &  &  &  &  &  &  &  &  \\
116&&H$_2$O & 10(2,9)--9(3,6) &321225.64&  &20.8&0.2&50.4&2254.3&851.11&2&  \\
117&&SO$_2$ & 18(0,18)--17(1,17) &321330.1661&  &22.6&--18.4&67.4&31.6&1.72&81&  \\
118&&Si$^{18}$O & 8--7 &322770.135&  &22.1&2.8&52.0&22.0&1.98&43&  \\
\multicolumn{3}{l}{\it gap 323041--323072\,MHz} &  &  &  &  &  &  &  &  &  \\
\multicolumn{3}{l}{\it gap 325041--325073\,MHz} &  &  &  &  &  &  &  &  &  \\
119&&H$_2$O & 5(1,5)--4(2,2) &325152.919&  &20.9&--6.9&56.3&122.0&8.09&41&  \\
\multicolumn{3}{l}{\it gap 327041--327073\,MHz} &  &  &  &  &  &  &  &  &  \\
120&&PO&$^2\Pi_{1/2}\,J$=15/2--13/2,\,$e$ &327221.5&  &24.6&--14.2&67.7&32.8&1.23&24& b\\
121&&PO&$^2\Pi_{1/2}\,J$=15/2--13/2,\,$f$ &327411.93&  &23.8&--8.8&74.2&28.4&1.06&25& b\\
122&&$^{34}$SO & 10(11)--10(10) &327788.88&  &13.1&--22.8&62.0&3.3&0.33&\nodata& \\
123&&TiO$_2$ & 29(8,22)--29(7,23) &327898.2095&  &11.1&--9.9&46.7&3.9&0.30&\nodata&  \\
124&&Na$^{37}$Cl $v$=1 & 26--25 &328370.583&  &21.8&--3.9&52.9&4.9&0.34&14&  \\
125&&PN & 7--6 &328888.006&  &20.0&--6.3&49.0&33.2&1.84&14&  \\
\multicolumn{3}{l}{\it gap 329041--329073\,MHz} &  &  &  &  &  &  &  &  &  \\
126&&SO? & 1(2)--0(1) &329385.477&  &9.6&--14.7&49.4&2.9&0.24&\nodata&  \\
127&&PO & $^2\Pi_{3/2}\,J$=15/2--13/2 &329564.1131&  &19.1&--17.6&60.5&22.5&0.81&30& b\\
\multirow{2}*{128}&\multirow{2}*{$\{$} &SO$_2$? & 32(10,22)--33(9,25) &329645.9196& \multirow{2}*{$\}$} & \multirow{2}*{--12.4} & \multirow{2}*{--63.9} & \multirow{2}*{52.6} & \multirow{2}*{9.8} & \multirow{2}*{0.42} & \multirow{2}*{\nodata} &  \\
                                &&SO$_2$? & 44(5,39)--44(4,40) &329689.2964&  &  &  &  &  &  &  &  \\
\multirow{2}*{129}&\multirow{2}*{$\{$} &NaCl $v$=3 & 26--25 &330524.0848& \multirow{2}*{$\}$} & \multirow{2}*{--32.2} & \multirow{2}*{--80.7} & \multirow{2}*{42.8} & \multirow{2}*{42.5} & \multirow{2}*{0.89} & \multirow{2}*{70} &  \\
                                 &&$^{13}$CO & 3--2 &330587.9653&  &  &  &  &  &  &  &  \\
\nodata &&{\it Instr.330742}& \nodata & \nodata &  & \nodata & \nodata & \nodata & \nodata & \nodata & \nodata & a \\
130&&Na$^{37}$Cl & 26--25 &330805.7492&  &21.3&--0.2&43.7&6.3&0.44&20&  \\
\nodata &&{\it Instr.330852}& \nodata & \nodata &  & \nodata & \nodata & \nodata & \nodata & \nodata & \nodata & a \\
\multicolumn{3}{l}{\it gap 331041--331071\,MHz} &  &  &  &  &  &  &  &  &  \\
131&&H$_2$O $\nu_2$=2 & 3(2,1)--4(1,4) &331123.815&  &19.3&2.1&37.2&4.2&0.39&10&  \\
\multirow{2}*{132}&\multirow{2}*{$\{$} &TiO$_2$ & 27(8,20)--27(7,21) &331211.7639& \multirow{2}*{$\}$} & \multirow{2}*{7.3} & \multirow{2}*{--19.0} & \multirow{2}*{43.7} & \multirow{2}*{3.9} & \multirow{2}*{0.18} & \multirow{2}*{41} &  \\
                               &&TiO$_2$ & 26(8,18)--26(7,19) &331240.1486&  &  &  &  &  &  &  &  \\
133&&TiO$_2$? & 39(8,32)--39(7,33) &331381.8392&  &17.7&--8.3&43.8&1.4&0.19&23&  \\
134&&$^{30}$SiO $v$=3 & 8--7 &331955.4496&  &27.3&--6.1&68.3&1.9&0.19&\nodata&  \\
135&&SO$_2$ & 21(2,20)--21(1,21) &332091.4311&  &35.7&--25.6&102.9&11.5&0.32&52&  \\
\multirow{2}*{136}&\multirow{2}*{$\{$} &SO$_2$ & 4(3,1)--3(2,2) &332505.2415& \multirow{2}*{$\}$} & \multirow{2}*{--2.7} & \multirow{2}*{--57.2} & \multirow{2}*{81.6} & \multirow{2}*{22.1} & \multirow{2}*{0.54} & \multirow{2}*{92} &  \\
                                  &&$^{30}$SiS & 19--18 &332550.3096&  &  &  &  &  &  &  &  \\
137&&NaCl $v$=2 & 26--25 &333007.2742&  &19.8&--2.8&38.8&4.3&0.29&19&  \\
\multicolumn{3}{l}{\it gap 333039--333071\,MHz} &  &  &  &  &  &  &  &  &  \\
138&&$^{34}$SO & 7(8)--6(7) &333900.9827&  &21.8&--14.0&64.2&8.2&0.36&18&  \\
\multirow{2}*{139}&\multirow{2}*{$\{$} &TiO$_2$ & 25(8,18)--25(7,19) &334261.126& \multirow{2}*{$\}$} & \multirow{2}*{--13.3} & \multirow{2}*{--53.2} & \multirow{2}*{33.8} & \multirow{2}*{30.0} & \multirow{2}*{0.96} & \multirow{2}*{29} &  \\
                                  &&$^{30}$SiO $v$=2 & 8--7 &334278.1299&  &  &  &  &  &  &  &  \\
140&&SO$_2$ & 8(2,6)--7(1,7) &334673.3526&  &40.4&--15.3&82.4&8.6&0.34&\nodata&  \\
141&&AlCl & 23--22 &334916.82&  &19.0&2.2&36.5&3.6&0.23&19& b \\
\multicolumn{3}{l}{\it gap 335039--335071\,MHz} &  &  &  &  &  &  &  &  &  \\
142&&SO$_2$ $\nu_2$=1 & 20(4,16)--20(3,17) &335128.5301&  &16.5&6.1&25.5&2.1&0.24&11&  \\
143&&SiO $v$=5 & 8--7 &335281.98&  &22.8&1.4&39.0&6.1&1.38&2&  \\
144&&Si$^{34}$S & 19--18 &335341.951&  &20.5&--6.4&44.7&11.7&0.91&12&  \\
145&&NaCl $v$=1 & 26--25 &335506.5502&  &21.2&--8.5&49.8&11.3&0.62&17&  \\
146&&$^{29}$SiO $v$=3 & 8--7 &335880.6672&  &12.7&--3.2&49.4&88.5&9.75&7&  \\
\multirow{2}*{147}&\multirow{2}*{$\{$} &$^{50}$TiO$_2$?? & 25(1,25)--24(0,24) &336011.8557& \multirow{2}*{$\}$} & \multirow{2}*{37.3} & \multirow{2}*{14.5} & \multirow{2}*{53.4} & \multirow{2}*{1.3} & \multirow{2}*{0.22} & \multirow{2}*{15} &  \\
                                &&$^{50}$TiO$_2$?? & 24(1,23)--23(2,22) &336050.6078&  &  &  &  &  &  &  &  \\
148&&SO$_2$ & 23(3,21)--23(2,22) &336089.2284&  &32.6&--28.3&81.9&20.2&0.59&62&  \\
149&&H$_2$O $\nu_2$=1 & 5(2,3)--6(1,6) &336227.62&  &20.1&--7.2&45.6&45.9&3.08&13&  \\
150&&TiO$_2$ & 24(3,21)--23(4,20) &336469.4033&  &33.9&--3.9&59.9&3.6&0.20&\nodata&  \\
\multirow{2}*{151}&\multirow{2}*{$\{$} &SO & 10(11)--10(10) &336553.8112& \multirow{2}*{$\}$} & \multirow{2}*{--18.7} & \multirow{2}*{--60.1} & \multirow{2}*{18.1} & \multirow{2}*{3234.5} & \multirow{2}*{295.45} & \multirow{2}*{21} & \multirow{2}*{g} \\
                                &&$^{30}$SiO $v$=1 & 8--7 &336602.9763&  &  &  &  &  &  &  &  \\
\multirow{2}*{152}&\multirow{2}*{$\{$} &SO$_2$ $\nu_2$=1 & 20(1,19)--19(2,18) &336760.678& \multirow{2}*{$\}$} & \multirow{2}*{--27.2} & \multirow{2}*{--57.8} & \multirow{2}*{25.2} & \multirow{2}*{7.5} & \multirow{2}*{0.30} & \multirow{2}*{39} & \multirow{2}*{d} \\
                                &&$^{29}$SiS $v$=1 & 19--18 &336814.9542&  &  &  &  &  &  &  &  \\
\multicolumn{3}{l}{\it gap 337039--337071\,MHz} &  &  &  &  &  &  &  &  &  \\
\multirow{2}*{153} &\multirow{2}*{$\{$}&TiO$_2$ & 24(1,23)--23(2,22) &337196.0954& \multirow{2}*{$\}$} & \multirow{2}*{17.8} & \multirow{2}*{--23.3} & \multirow{2}*{61.2} & \multirow{2}*{19.5} & \multirow{2}*{0.55} & \multirow{2}*{38} &  \\
                                 &&TiO$_2$ & 22(8,14)--22(7,15) &337206.1149&  &  &  &  &  &  &  &  \\
154&&C$^{34}$S & 7--6 &337396.459&  &29.9&--3.0&50.4&2.8&0.22& \nodata &  \\
155&&$^{34}$SO & 8(8)--7(7) &337580.1467&  &24.6&--2.7&63.2&12.0&0.44&30&  \\
\multirow{2}*{156}&\multirow{2}*{$\{$} &SiO $v$=4 & 8--7 &337687.29& \multirow{2}*{$\}$} & \multirow{2}*{6.7} & \multirow{2}*{--44.2} & \multirow{2}*{31.7} & \multirow{2}*{4.2} & \multirow{2}*{0.31} & \multirow{2}*{\nodata} &  \\
                                &&TiO$_2$ & 2(8,14)--22(7,15) &337732.9754&  &  &  &  &  &  &  &  \\
\multirow{2}*{157}&\multirow{2}*{$\{$} &SO $v$=1 & 7(8)--6(7) &337885.7083& \multirow{2}*{$\}$} & \multirow{2}*{19.9} & \multirow{2}*{2.4} & \multirow{2}*{40.5} & \multirow{2}*{7.4} & \multirow{2}*{0.48} & \multirow{2}*{15} &  \\
                                 &&$^{34}$SO & 3(3)--3(2) &337892.2466&  &  &  &  &  &  &  &  \\
158&&NaCl & 26--25 &338021.9126&  &21.8&--1.4&43.5&15.8&0.85&20&  \\
159&&U338173& \nodata & \nodata &  & \nodata & \nodata & \nodata &3.4&0.28&23&  \\
\multirow{2}*{160}&\multirow{2}*{$\{$} &$^{29}$SiO $v$=2 & 8--7 &338245.1561& \multirow{2}*{$\}$} & \multirow{2}*{--5.0} & \multirow{2}*{--55.6} & \multirow{2}*{45.6} & \multirow{2}*{34.0} & \multirow{2}*{0.77} & \multirow{2}*{86} &  \\
                                &&SO$_2$ & 18(4,14)--18(3,15) &338305.9931&  &  &  &  &  &  &  &  \\
161&&$^{29}$SiS & 19--18 &338447.36&  &21.6&--5.7&44.9&16.5&0.87&15&  \\
162&&SO$_2$ & 20(1,19)--19(2,18) &338611.8103&  &29.8&--14.4&90.5&24.4&1.35&50&  \\
163&&$^{30}$SiO & 8--7 &338930.0437&  &25.9&--25.3&78.8&437.6&11.32&42&  \\
\multicolumn{3}{l}{\it gap 339039--339071\,MHz} &  &  &  &  &  &  &  &  &  \\
164&&SO & 3(3)--3(2) &339341.459&  &26.4&--0.2&65.7&4.0&0.35&40&  \\
165&&$^{34}$SO & 9(8)--8(7) &339857.2694&  &19.5&--36.1&74.5&14.1&0.49&20&  \\
166&&CN & $N,J$=3,5/2--2,3/2 &340033.4841&  &31.7&6&65.2&18&0.83&51& b, k \\
167&&SiO $v$=3 & 8--7 &340094.734&  &21.6&--2.4&38.7&11.8&1.44&14&  \\
168&&CN & $N,J$=3,7/2--2,5/2 &340247.952&  &13.5& --29.3 &41.1&21.8&0.76&54& b, k \\
\multirow{2}*{169}&\multirow{2}*{$\{$} &SO$_2$    & 28(2,26)--28(1,27) &340316.4059& \multirow{2}*{$\}$} & \multirow{2}*{33.5} & \multirow{2}*{--4.8} & \multirow{2}*{68.6} & \multirow{2}*{10.1} & \multirow{2}*{0.41} & \multirow{2}*{51} &  \\
                                &&SO $v$=1? & 3(3)--3(2)         &340323.39&  & &  &  & &  & &  \\
170&&TiO$_2$ & 19(8,12)--19(7,13) &340367.5649&  &14.7&--2.9&38.4&3.6&0.26&\nodata&  \\
171&&$^{29}$SiO $v$=1 & 8--7 &340611.8623&  &20.6&--9.7&53.5&83.6&2.60&36&  \\
172&&SO & 7(8)--6(7) &340714.155&  &24.1&--20.2&77.1&138.4&4.04&31&  \\
\multirow{2}*{173}&\multirow{2}*{$\{$} &Na$^{37}$Cl $v$=1 & 27--26 &340949.1317& \multirow{2}*{$\}$} & \multirow{2}*{19.3} & \multirow{2}*{5.2} & \multirow{2}*{35.1} & \multirow{2}*{4.5} & \multirow{2}*{0.39} & \multirow{2}*{20} &  \\
                                &&TiO$_2$ & 18(8,10)--18(7,11) &340952.6081&  &  &  &  &  &  &  &  \\
\multicolumn{3}{l}{\it gap 341039--341071\,MHz} &  &  &  &  &  &  &  &  &  \\
174&&SO$_2$? & 21(8,14)--22(7,15) &341275.5244&  &--16.8&--34.5&5.1&9.6&0.54&21&  \\
\multirow{2}*{175}&\multirow{2}*{$\{$} &SO$_2$ & 40(4,36)--40(3,37) &341403.0685& \multirow{2}*{$\}$} & \multirow{2}*{8.4} & \multirow{2}*{--42.4} & \multirow{2}*{55.3} & \multirow{2}*{10.7} & \multirow{2}*{0.43} & \multirow{2}*{39} &  \multirow{2}*{l} \\
                                 &&TiO$_2$ & 17(8,10)--17(7,11) &341463.1062&  &  &  &  &  &  &  &  \\
176&&SO $v$=1 & 8(8)--7(7) &341559.3721&  &18.8&2.1&37.1&3.8&0.50&8&  \\
177&&SO$_2$ & 36(5,31)--36(4,32) &341673.9613&  &21.4&--13.1&49.4&7.9&0.54&15&  \\
\multirow{2}*{178}&\multirow{2}*{$\{$} &TiO$_2$ & 15(8,8)--15(7,9) &342217.332& \multirow{2}*{$\}$} & \multirow{2}*{33.6} & \multirow{2}*{5.7} & \multirow{2}*{77.8} & \multirow{2}*{2.3} & \multirow{2}*{0.32} & \multirow{2}*{\nodata} &  \\
                                &&$^{34}$SO$_2$? & 20(1,19)--19(2,18) &342231.63&  &  &  & & & & &\\
179&&SO$_2$ $\nu_2$=1 & 23(3,21)--23(2,22) &342435.9316&  &15.3&--0.9&26.6&2.0&0.29& \nodata &  \\
180&&SiO $v$=2 & 8--7 &342504.383&  &19.7&--3.5&39.9&28.1&1.74&16& d \\
181&&U342638& \nodata & \nodata &  & \nodata & \nodata & \nodata &0.4&0.45& \nodata &  \\
182&&SO$_2$ & 34(3,31)--34(2,32) &342761.6254&  &21.0&6.5&59.8&6.7&0.52&13&  \\
183&&CS & 7--6 &342882.8503&  &22.7&1.1&59.1&16.3&0.75&35& d \\
184&&$^{29}$SiO & 8--7 &342980.8425&  &23.1&--22.9&73.5&343.9&9.28&35&  \\
\multicolumn{3}{l}{\it gap 343040--343072\,MHz} &  &  &  &  &  &  &  &  &  \\
185&&SiS $v$=1? & 19--18 &343100.9841&  &21.4&--9.0&44.1&26.8&1.36&18& c \\
186&&NaCl $v$=3 & 27--26 &343183.4501&  &25.0&--1.6&50.9&3.2&0.19&19&  \\
187&&Na$^{37}$Cl & 27--26 &343477.8437&  &21.7&1.4&39.8&8.0&0.48&18&  \\
188&&SO $v$=1 & 9(8)--8(7) &343829.4217&  &19.7&--3.9&43.9&8.0&0.51&15&  \\
189&&SO & 8(8)--7(7) &344310.612&  &24.7&--22.4&82.1&154.4&4.08&46&  \\
190&&AlO & $N$=9--8 &344453.995&  &26& --20.7 &76.9&37.2&0.88&45& b \\
191&&$^{34}$SO$_2$ & 19(1,19)--18(0,18) &344581.0445&  &10.0&--24.2&46.6&4.4&0.19&50&  \\
192&&SiS & 19--18 &344779.481&  &20.9&--28.5&67.3&132.5&4.19&31&  \\
193&&SiO $v$=1 & 8--7 &344916.247&  &20.6&--17.8&59.1&155.4&5.56&26&  \\
\multicolumn{3}{l}{\it gap 345041--345073\,MHz} &  &  &  &  &  &  &  &  &  \\
194&&H$^{13}$CN $\nu_2$=1 & $J$=4--3 $l$=1$e$ &345238.7103&  &18.0&4.1&29.7&1.8&0.32&9&  \\
\multirow{2}*{195}&\multirow{2}*{$\{$} &SO$_2$ & 13(2,12)--12(1,11) &345338.5377& \multirow{2}*{$\}$} & \multirow{2}*{15.7} & \multirow{2}*{--23.8} & \multirow{2}*{86.4} & \multirow{2}*{76.2} & \multirow{2}*{1.76} & \multirow{2}*{50} &  \\
                                &&H$^{13}$CN & 4--3 &345339.7694&  &  &  &  &  &  &  &  \\
\multirow{2}*{196} &\multirow{2}*{$\{$}&NaCl $v$=2 & 27--26 &345762.0205& \multirow{2}*{$\}$} & \multirow{2}*{--3.1} & \multirow{2}*{--65.6} & \multirow{2}*{56.4} & \multirow{2}*{467.8} & \multirow{2}*{8.73} & \multirow{2}*{50} & \multirow{2}*{h} \\
                                 &&CO & 3--2 &345795.9899&  &  &  &  &  &  &  &  \\
\multirow{2}*{197} &\multirow{2}*{$\{$}&AlOH & 11--10 &346155.538& \multirow{2}*{$\}$} & \multirow{2}*{--0.7} & \multirow{2}*{--77.7} & \multirow{2}*{68.5} & \multirow{2}*{18.7} & \multirow{2}*{0.40} & \multirow{2}*{124} &  \\
                                &&NS & $^2\Pi_{1/2}J$=15/2--13/2, $f$ &346220.137&  &  &  &  &  &  &  &  \\
198&&SO$_2$ $\nu_2$=1 & 19(1,19)--18(0,18) &346379.185&  &24.9&9.5&46.5&3.2&0.33&15&  \\
\multirow{2}*{199}&\multirow{2}*{$\{$} &SO$_2$ & 16(4,12)--16(3,13) &346523.8784& \multirow{2}*{$\}$} & \multirow{2}*{21.9} & \multirow{2}*{--20.8} & \multirow{2}*{82.6} & \multirow{2}*{206.1} & \multirow{2}*{5.06} & \multirow{2}*{49} &  \\
                                &&SO & 9(8)--8(7) &346528.481&  &  &  &  &  &  &  &  \\
200&&SO$_2$ & 19(1,19)--18(0,18) &346652.1691&  &29.9&--19.4&72.5&38.4&1.03&70&  \\
201&&H$^{13}$CN $\nu_2$=1 & $J$=4--3 $l$=1$f$ &346956.7886&  &15.8&--20.4&41.2&3.4&0.23&35&  \\
\multicolumn{3}{l}{\it gap 347041--347072\,MHz} &  &  &  &  &  &  &  &  &  \\
202&&SiO & 8--7 &347330.5786&  &22.6&--39.9&80.7&926.6&23.37&36&  \\
203&&TiO$_2$ & 24(2,22)--23(3,21) &347788.1307&  &28.5&--17.1&81.7&5.5&0.32&27&  \\
\multirow{2}*{204}&\multirow{2}*{$\{$} &TiO & $^3\Delta_1$ $J$=11--10 &348159.786& \multirow{2}*{$\}$} & \multirow{2}*{25.4} & \multirow{2}*{--0.6} & \multirow{2}*{82.4} & \multirow{2}*{10.5} & \multirow{2}*{0.53} & \multirow{2}*{34} & \multirow{2}*{} \\
                                 &&$^{34}$SO$_2$ & 19(4,16)--19(3,17) &348117.4691&  &  &  &  &  &  &  &  \\
\multirow{2}*{205}&\multirow{2}*{$\{$} &HN$^{13}$C? & 4--3 &348340.8139& \multirow{2}*{$\}$} & \multirow{2}*{3.7} & \multirow{2}*{--33.2} & \multirow{2}*{37.5} & \multirow{2}*{21.4} & \multirow{2}*{0.86} & \multirow{2}*{21} & \multirow{2}*{g} \\
                                &&NaCl $v$=1 & 27--26 &348357.2992&  &  &  &  &  &  &  &  \\
206&&U348672& \nodata & \nodata &  & \nodata & \nodata & \nodata &3.8&0.21& \nodata &  \\
\multicolumn{3}{l}{\it gap 349041--349073\,MHz} &  &  &  &  &  &  &  &  &  \\
207&&AlCl & 24--23 &349444.2713&  &18.0&--7.1&40.2&5.7&0.47&19& b\\
208&&SO$_2$ & 46(5,41)--46(4,42) &349783.3171&  &18.8&--7.0&36.8&5.3&0.50&17&  \\
209&&$^{30}$SiS & 20--19 &350035.636&  &20.7&--0.9&47.1&12.4&0.75&14&  \\
210&&U350271& \nodata & \nodata &  & \nodata & \nodata & \nodata &0.8&0.22& \nodata &  \\
211&&TiO$_2$ & 26(0,26)--25(1,25) &350398.9509&  &34.2&--3.3&74.9&10.3&0.38&59&  \\
212&&TiO$_2$ & 25(2,24)--24(1,23) &350707.9333&  &27.7&1.8&64.4&7.9&0.35&41&  \\
213&&SO$_2$? & 10(6,4)--11(5,7) &350862.756&  &--18.9&--29.2&--6.6&1.8&0.21&9&  \\
214&&NaCl & 27--26 &350969.2862&  &22.9&2.3&41.4&19.4&1.28&20&  \\
\multicolumn{3}{l}{\it gap 351041--351073\,MHz} &  &  &  &  &  &  &  &  &  \\
\multirow{2}*{215}&\multirow{2}*{$\{$} &SO$_2$ & 5(3,3)--4(2,2) &351257.2233& \multirow{2}*{$\}$} & \multirow{2}*{18.3} & \multirow{2}*{--18.4} & \multirow{2}*{67.4} & \multirow{2}*{13.6} & \multirow{2}*{0.68} & \multirow{2}*{76} &  \\
                                &&Si$^{34}$S $v$=1 & 20--19 &351279.205&  &  &  &  &  &  &  &  \\
216&&SO$_2$ & 14(4,10)--14(3,11) &351873.8732&  &30.2&--19.3&109.3&18.0&0.58&65&  \\
217&&TiO & $^3\Delta_2$ $J$=11--10 &352157.636&  &19.6&3.8&38.6&2.2&0.30& \nodata &  \\
218&&Si$^{34}$S & 20--19 &352973.907&  &20.8&--3.1&47.0&11.7&0.66&18&  \\
\multicolumn{3}{l}{\it gap 353041--353073\,MHz} &  &  &  &  &  &  &  &  &  \\
219&&Na$^{37}$Cl $v$=1?& 28--27 &353521.8958& &24.0&5.3&38.8&1.9&0.30&6& \\
220&&U353574& \nodata & \nodata &  & \nodata & \nodata & \nodata &1.2&0.4&4& a?  \\
\multirow{2}*{221}&\multirow{2}*{$\{$} &HCN $\nu_2$=1? & 4--3 &354460.4346& \multirow{2}*{$\}$} & \multirow{2}*{--19.4} & \multirow{2}*{--71.7} & \multirow{2}*{48.1} & \multirow{2}*{406.6} & \multirow{2}*{9.55} & \multirow{2}*{35} & \multirow{2}*{m} \\
                                &&HCN & 4--3 &354505.4773&  &  &  &  &  &  &  &  \\
222&&SO$_2$ $\nu_2$=1 & 16(4,12)--16(3,13) &354799.9934&  &15.5&4.9&27.6&3.6&0.48&15&  \\
223&&SO$_2$ & 12(4,8)--12(3,9) &355045.5169&  &48.2&21.3&75.8&10.1&0.54&21& c \\
\enddata
\tablecomments{Instrument artifacts marked in column~(2) as {\it Instr.} and the approximate frequency of the feature in MHz (see also comment a) were not designated in column~(1). Unidentified features are marked in column~(2) as U and the approximate observed central frequency of the feature in MHz. The centroid velocity (column~5) and velocity range of blended features (column~6) is estimated for the lowest rest (laboratory) frequency specified in column~(4). The flux (column~7) was determined by numerically integrating the line profile over the velocity range in column~(6). Max (column~8) is the maximum flux density within the line profile and given per 0.8\,km\,s$^{-1}$ bin.}
\tablenotetext{a}{Instrumental artifact.}
\tablenotetext{b}{Frequency of the center of a simulated profile with multiple hyperfine components -- see Sects.\,\ref{sect-PO-PN} and \ref{sect-al}.}
\tablenotetext{c}{In/Near a gap.}
\tablenotetext{d}{Contribution from TiO$_2$ expected.}
\tablenotetext{e}{Contribution from $^{29}$SiO possible.}
\tablenotetext{f}{Weak contribution from SO expected.}
\tablenotetext{g}{Weak contribution from SO$_2$ expected.}
\tablenotetext{h}{Possible weak contribution from NS.}
\tablenotetext{i}{This may be a line of vibrationally excited H$_2$O -- see Sect.\,\ref{Sect-water} for a detailed discussion.}
\tablenotetext{j}{Weak contribution from $^{34}$SO$_2$ expected.}
\tablenotetext{k}{This feature is only seen in the outer nebula.  The measurements were obtained from a spectrum extracted with a square aperture of 4\arcsec$\times$4\arcsec. }
\tablenotetext{l}{Contribution from SiS $v$=2?}
\tablenotetext{m}{Weak contribution from $^{29}$SiS.}
\end{deluxetable}
\begin{deluxetable}{lcp{12cm}} 
\tabletypesize{\scriptsize}
\tablecaption{Summary of molecules detected in the SMA survey of VY\,CMa\label{Tab-summary}}
\tablewidth{0pt}
\tablehead{
  \multicolumn{1}{l}{Species}
& \multicolumn{1}{c}{No. of}   
& \multicolumn{1}{c}{Line No. in column (1) of Table\,\ref{Tab-main}}\\ 
 \multicolumn{1}{l}{} & 
 \multicolumn{1}{c}{lines} &\\
}
\startdata
AlCl      & 5  &  41 87 114 141 207   \\      
AlO       & 2  &  88 190    \\        
AlOH      & 2  &  12 197   \\          
CN        & 2  &  166 167    \\
CO        & 1  &  196     \\
$^{13}$CO & 1  &  129   \\
CS        & 2  &  48 183      \\              
C$^{34}$S & 1  &  154        \\     
SO        &17  &  29 49 57 64 73 74 81 126 151 157 164 169 172 176 188 189 199   \\     
$^{34}$SO & 9  &  37 53 62 66 122 138 155 157 165  \\          
SO$_2$    &50  &  4 5 6 8 9 11 13 26 30 34 42 45 56 63 67 70 76 83 90 98  
                  99 100 103 110 117 128 135 136 140 142 148  152 160 
                  162 169 174 175 177 179 182 195 198 199 200 208 213 215  
                  216 222 223 \\                                       
$^{34}$SO$_2$&3&  178 191 204 \\
H$_2$S    & 1  &  71        \\ 
TiO       & 7  &  22 33 39 105 112 204 217    \\     
TiO$_2$   & 33 &  14 18 20 21 23 27 31 38 40 46 51 52 63 72 82 94 96 106   
                  108 123 132 133 139 150 153 156 170 173 175 178 203 211 212\\ 
$^{50}$TiO$_2$&1& 147       \\          
H$_{2}$O  & 6  &  47 59 116 119 131 149 \\ 
NaCl      & 21 &  1 7 19 28 43 51 58 65 84 93 97 108 113 129 137 145 158     
                  186 196 205 214 \\         
Na$^{37}$Cl&12 &  2 37 44 78 85 102 109 124 130 173 187 220    \\         
PN        & 2  &  7 125        \\     
PO        & 6  &  14 16 25 120 121 127 \\      
SiO       & 11 &  54 60 68 75 80 143 156 167 180 193 202 \\          
$^{29}$SiO& 7  &  55 61 69 146 160 171 184 \\        
$^{30}$SiO& 7  &  43 50 57 134 139 151 163   \\                     
Si$^{18}$O& 4  &  3 10 115 118       \\           
SiS       & 9  &  32 35 36 86 89 92 175l 185 192 \\        
$^{29}$SiS& 6  &  15 24 77 115 152 161    \\                  
$^{30}$SiS& 4  &  2 60 136 209     \\                  
Si$^{34}$S& 7  &  10 69 107 144 215 218 221m    \\             
HCN       & 1  &  221      \\
H$^{13}$CN& 3  &  194 195 201  \\
HN$^{13}$C& 1  &  205    \\
NS        &  1 &  197    \\           
U         & 12 &  17 79 91 95 101 104 111 159 181 206 210 219   \\          
\enddata
\end{deluxetable}
\clearpage
\subsection{Molecular species \label{Sect-species}}
\subsubsection{SO$_2$ \label{mole-SO2}} 

\begin{figure}[h!]\centering
\includegraphics[angle=270, width=0.45\textwidth]{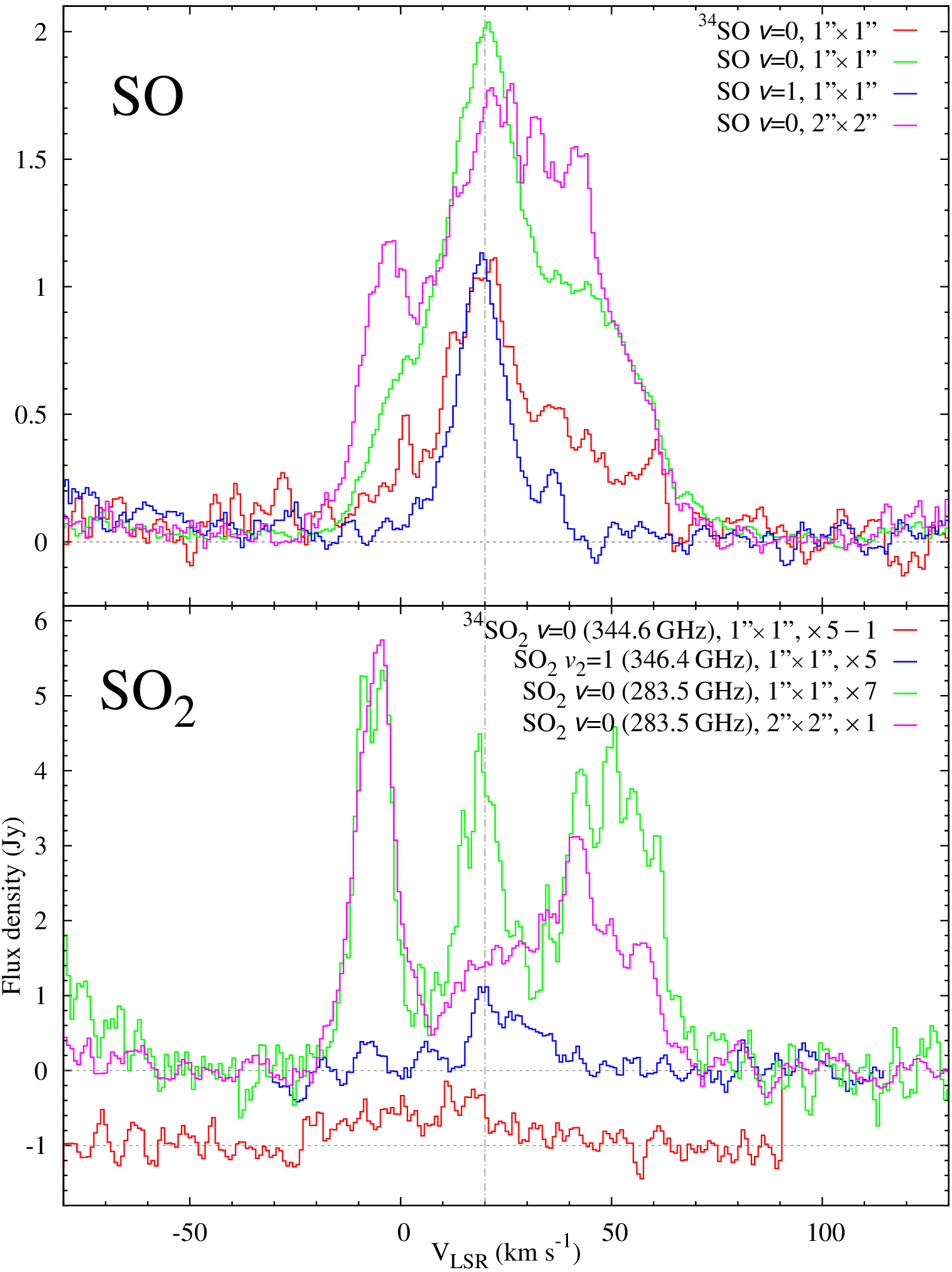}
\caption{Representative profiles of SO and SO$_2$. 
{\bf Top:} Profiles of SO in the ground and first excited vibrational level, and $^{34}$SO in the ground state observed with a $1\arcsec \times 1\arcsec$ region.  Also shown is a profile of SO in the ground  vibrational state for a $2\arcsec \times 2\arcsec$ aperture. The spectra of SO were obtained by combining several transitions in the specified aperture and vibrational level. For clarity, spectra  are scaled with different factors. 
{\bf Bottom:} Profiles of the SO$_2$ $16_{0,16}-15_{1,15}$ transition in $v=0$, SO$_2$ $19_{1,19}-18_{0,18}$ in $\nu_2$=1, and $^{34}$SO$_2$ $19_{1,19}-18_{0,18}$ in the ground vibrational state observed in a $1\arcsec \times 1\arcsec$ region. A spectrum of SO$_2$ $16_{0,16}-15_{1,15}$ in $v=0$ is also shown for an aperture of $2\arcsec \times 2\arcsec$. Spectra were scaled as indicated in the legend and the spectrum of $^{34}$SO$_2$ was additionally shifted in intensity.
(See the electronic edition of the Journal for a color version of this figure.)
\label{Fig-SO-spectra}}
\end{figure}

Sulfur dioxide is known to be omnipresent in the (sub-)millimeter spectrum of VY\,CMa \citep[e.g.,][]{sahai_sulfur,tenen_survey}. In the SMA survey, SO$_2$ is the molecule with the highest number of observed transitions (50), which constitute nearly a quarter of all observed features. At a resolution higher than a few arcsec, the SO$_2$ emission was shown to be extended \citep{muller,fu}. Line profiles observed with an angular resolution of a few arcsec or lower (typical of most single-antenna observations), consist of a red- and a blue-shifted components with respect to the stellar velocity. 
This led \citet{ziurys_nat} and \citet{tenen_survey} to conclude that SO$_2$ is present only in the red- and blue-shifted outflows of VY\,CMa. Our observations show that there is a significant component centered close to the stellar velocity, when SO$_2$ is observed at a higher angular resolution of 0\farcs9 (see Fig.\,\ref{Fig-SO-spectra}). Our interferometric observations also show that the central, red-, and blue-shifted components do not correspond to separate spatial components (as was assumed in the earlier studies), but rather arise from the superposition of emission from several spatially  separated regions (Sect.\,\ref{morph}, Fig.\,\ref{Fig-HST}).          

To aid in the identification of lines of SO$_2$ in the SMA survey, and to avoid misidentifying lines of other species, rotational temperature diagrams for the red-, blue-shifted, and central components of the line profile were constructed for about 33 unblended lines of SO$_2$ from levels between 28 and 1071\,K above ground (Appendix\,\ref{sect-RDintro}). From the derived rotational temperatures and column densities of the main species ($^{32}$S$^{16}$O$_2$; Table\,\ref{Tab-RD-results}), it was concluded that only a few transitions of $^{34}$SO$_2$ are intense enough to be observed with the present sensitivity, and the abundances of the rarer sulfur--33, oxygen--18, and oxygen--17 isotopic species are too low to produce lines observable for us. As expected, only a few lines in $^{34}$SO$_2$ and a couple in the $\nu_2$ excited vibrational level (745\,K above ground) were tentatively assigned in spectra extracted with the $1\arcsec\times1\arcsec$ aperture. Prior to this work, a tentative detection of a line of SO$_2$ in the $\nu_2=1$ level was reported in a spectrum obtained with the Plateau de Bure Interferometer (PdBI) at 222.4\,GHz \citep{kami_tio}.

\subsubsection{SO\label{mole-SO}}

Among 17 identified lines of sulfur monoxide, five lines in the ground vibrational level ($v=0$) are very prominent and  observed at a very high S/N ($\lesssim150$).  Six lines in $v=1$ and nine lines of $^{34}$SO in $v=0$ were also observed. The emission of SO in the ground vibrational state is extended by 2\arcsec--3\arcsec, nearly as much as that of SO$_2$. Several features identified in the HST images are seen in the maps of SO $v=0$ emission (Fig.\,\ref{Fig-morph}), including the SW, W, and NW\,Arcs (Sect.\,\ref{morph}). The line profiles in the $v = 0$ state extend over the same velocity range as that of SO$_2$, but the velocity components are not as well resolved as those in SO$_2$. Shown in Fig.\,\ref{Fig-SO-spectra} are average profiles of SO in the $v=0$ state extracted from regions of $1\arcsec \times 1\arcsec$ and $2\arcsec \times 2\arcsec$. The emission of the rarer $^{34}$SO species at $v=0$ and the main isotopic species in $v=1$ are confined to a central source with a deconvolved size of 0\farcs6 (FWHM) or less. The $^{34}$SO profiles  are narrow (FWHM of 13--18\,\kms; Fig.\ref{Fig-SO-spectra}), and exhibit a ``pedestal'' that is nearly as broad as profiles of SO at $v=0$ when averaged over the same region. The main peak of the SO $v=1$ emission is at 19.2$\pm$0.4\,\kms, while that of $^{34}$SO is at 20.7$\pm$0.5\,\kms. 

\subsubsection{SiS}

Covered in the survey are three rotational transitions of SiS in the ground and three in each of the two lowest excited vibrational levels ($v = 1$ and 2). All nine transitions in the main isotopic species ($^{28}$Si$^{32}$S) were observed: lines in the $v=0$ and 1 levels are fairly intense (3--6\,Jy and $\gtrsim 1$\,Jy at a resolution of 0.8\,km\,s$^{-1}$), while those in $v=2$ are about five times less intense than in $v=1$. Also present in the survey are several lines of the rare isotopic species in $v=0$: five of Si$^{34}$S, three of $^{29}$SiS, and four in $^{30}$SiS. In addition, in $v=1$ one line of Si$^{34}$S was detected and two of $^{29}$SiS were tentatively identified. 

\begin{figure}[h!]\centering\includegraphics[angle=270, width=0.6\textwidth]{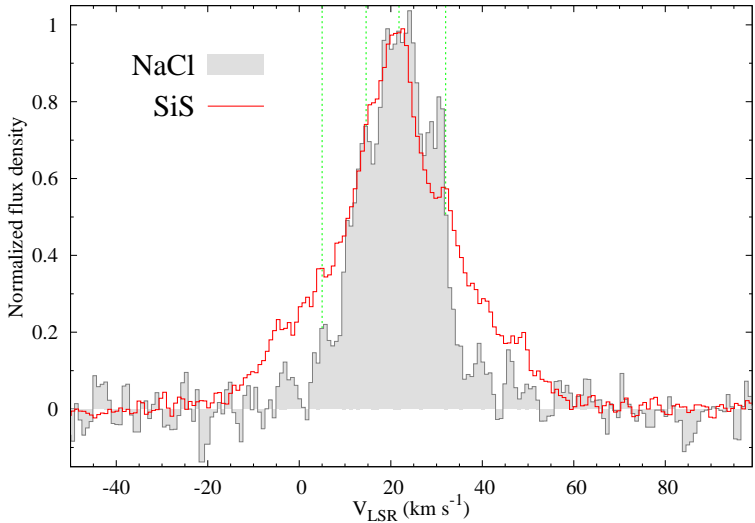}
\caption{Profiles of SiS and NaCl obtained by averaging and normalizing several spectra of the two species in the ground vibrational state. The dashed (green) vertical lines at 5.0, 14.6, 21.8, and 32.0\,km\,s$^{-1}$ mark the velocity components that appear in both profiles. (See the electronic edition of the Journal for a color version of this figure.)\label{Fig-SiS-NaCl}}
\end{figure}

The SiS lines are centered at $V_{\rm{LSR}} = 21.0 \pm 0.5$\,km\,s$^{-1}$ and are well reproduced by a  Lorentzian  profile.  Superposed on the profiles is a ``substructure'' with extra discrete components at about $V_{\rm{LSR}} =15$, 32, and 48\,km\,s$^{-1}$, which correspond very well to the substructure in the average profiles of SiS and NaCl in $v=0$. As Fig.\,\ref{Fig-SiS-NaCl} shows, there is some correspondence between the substructure in the average profiles of SiS and NaCl at $v=0$. 
The FWHM of the rotational lines of the main species in $v=0$ is about  22\,km\,s$^{-1}$, or about two times greater than the  lines in the excited vibrational levels and the rare isotopic species.  


The spatial distribution of  the main isotopic species consists of an intense central component and a second source in the SW direction (SW\,Clump) whose flux is about five times lower than that of the central component. Maps of all the rotational transitions in the ground vibrational state of the main isotopic species have a  structure that is best described as an elongated or cometary-like shape with a $\sim$1\arcsec\ tail directed toward the southwest. Maps of the emission in the excited vibrational levels and that of the rare isotopic species (in $v=0$) do not have this extra spatial component, possibly owing to the lower S/N. 

In the carbon-rich star IRC$+10\degr216$, several transitions from the rotational spectrum of SiS in the ground vibrational state consists of intense emission lines, with fluxes of a few hundreds Jy, which is interpreted as maser emission \citep{SiSmasers}. In oxygen-rich AGB stars, (sub-)millimeter emission of SiS in the ground vibrational state is much less intense (i.e., comparable to that observed in VY\,CMa),  thereby ruling out strong maser emission in these sources. However, in AGB stars the population of rotational levels is known to be influenced by non-LTE effects, e.g., radiative excitation from $v=0$ to $v=1$ by the rovibrational band at 13\,$\mu$m \citep{agb-sis}. Similar effects are expected in VY\,CMa  whose infrared flux is particularly intense. Our observation of emission from the $v=2$ level may indicate that the first rovibrational overtone at 6.7\,$\mu$m is also important in the excitation of SiS. 


\subsubsection{H$_2$S and CS} 

The spectra of H$_2$S and CS are very sparse, with only one intense line of H$_2$S and two of CS observed in the survey, all in $v=0$. Emission from both molecules is localized in the central region and the SW\,Clump. The H$_2$S emission from the unresolved central source consists of several velocity components (with the major peaks near 12, 32, and 56\,\kms), while the seven times weaker emission arising in the SW\,Clump is centered near 22\,\kms\ and has a rectangular profile with FWHM of 
15\,\kms. The column density of H$_2$S in the central region is two times higher than that of SO (Appendix\,\ref{sect-RDintro}), implying that H$_2$S is an important carrier of sulfur in the circumstellar material of VY\,CMa. The high observed column density of H$_2$S, and much lower angular extent than SO and SO$_2$, establishes important constraints on chemical models for this source.

In addition to the main isotopic species, one line of C$^{34}$S was also observed in the central region.  
Lines of CS in the two spatial components have comparable intensity, the profiles from the two regions  are similar, and the FWHM of both are about 60\,\kms.  The derived column density and $T_{\rm{rot}}$ of CS are very similar in the central region and SW\,Clump (Appendix\,\ref{sect-RDintro}).

\subsubsection{PO and PN\label{sect-PO-PN}}

\begin{figure}[h!]\centering
\includegraphics[angle=270, width=0.95\textwidth]{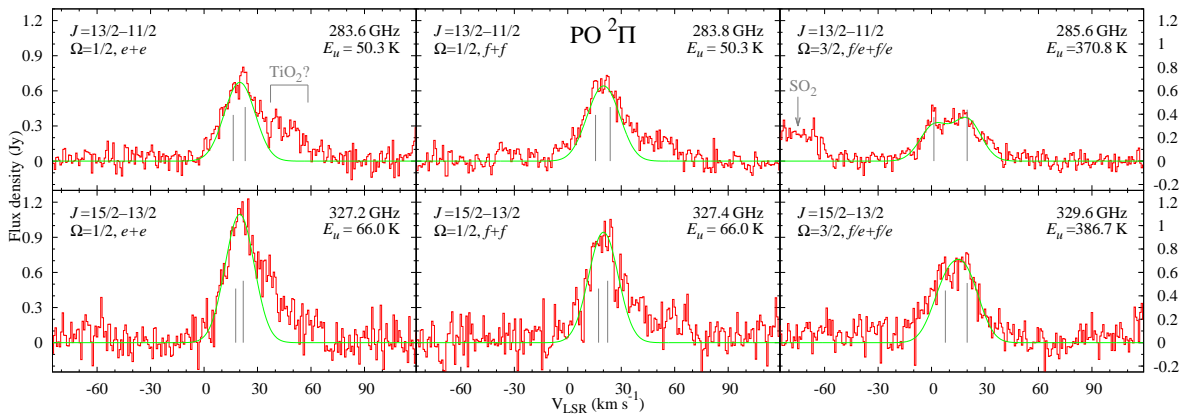}
\caption{Spectra of all transitions of PO covered in the SMA survey (red histograms). Overlayed on the observed lines are  simulated profiles on the assumption that each fine/hyperfine component is a Gaussian (green). The positions of the underlying fine/hyperfine components are marked with vertical lines whose height is proportional to the corresponding line strength. The simulation yielded a central velocity of $V_{\rm LSR}=20$\,\kms. It is apparent that the observed profiles contain an additional red-shifted component which was not accounted for in the simulations. (See the electronic edition of the Journal for a color version of this figure.)
\label{Fig-PO}}
\end{figure}

Two rotational transitions of the phosphorous monoxide radical in the $X^2\Pi$ electronic ground state ($J=$\,6.5--5.5 and 7.5--6.5) were observed with a S/N of 30--40. Each rotational transition consists of two spin components ($e$ and $f$) separated in frequency by about 2.05\,GHz. Transitions in the $^2\Pi_{1/2}$ component are further split by about 191\,MHz ($\sim$200\,km\,s$^{-1}$) owing to lambda doubling, while the two lambda components in $^2\Pi_{3/2}$ are separated by only 26\,MHz ($\sim$20\,km\,s$^{-1}$) and are not resolved. Hyperfine structure of $\sim$7\,km\,s$^{-1}$ is not resolved in either the $^2\Pi_{1/2}$ or $^2\Pi_{3/2}$ components. In all, six spectral features were observed, all shown in Fig.\,\ref{Fig-PO}. One, at 327.2\,GHz, is blended with a line of    SO$_2$ that is only important at angular scales larger than the 1\arcsec\ resolution used here. 
The observed profiles were simulated by combining Gaussian profiles at the frequencies of the fine and hyperfine components, weighted by their relative quantum-mechanical line strengths ($S_{ij}$). 
Shown in Fig.\,\ref{Fig-PO} are the simulated profiles superposed on the observed lines which
allowed us to determine the average central velocity ($V_{\rm LSR}=20.0\pm1.5$\,\kms) and typical FWHM ($18.5\pm1.5$\,\kms). Lines in the  $^2\Pi_{1/2}$ component, all at low values of $E_u$ (50--66~K), have wings in the red which could not be reproduced by the simple simulation, implying that there is an extra kinematical component that is red-shifted by about 25\,\kms\ with respect to the main component.  The maps of PO emission show that the source is compact and concentrated within a region of $0\farcs4\pm0\farcs2$ (deconvolved FWHM). No lines of PO at $v>0$ were detected.


Only two lines of PN ($J=6-5$ and $7-6$, both with good S/N) were observed in the survey, because PN is a closed shell molecule and the hyperfine splitting ($<4$\,MHz) is much less than the intrinsic line width in the source. Although the $J=7-6$ line at 281.9\,GHz is close in frequency to a rotational transition of NaCl in the $v=2$ level, the flux density of the PN line is unaffected. The line positions ($V_{\rm LSR}$=19.0$\pm$0.5\,\kms), widths (FWHM$\approx16$\,\kms), flux densities, and angular distributions of the PN emission are very similar to those of PO in the $^2\Pi_{1/2}$ component. The compact emission of PN has a deconvoled FWHM of $0\farcs5 \pm 0\farcs2$.

\subsubsection{NaCl}

Sodium chloride is a closed shell molecule with a rotational transition every 13\,GHz. Six transitions in the ground vibrational state were covered in the SMA survey, but only five  were observable because the frequency of one is only 90\,MHz lower than the center of the saturated telluric band of water  (see Sect.\,\ref{obs} and Fig.\,\ref{Fig-Full_survey}). Lines of NaCl are fairly intense with a S/N of about 9. Shown in Fig.\,\ref{Fig-SiS-NaCl} is the averaged line profile of the ground vibrational state. 
Symmetrically displaced about the most intense feature at $V_{\rm{LSR}} = 21.8$\,km\,s$^{-1}$, are partially resolved blue- (14.5\,km\,s$^{-1}$) and red-shifted (29.5\,km\,s$^{-1}$) features that are about 25\% less intense than the central component. A similar although less pronounced velocity substructure is also observed in SiS. The NaCl emission is approximately centered on the continuum (but see Sect.\,\ref{Sect-conti}), and is slightly elongated with mean deconvolved FWHMs of $(0\farcs42 \pm 0\farcs 05) \times (0\farcs30 \pm 0\farcs04)$  and a position angle of $-33\degr\pm22\degr$, where the indicated uncertainties are 3$\sigma$ \citep[see Fig.\,5 in][]{kami_tio}.  Weak emission is also seen $\sim$1\arcsec\ southwest from the center of the main emission region in the most intense transition. 

Several rotational transitions were also observed in the three lowest excited vibrational levels ($v =$\,1, 2, and 3). Similarly, lines of the less abundant isotopic species Na$^{37}$Cl were observed in the ground and first excited vibrational level. From these, a vibrational-temperature analysis was done for NaCl (Appendix\,\ref{sect-RDintro}). The high observed vibrational temperature ($T_{\rm{vib}}$) of $960 \pm 120$~K is consistent with the compact emission. 


Following the analysis of rotational lines of vibrationally excited SiO and H$_2$O  in the inner expansion zone of VY\,CMa \citep[][and references therein]{menten_water}, we examined whether the vibrationally excited NaCl we observe is collisionally or radiatively excited. We used the same collision rate coefficients ($\gamma_{ij}$) for NaCl as those calculated for SiO in ground vibrational state at $T = 1000$~K, on the assumption that $\gamma_{ij}$ is similar in both molecules. For the rate coefficients for collision from the vibrational ground to the first vibrationally excited state, we assume following Menten et al., that they are given by the values for excitation within the ground state multiplied by a factor $c_{\rm vg}$; as in Menten et al., we assume $c_{\rm vg} = 0.02$. 
The critical density for the rotational transitions in the $v = 1$ level of NaCl of $\ge 10^{10}$\,cm$^{-3}$ was estimated by comparing $n c_{\rm{vg}}  \gamma_{ij} $ with the Einstein $A$ coefficient of the $v = 1$ level ($A_{\rm{vib}}$), where  $A_{\rm{vib}}$ (2\,s$^{-1}$) was estimated from the change in dipole moment with internuclear distance ($\partial \mu/\partial r = 6.5$\,D/\AA; CDMS) and the force constant ($k =1.1$\,dyn\,cm$^{-1}$; \citealp{matcha}). By analogy with SiO \citep{lamda}, the collision rate is $\gamma_{ij} \sim c_{\rm{vg}}~1 \times 10^{-10}$~cm$^3$~s$^{-1}$. The derived critical density of $\ge 10^{10}$~cm$^{-3}$ implies that the vibrationally excited levels of NaCl that we observe are not excited by collisions for the densities in the inner expansion zone as estimated by \citet{menten_water}. As \citet{carroll} have shown, the criterion for excitation by radiation is $f/(e^{h\nu_{\rm{IR}}/T_{\rm vib}} -1) > A_{\rm{rot}}/A_{\rm{vib}}$, where $f$ is the factor of geometrical dilution of the radiation in the region of interest. For NaCl this criterion is satisfied, because the left hand side (1.3) is greater than $A_{\rm{rot}}/A_{\rm{vib}}$ of 0.01, on the assumption that $f = 1$ and $T_{\rm{vib}} = 960$~K. It therefore appears from our analysis, that the three lowest vibrational levels of NaCl are in equilibrium with the radiation.

\subsubsection{AlO, AlOH, and AlCl\label{sect-al}}

\begin{figure}[h!]\centering
\includegraphics[angle=0, height=0.5\textheight]{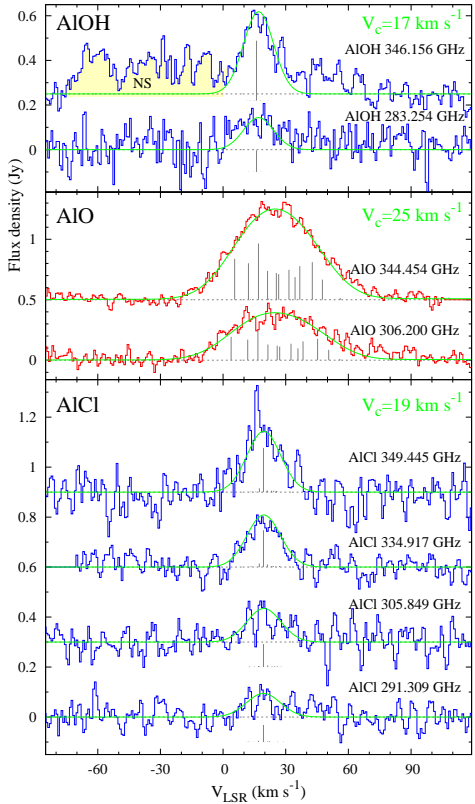}
\caption{Spectra of three aluminum-bearing molecules in VY\,CMa. Shown in the three panels (top to bottom) are the rotational transitions of AlOH, AlO, and AlCl observed in the SMA survey. The positions of the unresolved hyperfine components are indicated by vertical lines below each spectrum with $V_{\rm LSR}=17$\,km\,s$^{-1}$ for AlOH, 25\,km\,s$^{-1}$ for AlO, and 19\,km\,s$^{-1}$ for AlCl. 
Shaded in yellow in the spectrum of AlOH at 346.16\,GHz is the upper lambda component of NS with partially resolved hyperfine structure ($^2\Pi_{1/2}~J = 15/2 - 13/2, f$). The spectra were extracted for a square aperture of $1\arcsec \times 1\arcsec$. (See the electronic edition of the Journal for a color version of this figure.)
\label{Fig-Al}}
\end{figure}

Three aluminum-bearing molecules (AlO, AlOH, and AlCl) were observed in the survey, all in a compact unresolved region. Both AlO and AlOH had been observed before with single antennas \citep[][and references therein]{tenen_survey}, but AlCl, although present in IRC$+10\degr216$, had not been observed previously in VY\,CMa. In addition to the four lines of AlCl shown in Fig.\,\ref{Fig-Al}, inspection of the SMA spectra reported in \citet{fu} reveals that two lines of AlCl ($J =$\,16--15 at 233.12\,GHz, and 17--16 at 247.67\,GHz) were detected at a very modest S/N in the earlier observations.  

Two of the aluminum-bearing molecules here (AlCl and AlOH) are closed shell molecules, with a very small contribution to the observed line width owing to unresolved quadrupole hyperfine structure from Al (nuclear spin $I = 5/2$) and Cl ($I = 3/2$). The AlO radical, however, has a $^2\Sigma$ electronic ground state with more than ten unresolved fine structure, and magnetic and electric quadrupole hyperfine structure components, each separated by several MHz (3--8\,km\,s$^{-1}$).  As a result, the observed rotational lines of AlO in VY\,CMa are much broader than those of AlOH and AlCl. The maximum flux density (0.9\,Jy at 0.8\,\kms\ resolution) and  full linewidth ($\sim$100\,km\,s$^{-1}$) of AlO are 2--3 times greater than those of AlOH and AlCl (both about $\lesssim$0.4\,Jy and full widths $\sim$40\,km\,s$^{-1}$).

Shown in Fig.\,\ref{Fig-Al} are spectra of aluminum-bearing molecules with synthetic profiles overlaid on the observed spectra. The synthetic profiles are the superposition of theoretical Gaussian profiles of each (unresolved) fine/hyperfine component. The velocity scale of the astronomical spectra of the aluminum-bearing molecules were estimated from rest frequencies derived from the peak intensities of the synthetic profiles. The intrinsic line widths derived from the simulations of AlOH (FWHM=17.0$\pm$1.5) and AlCl (FWHM=17.5$\pm1.5$\,\kms) are similar. For AlO, we determined that the width  of the unresolved components is $30 \pm 2$\,km\,s$^{-1}$ FWHM (see Fig.\,\ref{Fig-Al}). Our derived line width is 2--3 times larger than had been estimated from spectra observed at S/N of 3--4 with a single antenna \citep[FWHM=10--15\,km\,s$^{-1}$;][]{TZ09}.
Our profile simulations of the millimeter-wave lines, confirm that the intrinsic width of $30 \pm 2$\,km\,s$^{-1}$ obtained from the SMA spectra is consistent with the spectra in Tenenbaum \& Ziurys, when the much lower S/N and the baseline uncertainties in the single antenna spectra are taken into account.

The intrinsic line width of AlO derived from the SMA spectra is greater than that recently obtained from measurements of optically thin rotational components of the $B^2\Sigma-X^2\Sigma$ band of AlO in the optical \citep[FWHM=13.5$\pm$1.5\,km\,s$^{-1}$,][]{kami_alo}. The excitation of the electronic levels is mostly radiative, whereas the rotational levels in the ground $X^2\Sigma$ electronic state most likely are collisionally excited. Because the optical emission is preferentially located in the  expanding wind closer to the star, the rotational components in the electronic bands have narrower lines than the pure rotational transitions observed with the SMA. 

The simulations allowed us to constrain the central velocities of the emission of aluminum-bearing molecules. It appears that the central LSR velocity of AlOH ($17\pm3$\,km\,s$^{-1}$) and AlCl ($19\pm2$\,km\,s$^{-1}$) are lower than that of AlO ($V_{\rm LSR} = 25\pm2$\,km\,s$^{-1}$) and most other molecules localized in the central unresolved source, but more precise estimates of the $V_{\rm{LSR}}$ of the aluminum-bearing molecules await observations at higher sensitivity. 
 


\subsubsection{H$_2$O\label{Sect-water}}



Six transitions of water were observed in this survey: two in the ground state ($J_{K_{a},K_{c}}=10_{2,9}-9_{3,6}$ at
321.2\,GHz and $5_{1,5}-4_{2,2}$ lines at 325.2\,GHz), three in the $\nu_2 = 1$ vibrational level, and one in $\nu_2 = 2$. 
The H$_2$O emission is not resolved with our 0\farcs9 beam. The deconvolved size derived using the {\it imfit} procedure in Miriad  (FWHM\,$\lesssim0\farcs2$) is consistent with the expectation that the size of the emitting region is comparable to that of the SiO masers ($\sim0\farcs15$; see \citealp{menten_water}). The two strongest vibrationally exited lines are narrow (12--15\,\kms\ FWHM)  and the intense H$_2$O maser at 321.2\,GHz is only a few \kms\ wide.

\citet{menten_water} observed two {\it ortho} lines with the APEX telescope ($5_{2,3} - 6_{1,6}$ near 336.2\,GHz, and $6_{6,1}-7_{5,2}$ near 293.7\,GHz), but could only obtain an upper limit for the flux of the $6_{6,0}-7_{5,3}$ {\it para} transition near 297.4\,GHz. They concluded that the {\it para} line at 336.2\,GHz (2939~K above the ground) might be thermally excited. The lines at 293.7\,GHz and 297.4\,GHz are at identical levels above ground (3920~K) and have the same transition probabilities. 
On the assumption that these two lines are optically thin and are in LTE, their intensity ratio should be equivalent to the {\it ortho}/{\it para} ratio of 3. 
We derive a ratio of peak fluxes of $F_{\rm 293.7\,GHz}/F_{\rm 297.4\,GHz}\approx 11$, which may indicate that the line at 293.7\,GHz is a weak maser. 


We also observed a weak line at 303.42\,GHz that is within $\sim$8\,MHz of a rotational transition assigned to vibrationally  excited  H$_2$O in the JPL catalog. 
However, a recent analysis of vibrationally excited H$_2$O has called into question this assignment. The transition of H$_2$O at 303.417\,GHz that appears in the JPL catalog is a calculated (rather than measured) frequency for a transition between two high-lying rotational levels in two different vibrational levels 9780\,K above ground ($17_{3,15},~\nu_2 - 16_{2,14},~\nu_1$). Because the entries in the JPL were derived from laboratory measurements of H$_2$O from levels  $\le 6470$\,K above ground, the actual uncertainty in the frequency for this transition may be much higher than indicated (B.~Drouin, personal communication). Although we observed a line in VY\,CMa that is within $\sim$8\,MHz of the transition in question, we have designated it as U (unassigned) in Table\,\ref{Tab-main}.
The emission at 303.42\,GHz is very compact, similar to the emission features of water whose  identification is firm.

\subsubsection{SiO \label{Sect-SiO}}

\begin{figure}[h!]
  \centering
  \savebox{\myimage}{
                  \includegraphics[angle=0, width=0.6\textwidth]{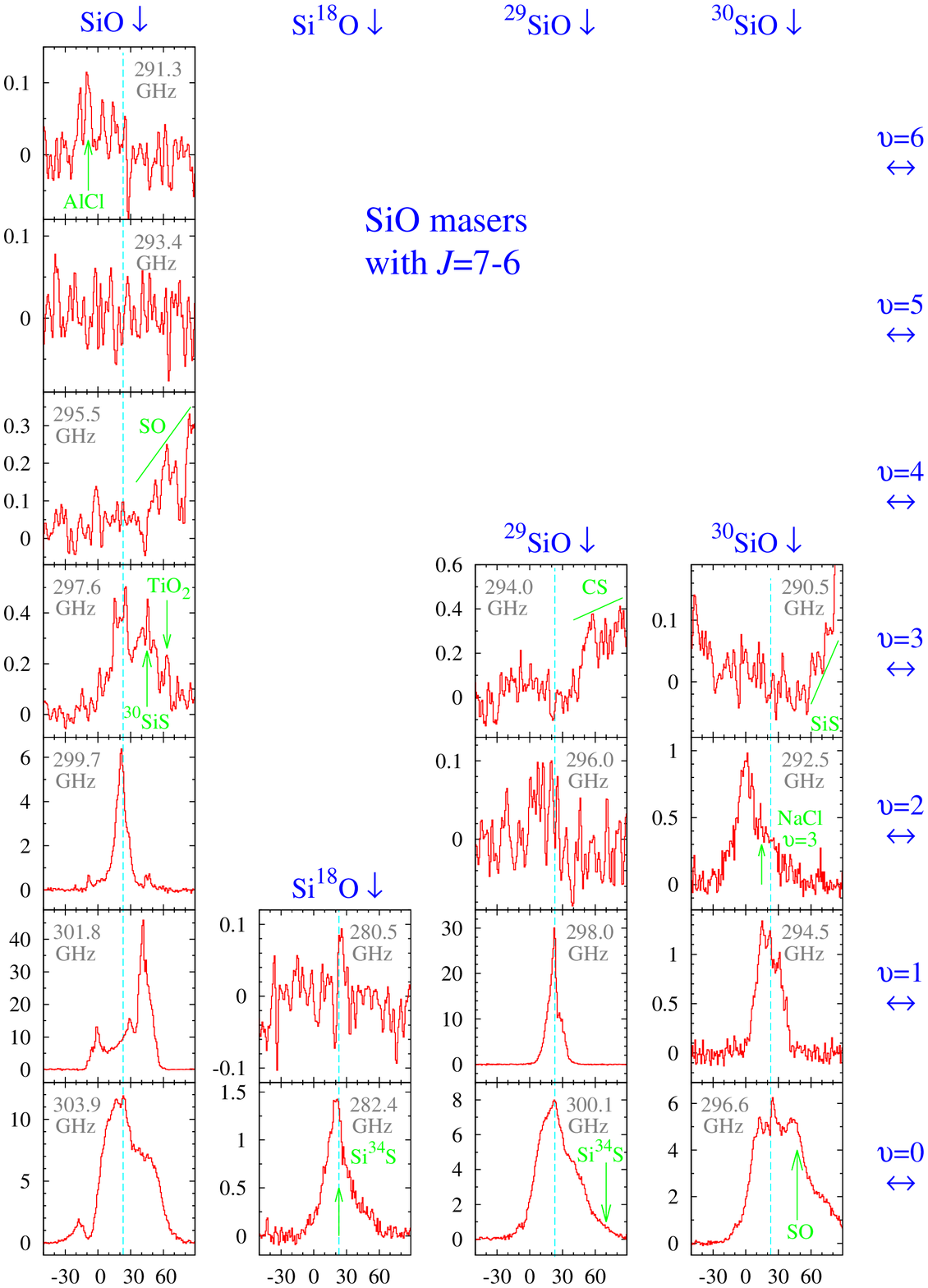}
                  \includegraphics[angle=0, width=0.582\textwidth]{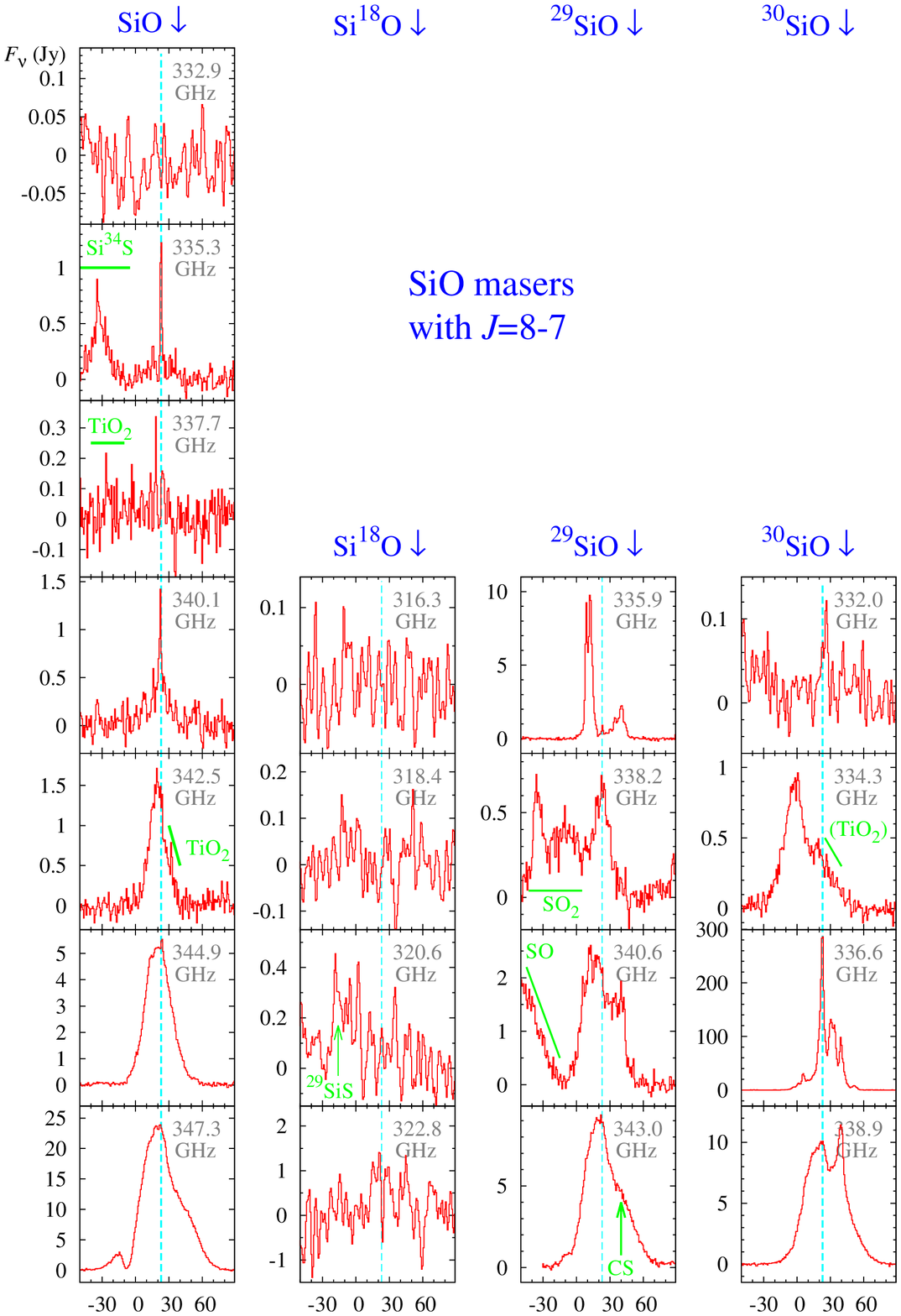}
                    }  
  \rotatebox{90}{
    \begin{minipage}{0.95\textheight}
      \usebox{\myimage}
      \caption{Spectra of SiO and three rare isotopic species in the ground and excited vibrational levels averaged over a $1\arcsec \times 1\arcsec$ region. 
{\bf Four left columns:} the $J = 7-6$ transition.  
{\bf Four right columns:} the $J=8-7$  transition.   
In both the left and right parts, the columns (from left to right) refer to $^{28}$SiO, Si$^{18}$O, $^{29}$SiO, and $^{30}$SiO; and the rows (from bottom to top) refer to $v=0$ to 6. 
Some of the spectra were smoothed. The central line frequency is given in each box (gray), and blends with other molecules are indicated (green). The vertical line refers to $V_{\rm LSR}=23$\,\kms.  The ordinate is flux density in Jy and abscissa is $V_{\rm LSR}$ in \kms. (See the electronic edition of the Journal for a color version of this figure.) \label{Fig-SiO}}
    \end{minipage}}
\end{figure}

Two rotational transitions of the main and three rare isotopic species of SiO were covered in our survey (see Fig.\,\ref{Fig-SiO}). For the main isotopic species, the $J=7-6$ transition was observed in the ground and excited vibrational levels at least up to $v=3$, and the $J=8-7$ transition was observed in vibrational levels as high as $v=5$. For the silicon--29 and silicon--30 species\footnote{
For Si$^{18}$O, $^{29}$SiO, $^{30}$SiO, our analysis is limited to lines at $v \leq 3$ because only those transitions are listed in CDMS. No spectroscopic information on rotational lines of Si$^{17}$O are available to us.}, the $J = 7-6$ transition in $v=0$,\,1,\,2, and $J= 8-7$ transition in $v=0$,$\dots$,3 were observed. 
For Si$^{18}$O, two transitions were covered in $J=7-6$ ($v$=0 and 1), while $v=0$,$\dots$,3 were covered in $J=8-7$, although those in $v$=0 and 1 are in the noisy part of the survey owing to telluric absorption. Only $v=1$ is definitely present in the $J=7-6$ transition in Si$^{18}$O, because the transition in the ground state is blended with an intense line of Si$^{34}$S. The $J=8-7$ line of Si$^{18}$O in the ground state is tentatively detected. Among all the observed features of isotopic species of SiO, the most intense is the $J=8-7$ transition in $v=1$ of $^{30}$SiO, whose flux is $\ge 10$ times higher than that of any other line of SiO (Figs.\,\ref{Fig-Full_survey} and \ref{Fig-SiO}).    

To our knowledge, the observation of SiO in $v=5$ (8732~K above ground) is the first detection of a transition in SiO from such a high level above ground. Prior to this, the highest rotational transition observed in SiO is $J=5-4$ in $v=4$ near 211.1\,GHz \citep[7004~K above ground;][]{cerni_sio}.
Another surprising result is our detection of the $J=7-6$ transition in the $v=1$ level of Si$^{18}$O, because emission in  Si$^{18}$O is rare in astronomical sources \citep[e.g.,][]{sio_search}, and this is the first observation of this rare molecule in VY\,CMa. 

Lines in the ground vibrational state of the main isotopic species have broad complex profiles, possibly owing to self-absorption. The blue-shifted absorption feature at $-8$\,km\,s$^{-1}$ nearly reaches the continuum level, whereas  the corresponding  feature in $^{29}$SiO and $^{30}$SiO is less pronounced, or possibly not present at all. Lines of SiO in $v=0$ are very broad, but the widths decrease with vibrational excitation. The FWHM of SiO in $v = 0$ is 35--50\,km\,s$^{-1}$, with wings that extend by about 60\,km\,s$^{-1}$ from line center. In the vibrationally excited levels, a very narrow feature centered at $V_{\rm{LSR}} = 22.8\pm0.5$\,km\,s$^{-1}$ and a  FWHM of $\sim$1.5\,km\,s$^{-1}$ is superposed on a broad pedestal ($\sim$30\,km\,s$^{-1}$ FWHM) that is increasingly weaker in levels from $v=3$,\,4,\,5 of the $J=8-7$ transition.  

Emission in the ground vibrational level is extended (Fig.\,\ref{Fig-morph}), however the flux is dominated by an intense unresolved component centered close to the continuum peak. The size of the emission region decreases with increasing vibrational number, e.g., the strong emission of $^{29}$SiO(8--7) at $v=3$ appears as a point source with our 0\farcs9 synthesized beam.

\subsubsection{NS and CN}

Several rotational transitions in the $^2\Pi_{1/2}$ and $^2\Pi_{3/2}$ spin components of NS were covered in the survey, but most are blended with intense lines of other species (e.g., SiO, $^{29}$SiO, CO). 
The only useful line of NS ($^2\Pi_{1/2}~J=15/2-13/2, f$ at 346.2~GHz) is partially blended with an  emission feature of AlOH (see Fig.\,\ref{Fig-Al}). This transition is dominated by three hyperfine components whose total splitting is $\sim$1\,\kms\ but also contains two weaker components lower in frequency by about 50\,MHz (43\,\kms), producing a broad but relatively weak spectral feature.
The spatial distribution of NS is limited to the central source and a source 1\arcsec\ SW from it (SW\,Clump). Both regions are of similar intensity. 

Two rotational lines of the CN ($X^2\Sigma$) radical were observed, i.e. the $J=5/2-3/2$ and $7/2-5/2$ spin-rotation components in $N=3-2$ near 340.0  and 340.2\,GHz.  In both lines, there is unresolved hyperfine splitting of a few MHz. Among all features observed in the survey, the emission of CN is distinct owing to its hollow ring-like morphology (see Fig.\,\ref{Fig-morph} and Sect.\,\ref{morph}). 
The spatial distribution of CN in VY\,CMa is similar to that in IRC$+10\degr216$ and other AGB stars \citep{cwleo_CN}, possibly indicating that CN is produced by photo-dissociation of HCN by the interstellar radiation field. 

Lines of CN and CS are not easily discernible in spectra extracted with a small ($\sim$1\arcsec) aperture (including the spectrum shown in Fig.\,\ref{Fig-Taco}), because it does not include a large fraction of the flux. Instead, lines of CN and CS are best observed in spectra averaged over a larger region of the nebula. The measurements of CN in Table\,\ref{Tab-main} were obtained in a spectrum extracted in a 4\arcsec$\times$4\arcsec\ region.

\subsubsection{CO  \label{Sect-CO} }

\begin{figure}[h!]\centering
\includegraphics[angle=270, width=0.3\textwidth]{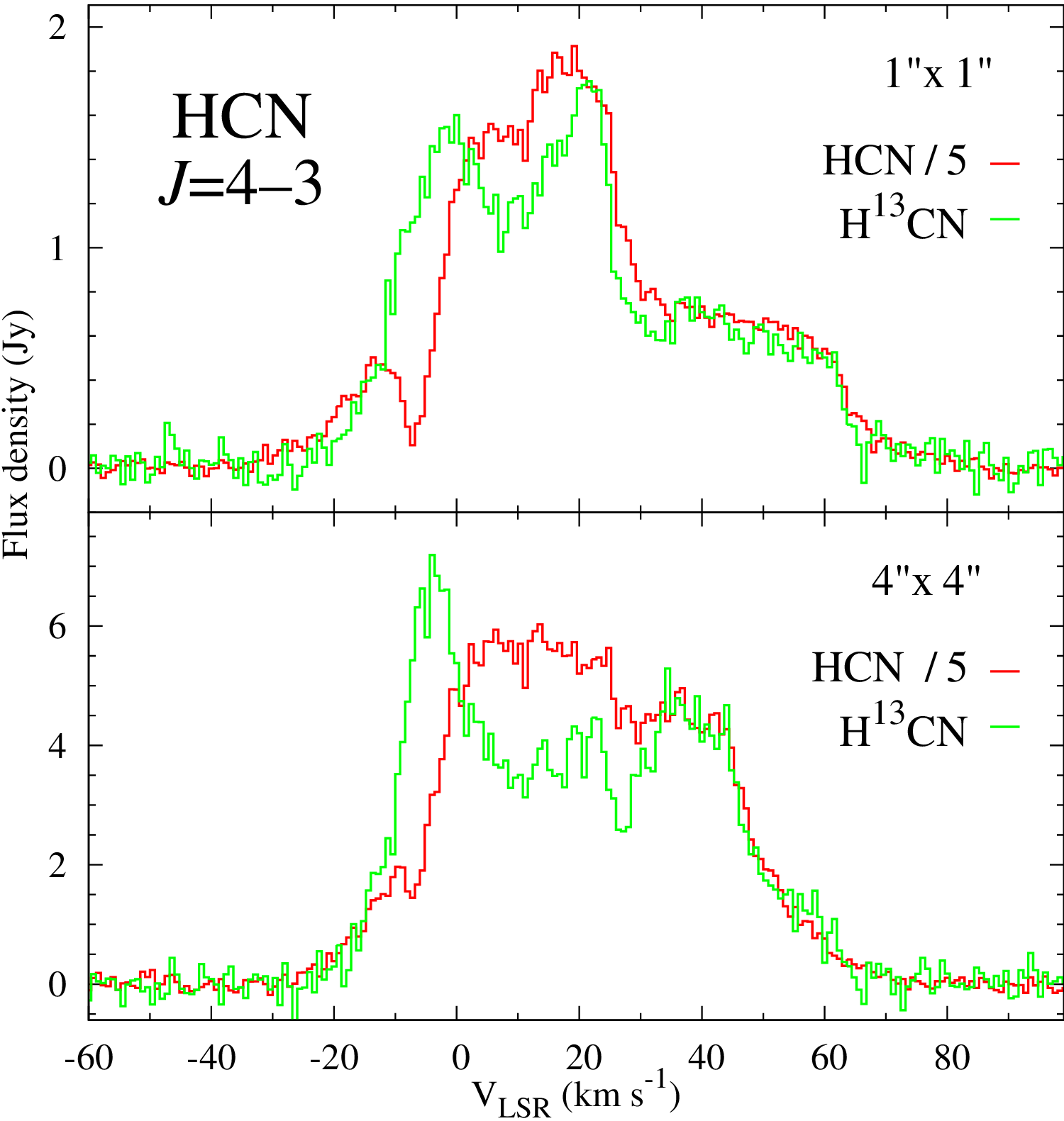}
\includegraphics[angle=270, width=0.3\textwidth]{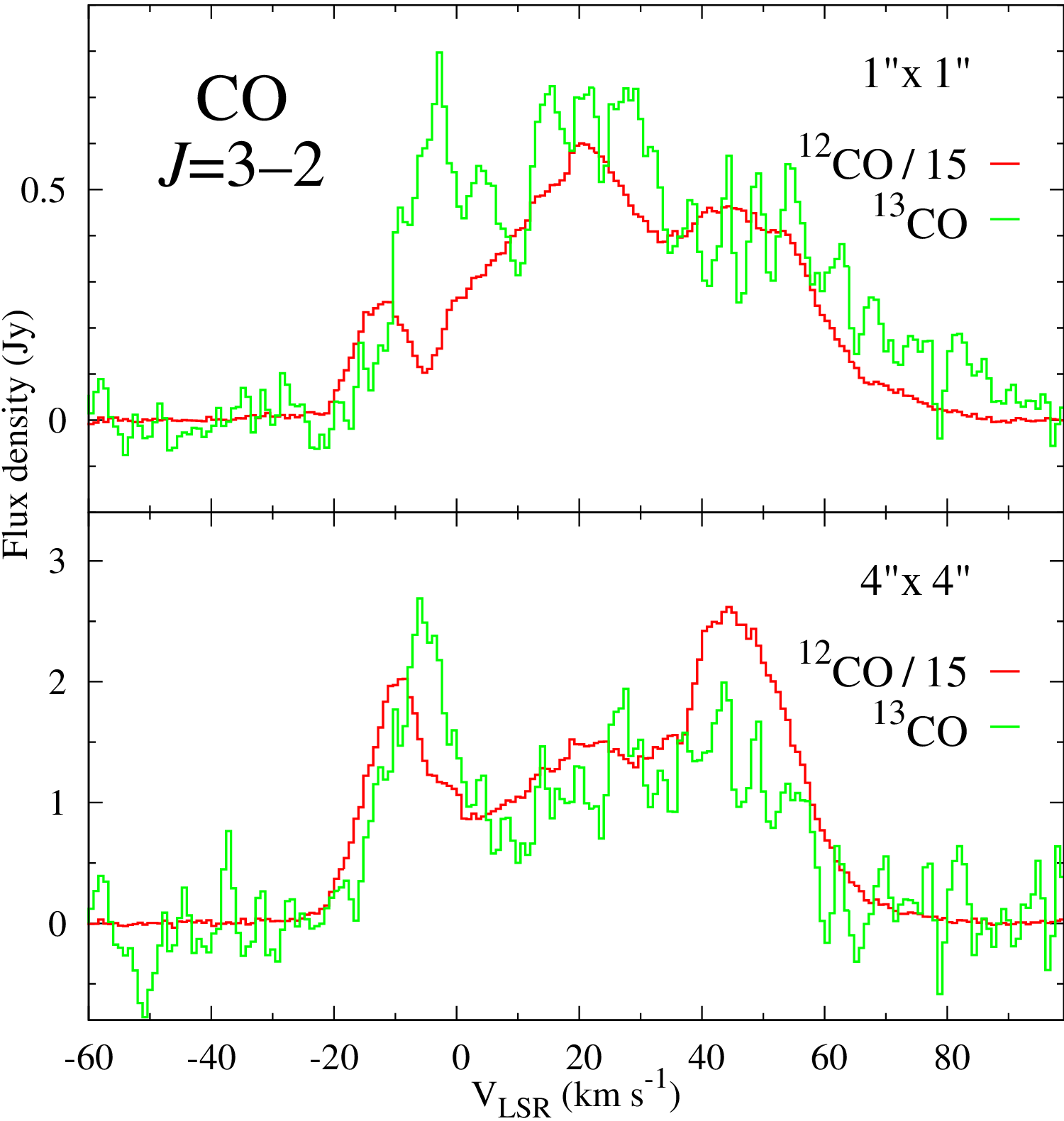}
\caption{Line profiles of HCN and H$^{13}$CN ({\it left}), and  CO and $^{13}$CO ({\it right}), averaged over square apertures of $1\arcsec \times 1\arcsec$ (top) and $4\arcsec \times 4\arcsec$ (bottom). The HCN profile was scaled by 1/5, CO by 1/15, and $^{13}$CO was smoothed. (See the electronic edition of the Journal for a color version of this figure.)
\label{Fig-CO-HCN}}
\end{figure}

The emission of the $J=$3--2 transition of both CO and $^{13}$CO is extended (Fig.\,\ref{Fig-morph}), with CO having the highest extent among all features in the survey (see Figs.\,\ref{Fig-morph} and \ref{Fig-HST}, and Sect.\,\ref{morph}). The spatial distribution of CO appears to be smoother than those of the other molecules, e.g. SO$_2$ and SO at $v=0$, whose angular extents are comparable to that of CO (Fig.\,\ref{Fig-HST}), suggesting that the $3-2$ transition of CO is optically thick. The line profiles of CO and $^{13}$CO have three main velocity components (Fig.\,\ref{Fig-CO-HCN}). For small apertures centered close to the morphological center, the CO profile has an absorption dip near $V_{\rm LSR}=-5$\,\kms, that might be a self-absorption feature. The profile of $^{13}$CO has an emission peak near that velocity (cf. Sect.\,\ref{Sect-HCN}), indicating that optical thickness effects are important for CO. Both lines coincide with transitions of NaCl whose contribution to the CO and $^{13}$CO profiles is negligible. 

\subsubsection{HCN, H$^{13}$CN, and HN$^{13}$C \label{Sect-HCN}}

The rotational transition of HCN in the ground vibrational state ($J=4-3$ at 354.5\,GHz) is one of the strongest features in the survey (Fig.\,\ref{Fig-Full_survey}). Hyperfine structure extending over 3.6\,MHz (3\,\kms) is unresolved in VY\,CMa. The same rotational transition in the vibrationally excited level $\nu_2= 1$ (1067~K above ground), if present, is blended with the intense ground state line, because it is only 45\,MHz lower in frequency and the full width of the ground state line is of $\sim$100\,\kms. The emission of HCN is extended and its morphology is similar to that of SO. As 
for other molecules with complex spatial distributions in VY\,CMa, the HCN line profile has several velocity components and  possible self-absorption at $V_{\rm LSR}=-7$\,\kms\ (see Fig.\,\ref{Fig-CO-HCN}).  

The $J=4-3$ transition of H$^{13}$CN in the ground vibrational state (at 345.34\,GHz), and two features in the $\nu_2=1$ level ($l$-type doublets $e$ and $f$ at 345.24 and 346.96\,GHz) are present in the spectrum of VY\,CMa. The ground state line is very close in frequency to that of a weak high-excitation line of SO$_2$ ($E_u$=1161~K) whose predicted intensity is negligible. The shape of the H$^{13}$CN line  is similar to that of HCN (Fig.\,\ref{Fig-CO-HCN}), but instead of an absorption feature there is an emission feature near --7\,\kms, similarly to what is observed in CO and $^{13}$CO. The $e$ and $f$ components of H$^{13}$CN(4--3) $\nu_2=1$ are relatively weak with a S/N of 9 and 3.5. Their profiles are centered at $V_{\rm LSR}=19\pm1$\,\kms\, and maps of integrated intensity show a compact source centered close to the continuum peak.   

We did not cover HNC in the ground vibrational level, but there is possible evidence for the $J=4-3$ line of the rarer species HN$^{13}$C. Because the frequency of this line nearly coincidences with a fairly intense line of NaCl in the $v=1$ level and a broad line of SO$_2$, the contribution of HN$^{13}$C to the observed feature is uncertain. 

\subsubsection{Molecular ions and recombination lines}
Two positive molecular ions have been identified in spectral surveys of VY\,CMa with single antennas. Protonated CO (HCO$^+$) was observed in the millimeter band \citep{ziurys_carbon}, and protonated water (H$_3$O$^+$) was observed with {\it Herschel} \citep{royer}. Rotational transitions of HCO$^+$ lie outside the frequency band covered here. However, the $J=4-3$ transition of H$^{13}$CO$^+$ was covered (at 347.0\,GHz), but  was not detected at an rms of 0.4\,Jy\,\kms. One transition of H$_3$O$^+$  was also covered, but it nearly coincides with a line of SO$_2$.  At least seven other molecular ions observed in space at radio wavelengths have important transitions in our band. These include H$_2$O$^+$, SH$^+$, SO$^+$, HCS$^+$, CO$^+$, HOCO$^+$, and NNH$^+$.  In addition to these, the HOSO$^+$ ion recently measured in the laboratory may also be a plausible species in VY\,CMa \citep{lattanzi}. We were unable to identify any of the  positive molecular ions referred to here in our survey, nevertheless this band appears well suited for future searches for new molecular ions in this source.

Weak emission in Balmer lines of hydrogen has been reported toward VY\,CMa \citep[e.g.,][]{hump-spec}. The most intense recombination lines covered by the SMA survey are those of H$\alpha$ from $n$=26 to 28. The H26$\alpha$ line was not observed at the stellar position or in the field around it at an rms of 0.6\,Jy\,\kms.

\subsection{Isotopic ratios \label{Sect-iso}}

Very limited information  about isotopic ratios in VY\,CMa was obtained at this time because:  (1) very few isotopic species were observed; (2) intrinsic uncertainties in the derived ratios, owing to possible maser emission in some (e.g., SiO); and (3) in the fairly light species (e.g., CS, CO, HCN), only a limited number of lines were observed. Nevertheless,
 
Constraints on the isotopic abundances of Cl, S, and Si were derived from an analysis of population diagrams of  five molecules (Appendix\,\ref{sect-RDintro}). Because the lines we observed are optically thin, the ratio of the column densities of the different isotopic species provides a direct measure of the elemental ratios (but see below). 
From Na$^{35}$Cl and Na$^{37}$Cl, we obtained $^{35}$Cl/$^{37}$Cl\,$=4 \pm 1$ which is essentially the same as the solar value \citep[3.13;][]{lodders}. 
We also determined the $^{32}$S/$^{34}$S ratio in three molecules: $6\pm3$ and $10\pm3$ from two spatial components of SO$_2$; $14\pm3$ from SO; $18\pm5$ from SiS; and $\sim13.5$ from CS. The weighted mean  ($^{32}$S/$^{34}$S $= 11\pm$2) is about two times less than the solar value \citep[22.6;][]{lodders}. 
Similarly, from the column densities of SiS and $^{29}$SiS, we estimate a $^{28}$Si/$^{29}$Si ratio of $10\pm1$, again about two times lower than the  solar value \citep[19.6;][]{lodders}.  
Because in red supergiants the surface isotopic ratios of elements heavier than Al are not expected to be substantially changed with respect to the cosmic values\footnote{Products of oxygen burning, very unlikely to be manifested in VY\,CMa, would enhance the abundance of $^{28}$Si and $^{32}$S, making the discussed ratios larger than the cosmic/solar values, i.e. would not explain the ratios derived here.}  \citep[e.g.,][]{heger_langer}, our derived isotopic ratios for S and Si are perplexing. The anomalies in the isotopic abundances might indicate that the assumptions used to derive the column densities are not valid for sulfur and silicon-bearing species, but this will be explored in detail elsewhere.

Of particular interest are the elemental abundances of the CNO group, because they may help establish the evolutionary status (initial mass) of VY\,CMa \citep[e.g.,][]{heger_langer}:

\begin{itemize}
\item Because of saturation at high optical depth (and chemical fractionation/selective photo-dissociation), we were unable to obtain reliable $^{12}$C/$^{13}$C ratios of $^{12}$CO and $^{13}$CO, and HCN and H$^{13}$CN. However, ratios of the line intensities may be used to estimate limits of the $^{12}$C/$^{13}$C ratio ($\gtrsim$5 from  HCN and $\gtrsim$15 from CO; Fig.\,\ref{Fig-CO-HCN}), which are consistent with those derived from millimeter-wave observations of CO  in which optical-depth effects were taken into account \citep[$^{12}$C/$^{13}$C=25--46;][]{milam_iso}. Only one of the limits is consistent with that obtained from the fundamental band of CO at 4.6\,$\mu$m \citep[$^{12}$C/$^{13}$C$\lesssim6$;][]{smith_CO}. Moreover, at millimeter wavelengths, a ratio was obtained from the $J=1-0$ transition of H$^{12}$CN and  H$^{13}$CN \citep[36, uncorrected for saturation effects;][]{HCN} that is much higher than the ratio of 15 derived here, suggesting that the optical depth of the submillimeter-wave line of HCN we observe is very high.
\item  We did not detect any nitrogen-15 bearing molecule in the survey, but we can establish an upper limit of the $^{14}$N/$^{15}$N ratio. The $J=$4--3 line of HC$^{15}$N at 344.2\,GHz is at least $\sim$200 times less intense than that of H$^{14}$CN, implying that the isotopic ratio is greater than about 200.  Because this limit is close to the solar ratio \citep[272;][]{lodders}, there is no evidence of enhancement of $^{14}$N, while some enhancement is expected in red supergiants after the first dredge-up \citep[e.g.,][]{heger_langer}. Although the line of HCN is optically thick, it provides the best available constraint on $^{14}$N/$^{15}$N, because it is more intense than that of any other nitrogen-bearing molecule in the survey. 
\item Although  the oxygen-18 isotopic species Si$^{18}$O was detected in the survey (Sect.\,\ref{Sect-SiO}), the emission is most likely affected by strong maser effects and does not allow a determination of the $^{16}$O/$^{18}$O ratio. 
\end{itemize}

\subsection{Spatial and kinematical structure of molecular emission \label{morph}}

\begin{figure*}[h!]\centering
\includegraphics[angle=270,width=0.325\textwidth]{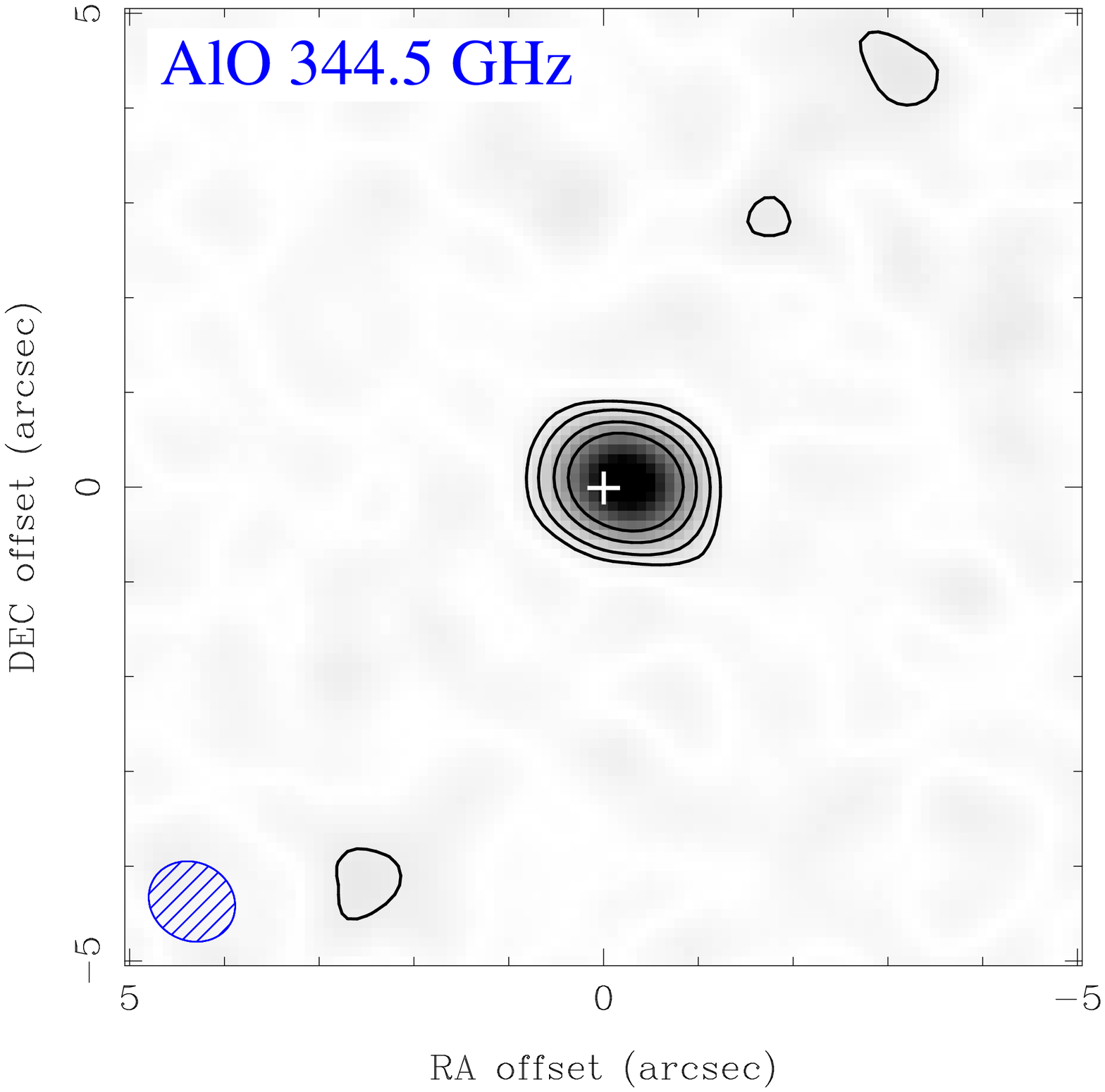}
\includegraphics[angle=270,width=0.325\textwidth]{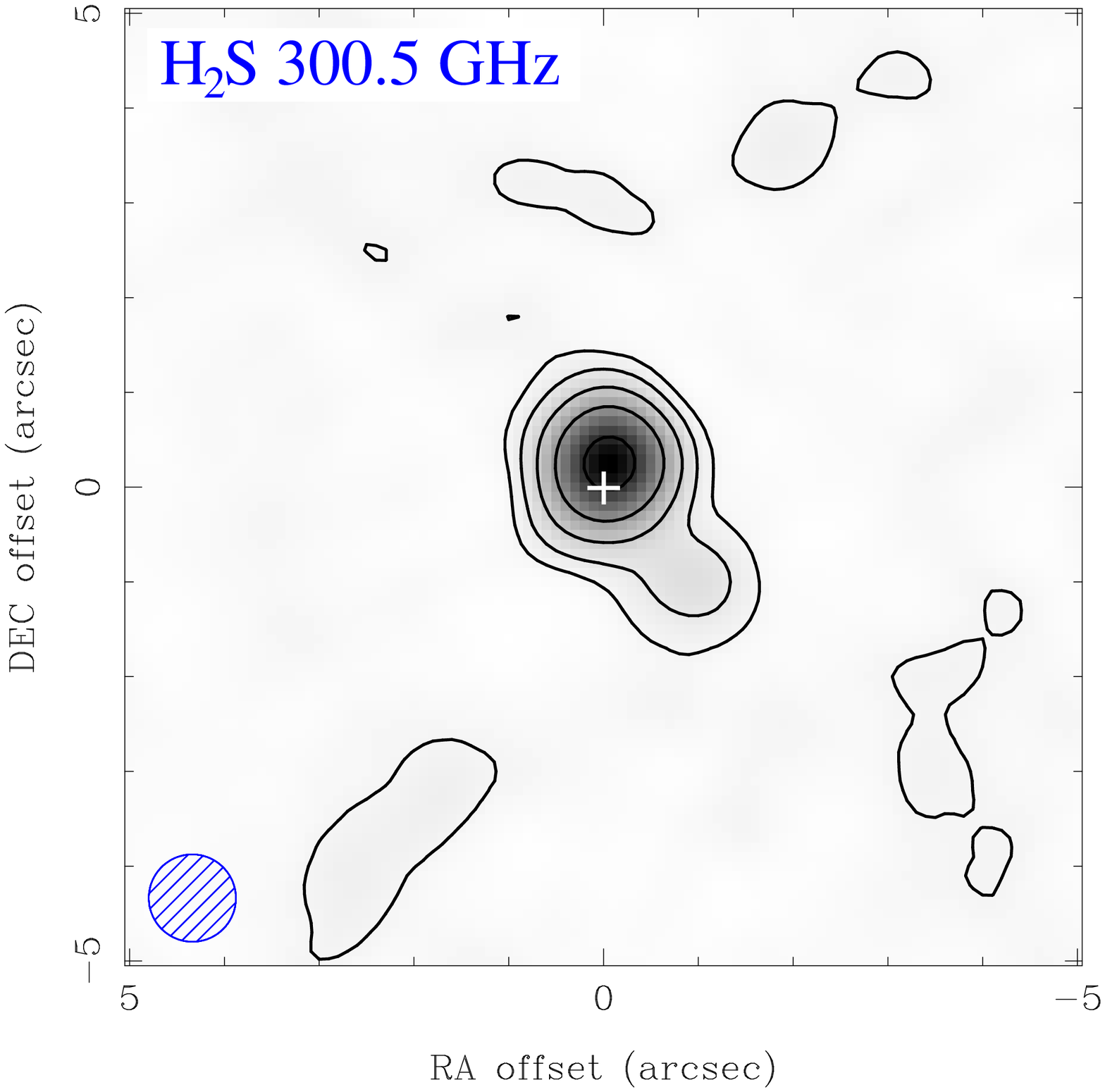}
\includegraphics[angle=270,width=0.325\textwidth]{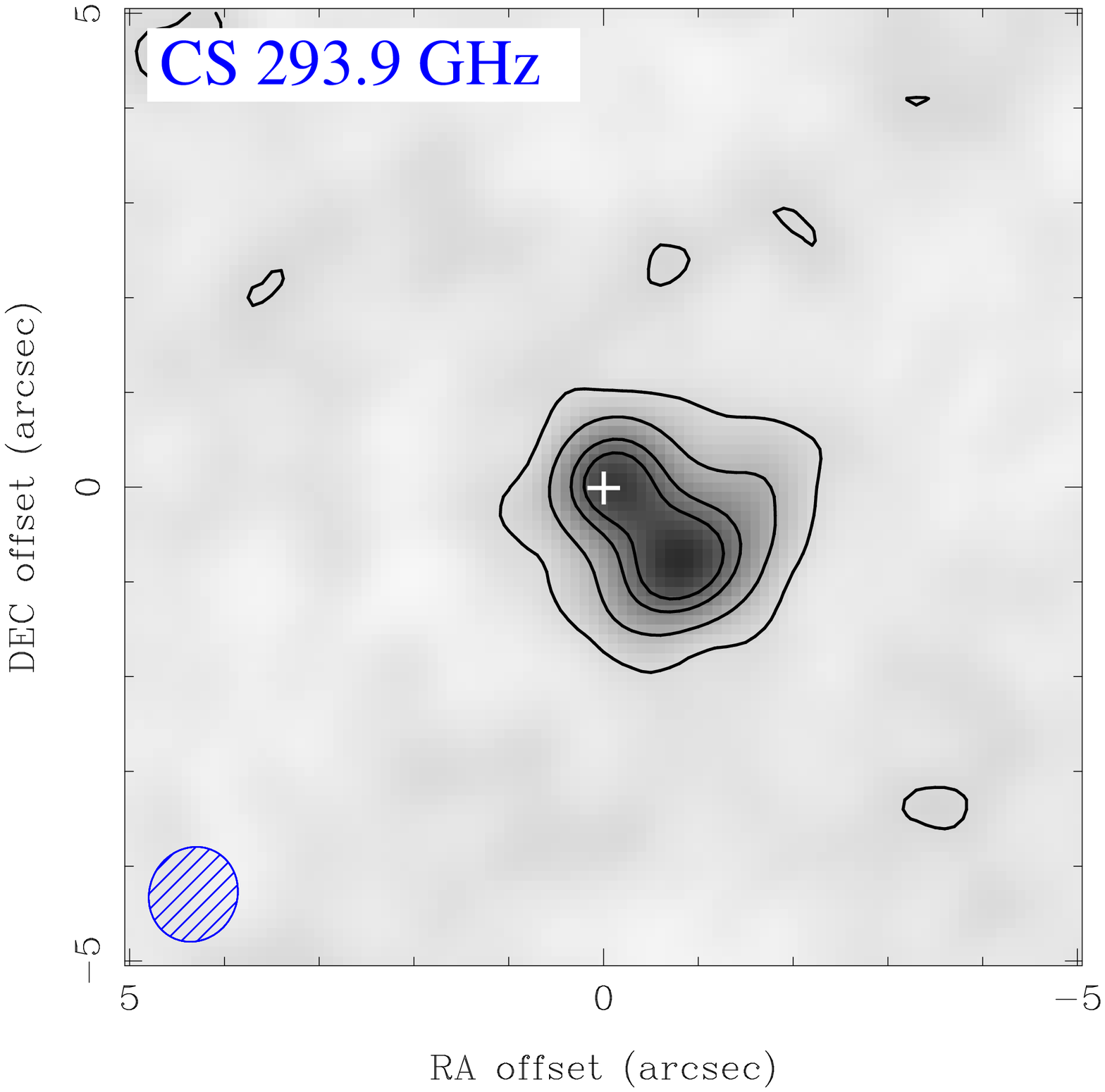}
\\
\includegraphics[angle=270,width=0.325\textwidth]{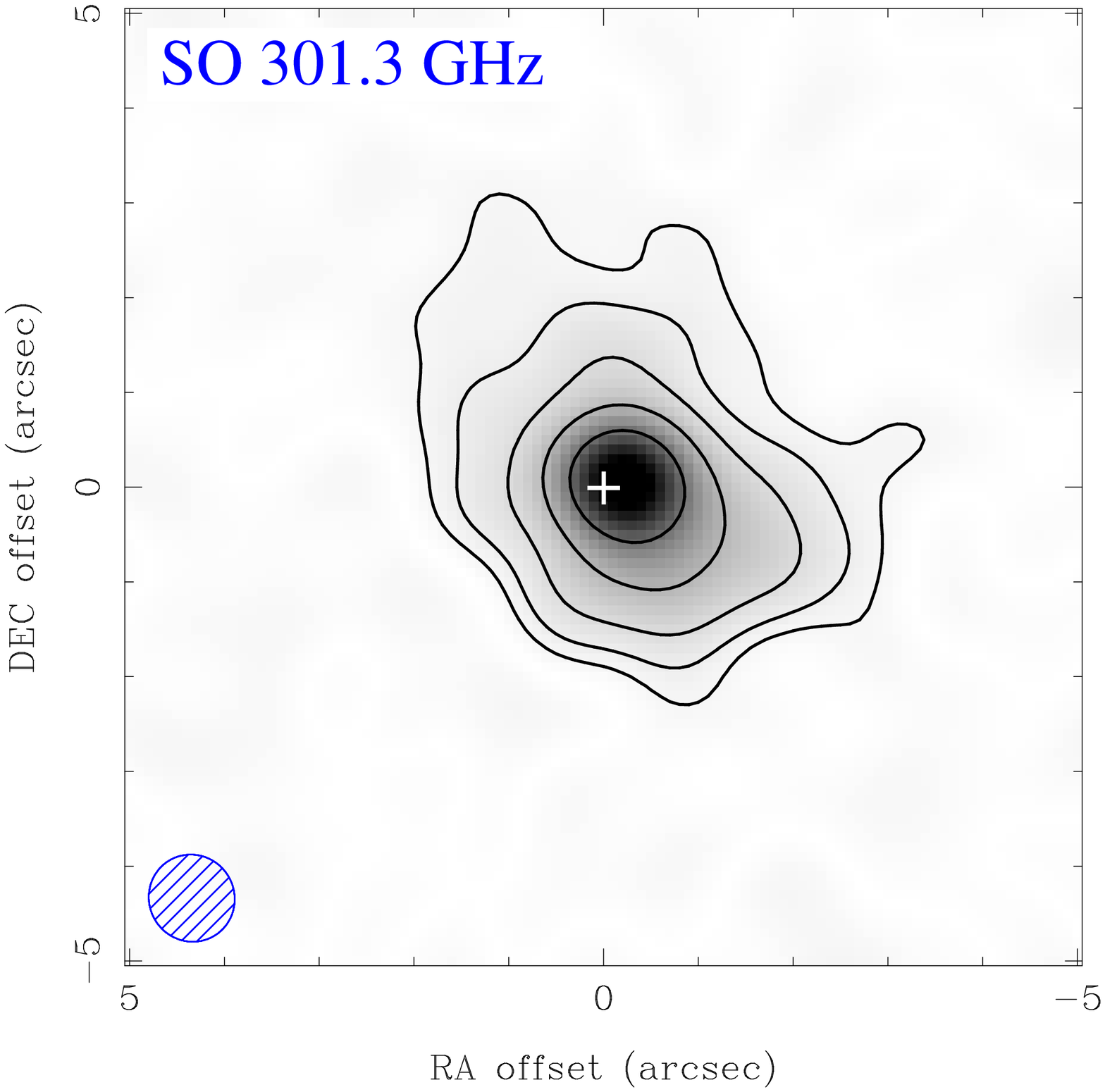}
\includegraphics[angle=270,width=0.325\textwidth]{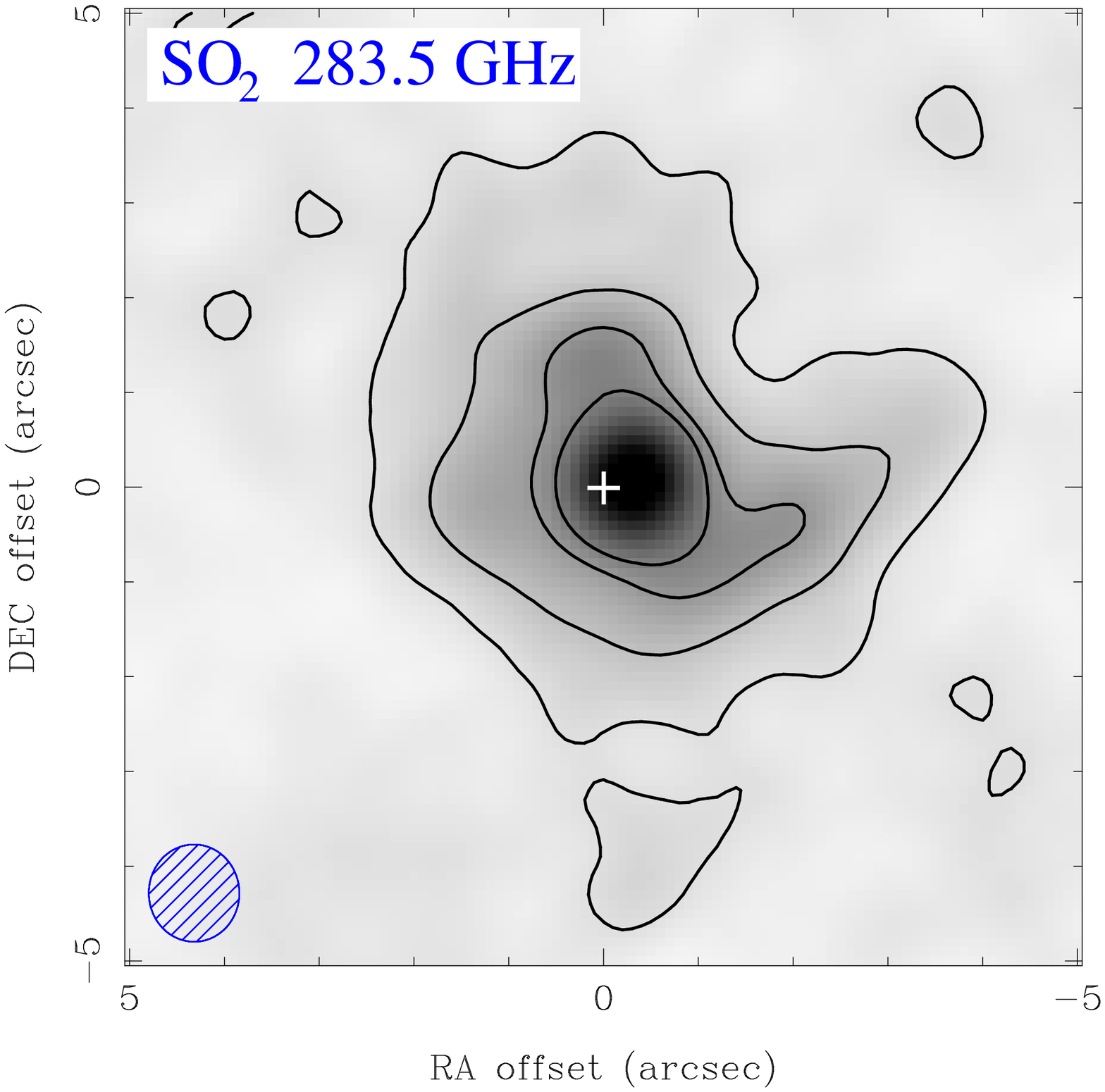}
\includegraphics[angle=270,width=0.325\textwidth]{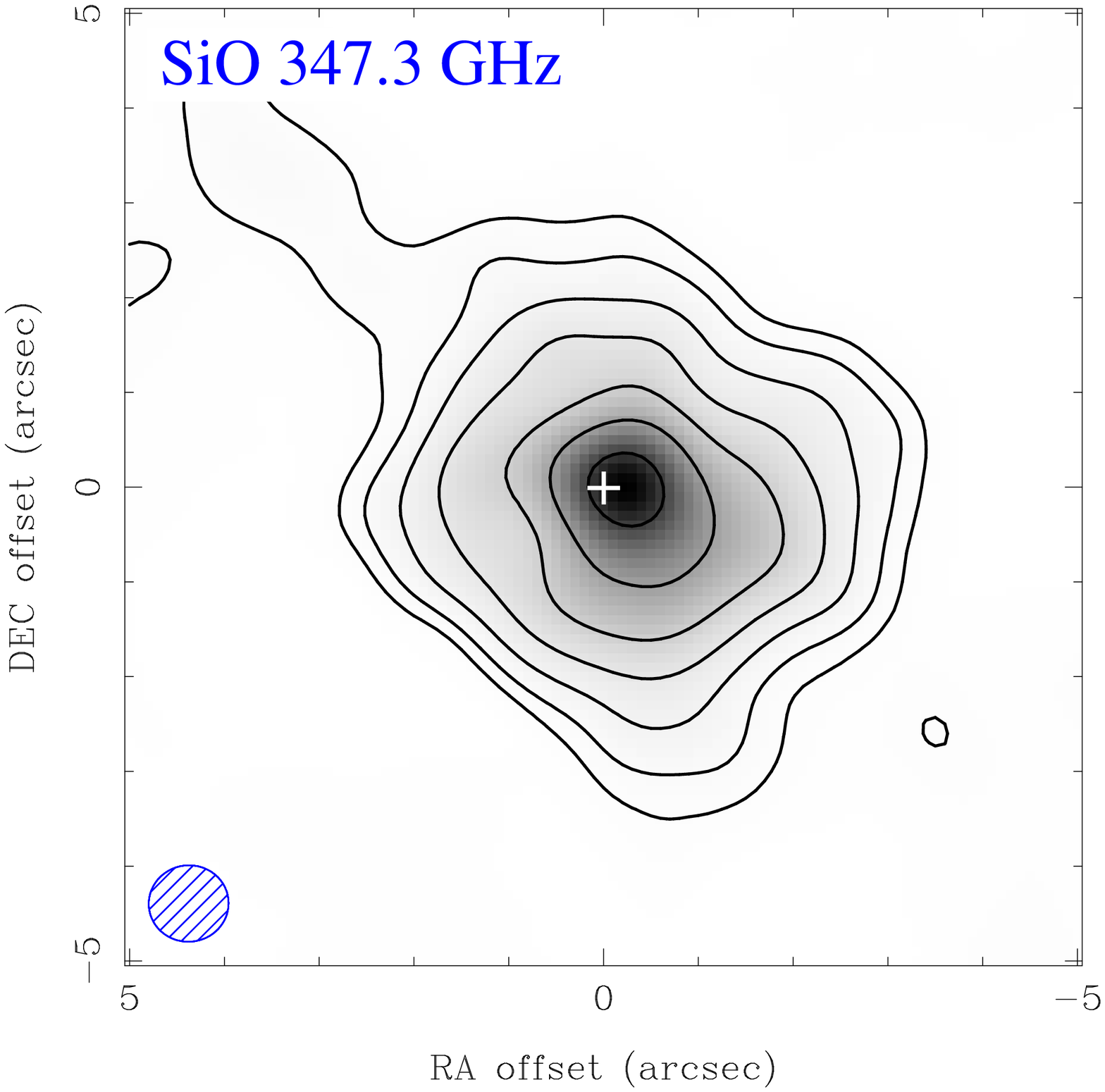}
\\
\includegraphics[angle=270,width=0.325\textwidth]{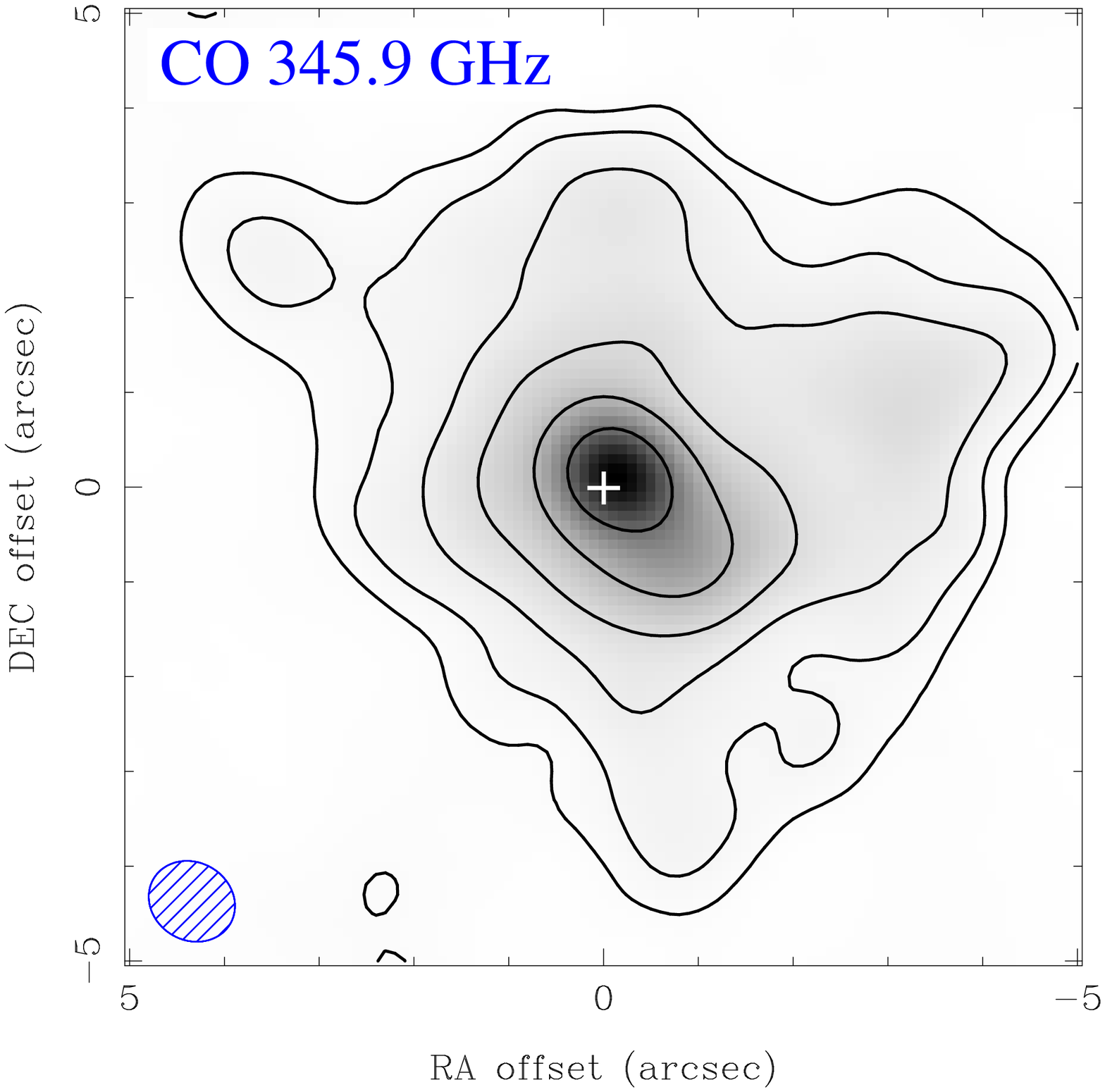}
\includegraphics[angle=270,width=0.325\textwidth]{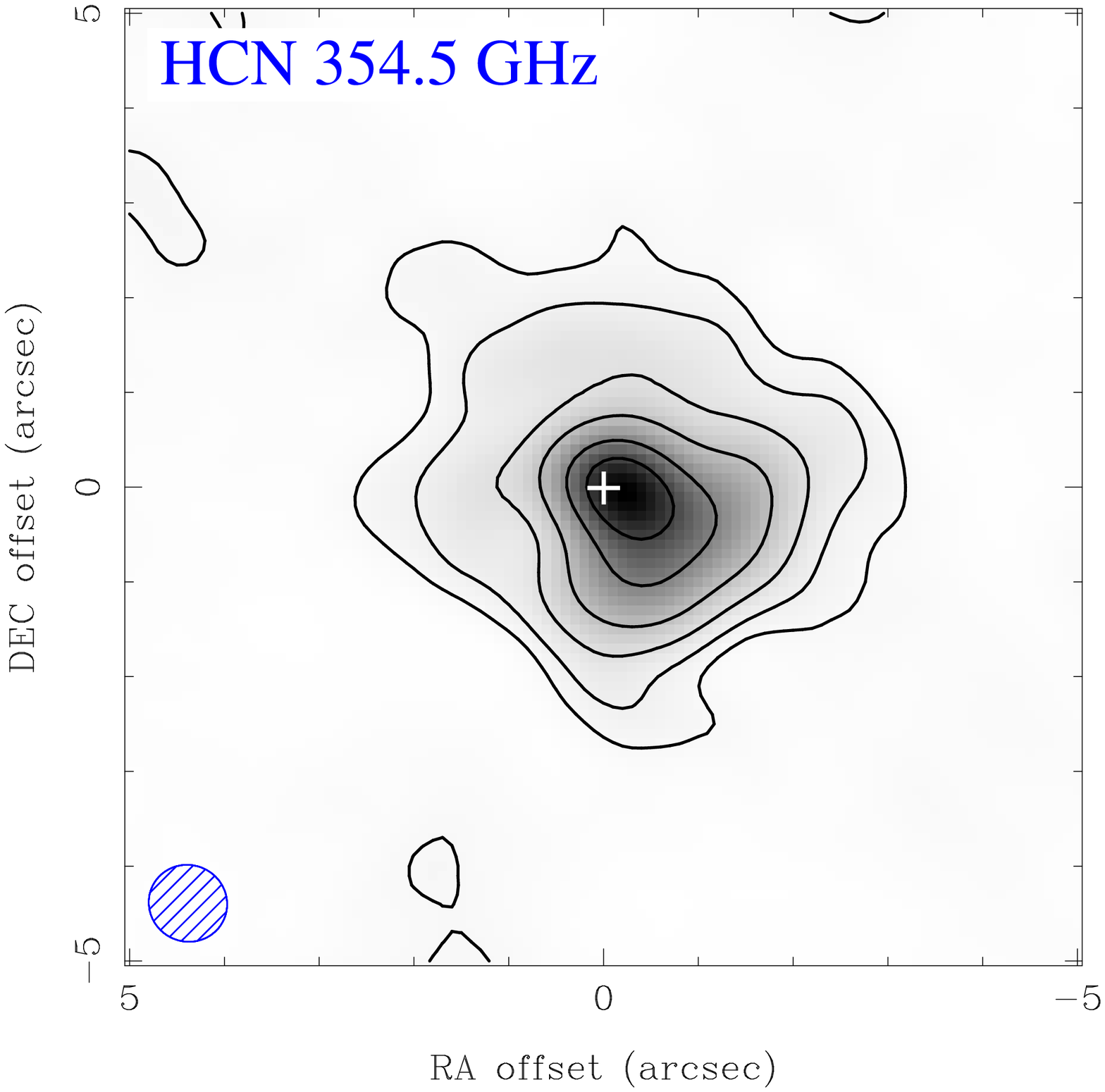}
\includegraphics[angle=270,width=0.325\textwidth]{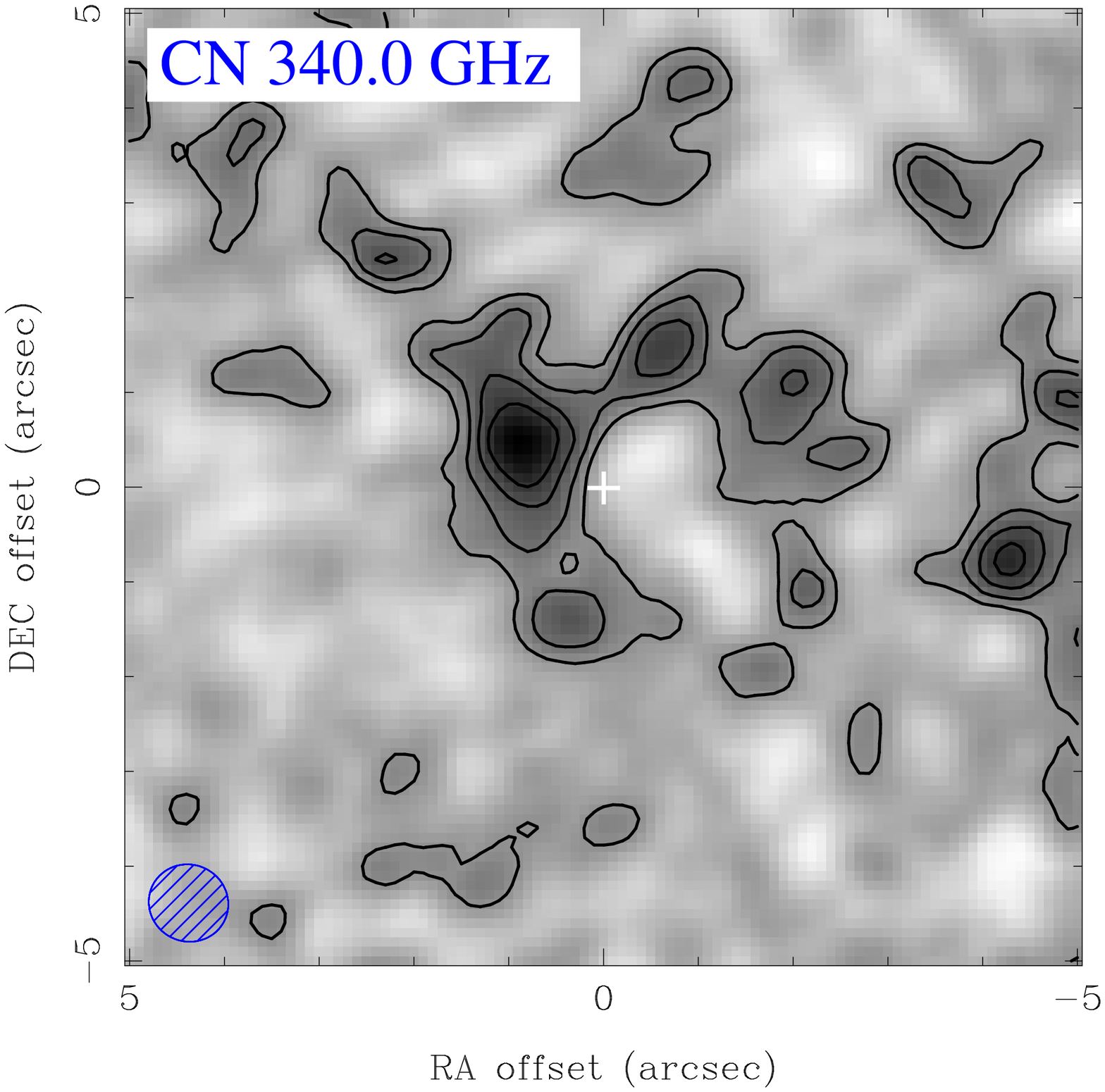}
\caption{Maps of the integrated emission of nine molecular species in VY\,CMa showing the wide range of spatial distributions. The peak of the continuum emission is indicated by the white cross (+). The corresponding beam is shown in the bottom left corner of each map.
The contour levels (in units of the rms) are:
3, 6, 12, and 25 for AlO;
6, 12, 25, 50, and 100 for H$_2$S and HCN;
6, 12, 20, and 25 for CS;
3, 6, 12, 25, and 50 for SO;
6, 12, 20, and 25 for SO$_2$;
6, 12, 25, 50, 100, 200, and 400 for SiO;
6, 12, 25, 50, 100, and 200 for CO; and
6, 7, 8, and 9 for CN.
\label{Fig-morph}}
\end{figure*}

Many of the molecules in the survey have multicomponent profiles and asymmetric angular distributions which reflect the complex morphology and kinematics of the circumstellar nebula of VY\,CMa. The molecular emission may be characterized as either: 
(1) ``point-like'', typical of  most metal-bearing species;
(2) ``double'', with a second source located to the southwest of the star; 
(3) ``ring-like'', of which CN is the only example in the survey;  or 
(4) ``extended'', with multiple components and a  complex angular distribution.
Using this classification, we summarized the morphology of 19 molecules in Table\,\ref{Tab-morph}. Examples of integrated intensity maps illustrating the different morphological types are shown in Fig.\,\ref{Fig-morph}.   

It was known from prior interferometric studies in the radio band that molecular emission of some species is within $\le 0.5\arcsec$ of the star (e.g., H$_2$O, \citealp{richards}; SiO, \citealp{shinnaga}; PN, \citealp{fu}), but  the emission from others may extend by up to a few arcsec  (e.g., OH at 1612\,MHz, \citealp{OH}; CO and SO, \citealp{muller}; and SO$_2$, \citealp{fu}). We find that the emission of a few are ``point-like'' (e.g., vibrationally excited H$_2$O and SiO, NaCl, AlO, AlOH, AlCl, PO, PN, TiO, and TiO$_2$) --- i.e., they are not resolved with our 0\farcs9 beam (see for example, Fig.\,\ref{Fig-morph}, and Figs.\,4 and 5 in \citealp{kami_tio}). The four sulfur-bearing molecules CS, NS, SiS, and H$_2$S are ``double'', because they show evidence of emission from a region displaced SW from the star (Fig.\,\ref{Fig-morph}), which might be associated with the SW\,Clump seen in continuum observations of \citet{hump-hst}. From our maps of CS which have good S/N in both spatial components, we determined that the SW~source is offset from the main knot of emission by $1\farcs00 \pm 0\farcs04$ with a position angle of $220\degr \pm 3\degr$. The same relative location was found for a dust clump at 1.25\,$\mu$m \citep[][see also \citealp{cruzalebes}]{monnier}. 
The SW component is also observed in the most intense line of NaCl ($J=26-25$ at 338.02\,GHz), but it is not observed in other transitions of NaCl owing to the lower S/N.  It is plausible that some emission referred to here as point-like, might contain the SW component if it were observed at higher sensitivity.   

\begin{figure*}
\includegraphics[trim=15 15 15 15,angle=270,width=0.45\textwidth]{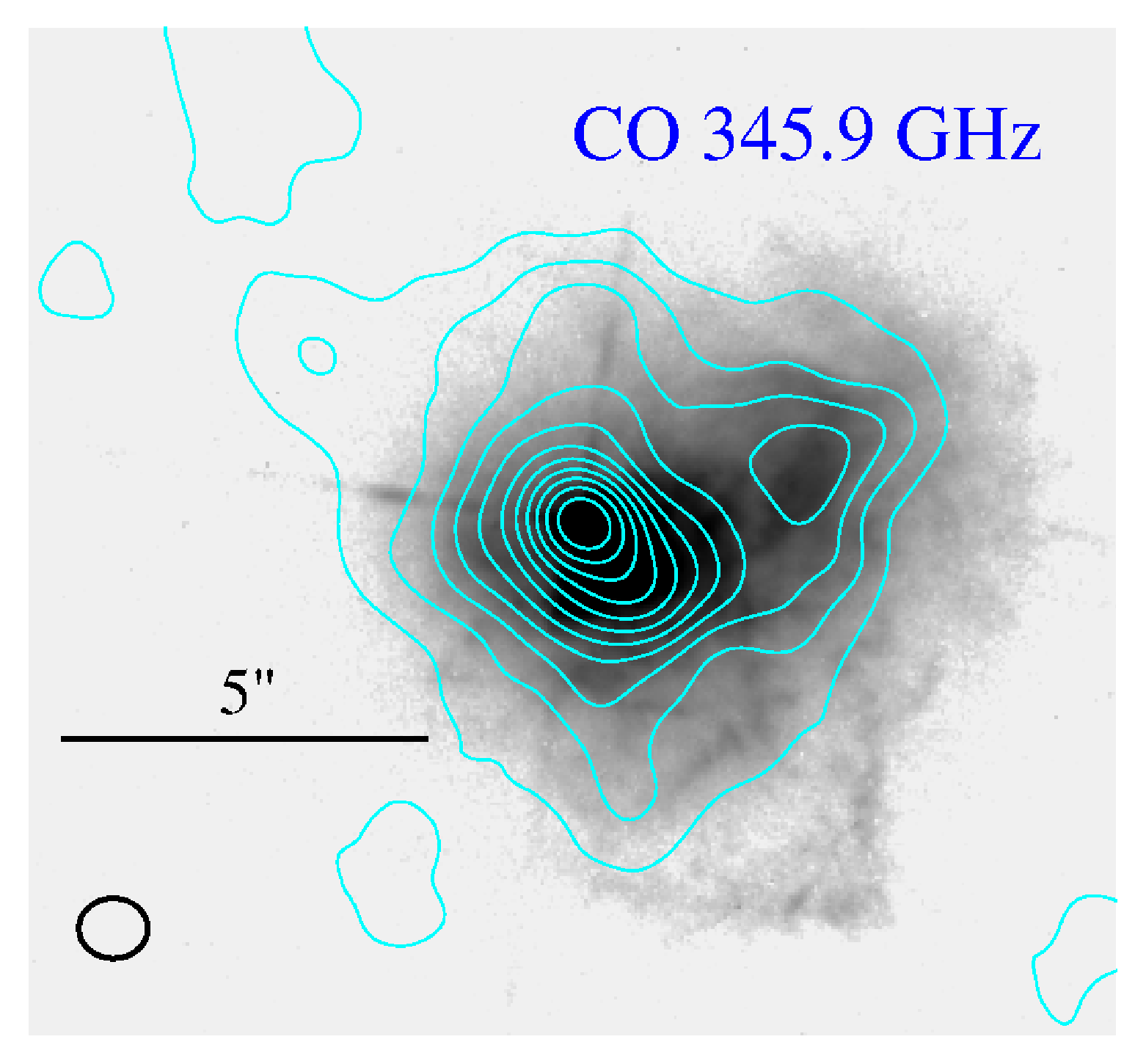}
\includegraphics[trim=15 15 15 15,angle=270,width=0.45\textwidth]{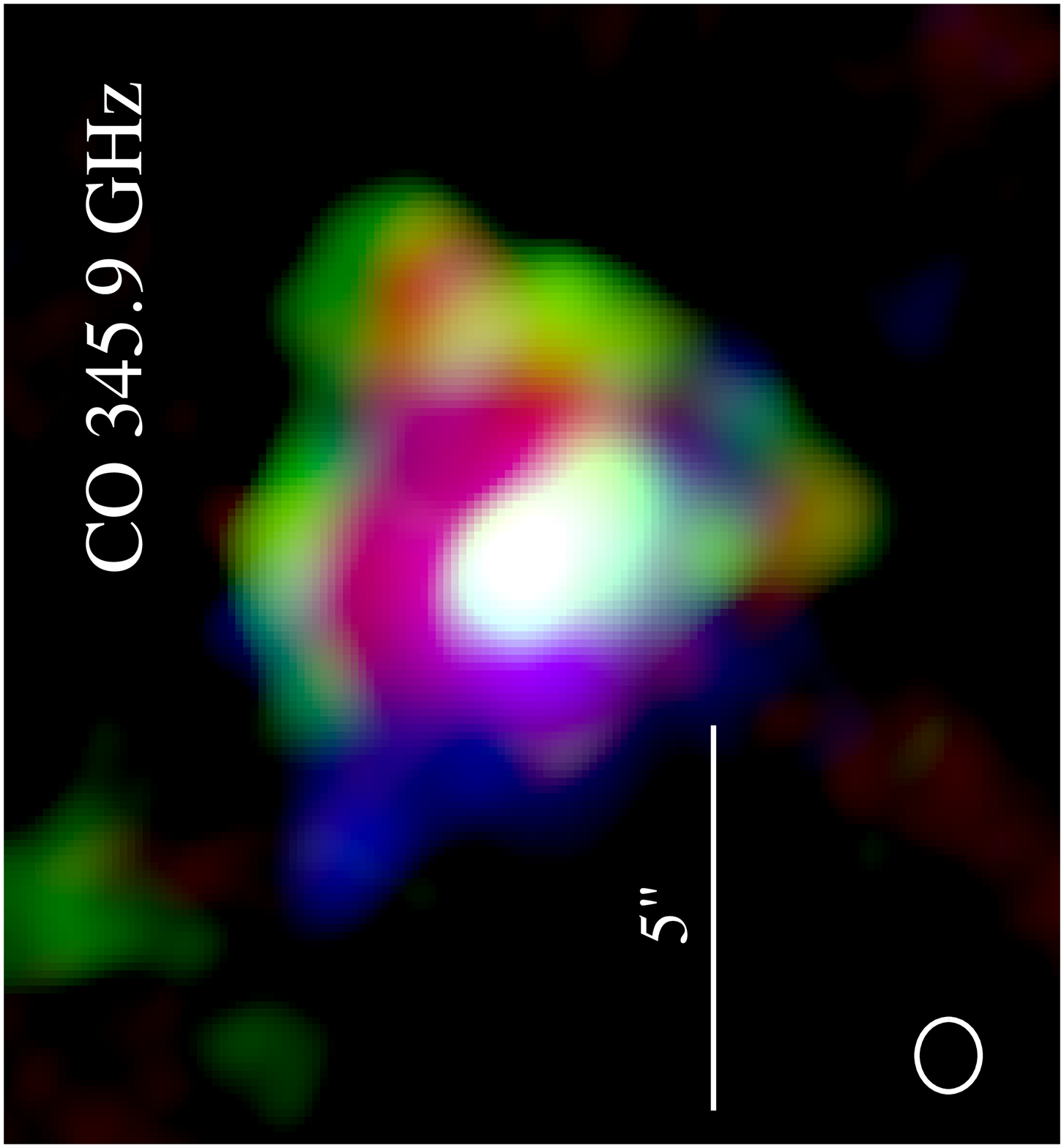}\\
\includegraphics[trim=15 15 15 15,angle=270,width=0.45\textwidth]{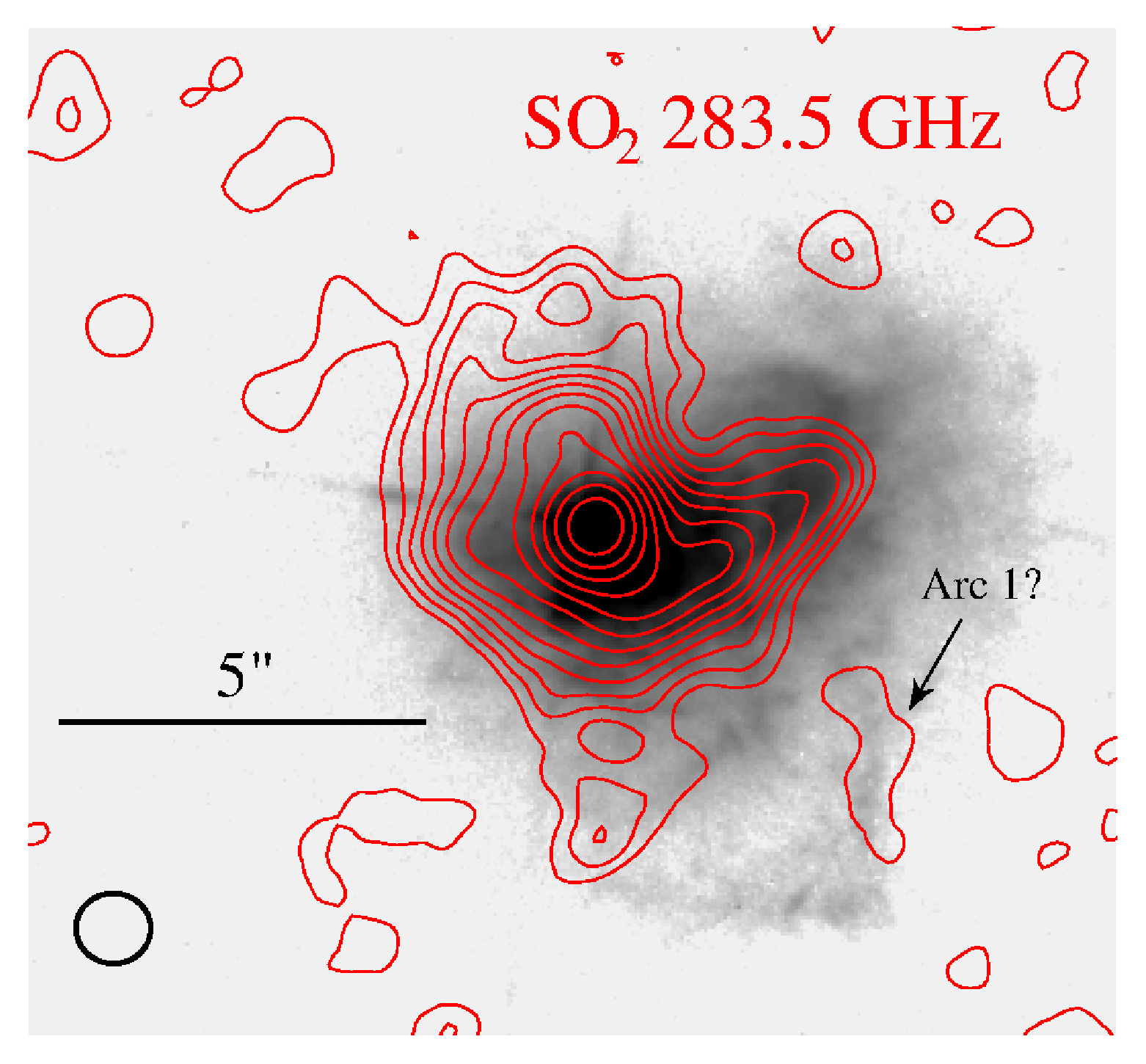}
\includegraphics[trim=15 15 15 15,angle=270,width=0.45\textwidth]{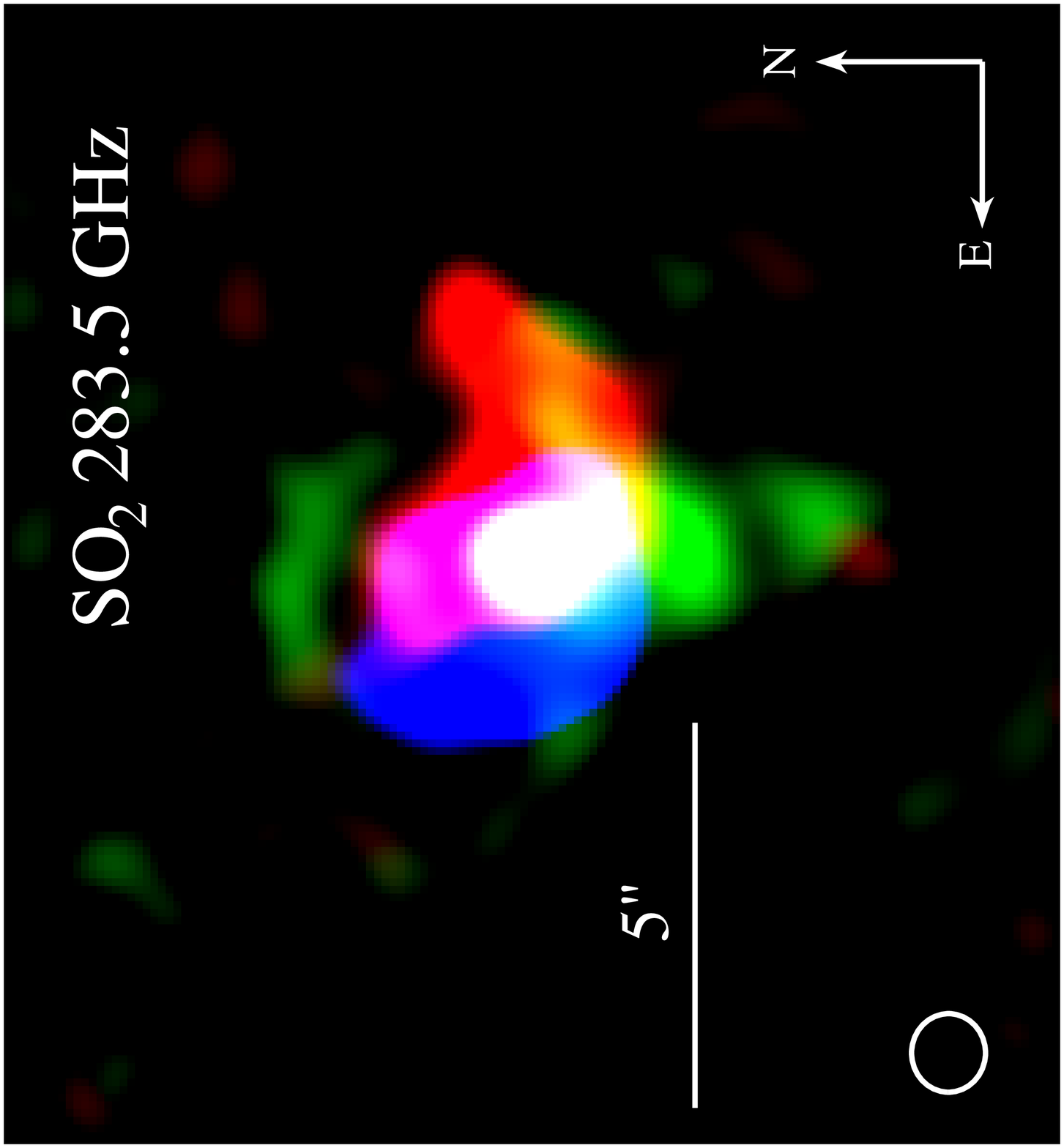}\\
\includegraphics[trim=15 15 15 15,angle=270,width=0.45\textwidth]{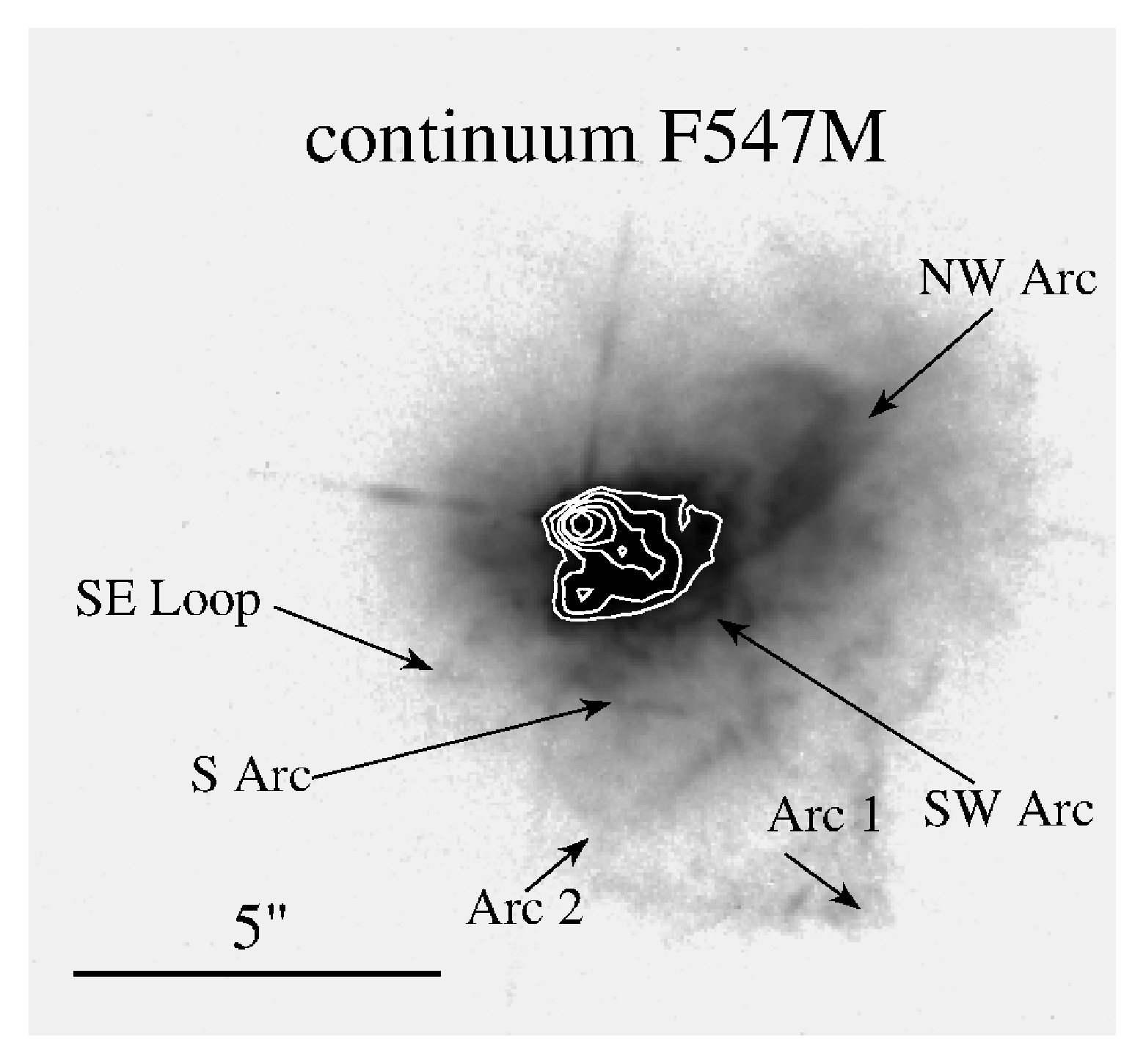}
\includegraphics[trim=15 15 15 15,angle=270,width=0.45\textwidth]{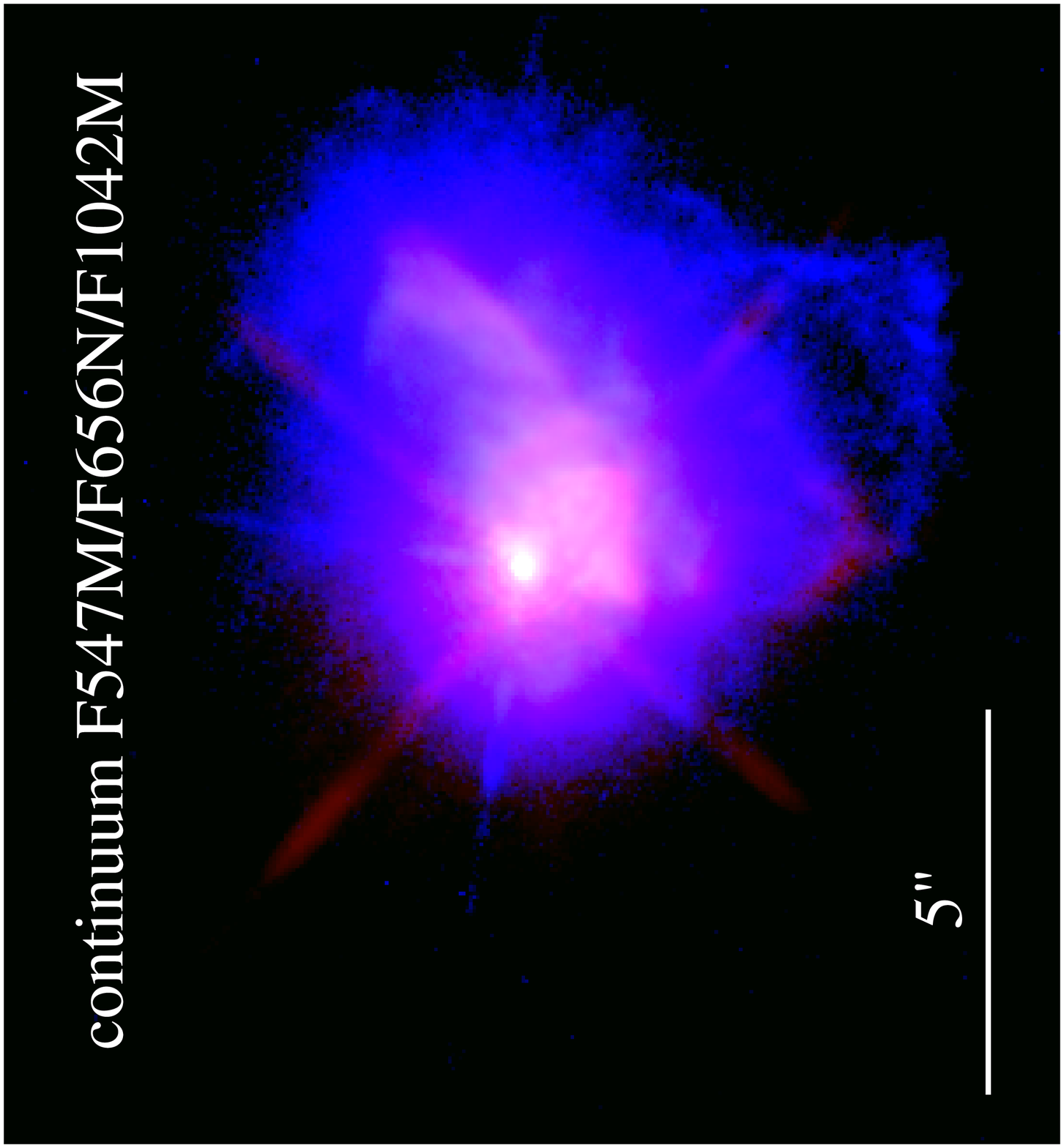}
\caption{See the extended caption on the previous page.}
\label{Fig-HST}
\end{figure*}

The angular distribution of the ``extended'' molecular emission is very complex.  
Typically it extends by 2\farcs5 to the west (red-shifted component), $<1\arcsec$ to the east (blue-shifted component), and is elongated in the north-south direction (i.e., similar to that observed in OH at 1612\,MHz by \citealp{OH}; see Fig.\,\ref{Fig-morph}). From the most prominent line of SO$_2$ ($16_{0,16}-15_{1,15}$ at 283.5\,GHz), we established that the emission extends by up to 3\farcs8 towards the west, up to about 3\arcsec\ to the east, nearly 4\farcs9 to the south, and 3\farcs8 to the north. Similarly, the emission in the more intense line CO(3--2) is seen up to 5\arcsec\ from the continuum peak (Fig.\,\ref{Fig-morph}), but there is also a hint of emission 7\farcs5 NE from the star (partially shown in Fig.\,\ref{Fig-HST}). As discussed in Sect.\,\ref{obs}, extended emission is filtered out in our interferometric observations. As a result, molecular emission around VY\,CMa might be even more extended than our maps show.

There is an apparent spatial correlation between the extended molecular emission we observe, and those in the optical and near-infrared images of the nebula \citep{monnier,smith-HST,hump-hst}. Here we refer mainly to the HST images  of optical scattered light, which reveal a very rich variety of nebular features at a resolution of $\sim50$\,mas. Some of the most characteristic features are indicated in the HST images in Fig.\,\ref{Fig-HST}. Throughout, we adhere to the designations of these features in \citet{hump-hst}. As the maps in Fig.\,\ref{Fig-HST} show, peaks in the spatial distribution of CO and SO$_2$ appear to coincide with several features in the HST images. The NW\,Arc and Arc\,2 (and the northern wall of Arc\,1) are more discernible in the maps of SO$_2$, but are also seen in the map of CO. The spatial distribution of the emission of CO appears to be generally smoother than that of SO$_2$, most likely owing to the higher optical depth of the CO line. The presence of extended molecular features that correspond to those seen in the  optical is indicated for different species in Table\,\ref{Tab-morph}. In addition to these features, there is emission to the north and northeast of the continuum peak which is especially prominent in CO, and to a lesser extent in SO, SiO, and SO$_2$. Only a weak halo is evident in the HST images at these positions \citep[``outer halo'' of][]{smith-HST}, most likely owing to high extinction close to the star that shadows the N and NE parts of the nebula \citep{monnier,smith-HST}.  


Prior attempts to assign velocity components in nearly featureless line profiles observed with single antennas to specific arcs and knots in HST images on the basis of radial velocities derived from slit spectra in the optical have been inconclusive (cf. Sect.\,\ref{mole-SO2}). Owing to resonance line scattering and scattering by the dust in the expanding nebula, velocities determined in the optical do not provide direct measurements of the radial motion of the emitting material and the line shapes are also modified \citep[][but see \citealp{smith_CO}]{vanBlerkom,hump-spec}. In principle, the kinematic structure of the circumstellar nebula might be inferred from tangential motions derived from multi-epoch observations of the continuum and from polarimetry at optical wavelengths \citep{hump-hst,polar}, but determination of the velocities by this method is not straightforward and does not provide the full kinematics of the nebula. Observations at submillimeter wavelengths are free from these distorting effects and  provide direct and full information of the kinematics of the resolved features. 

\begin{figure}\centering
\includegraphics[angle=270,width=0.85\textwidth]{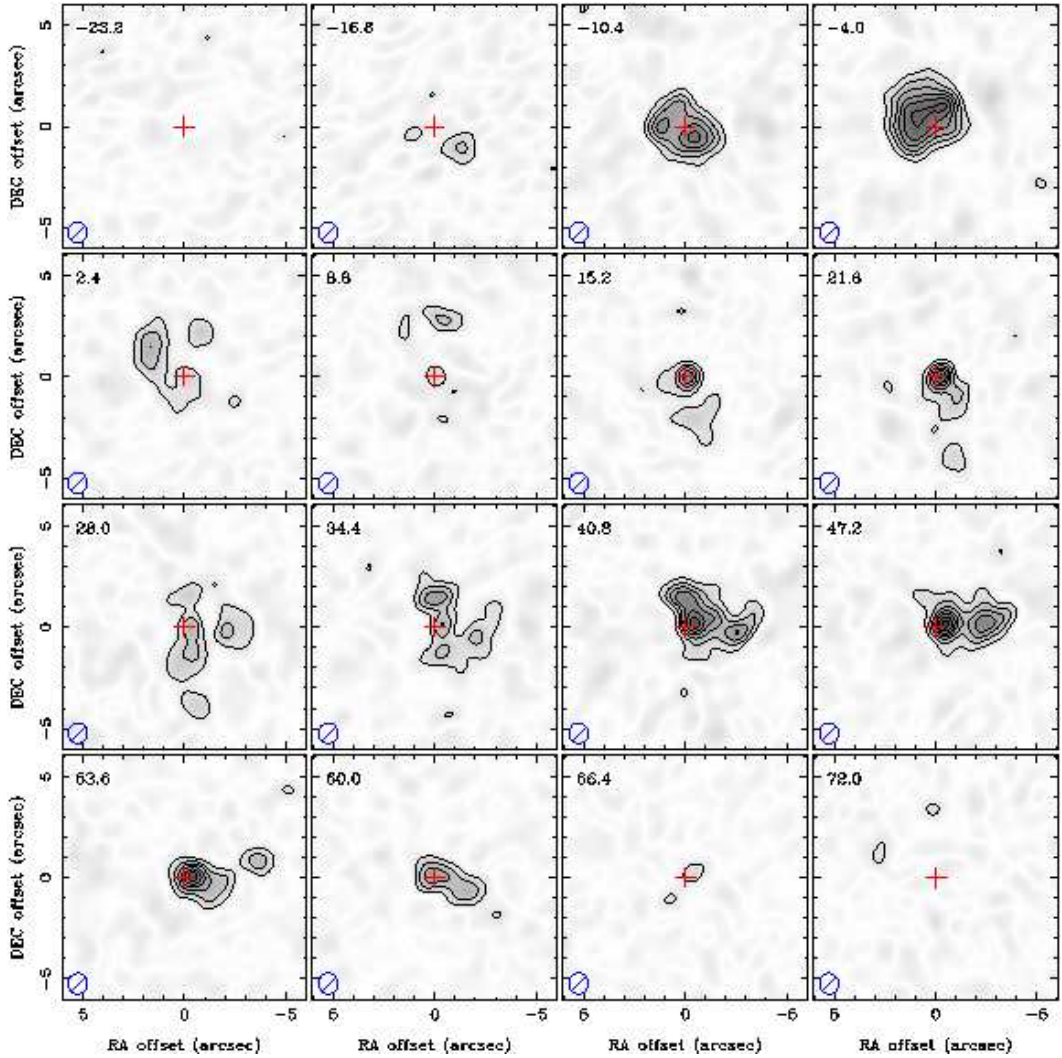}
\caption{Channel maps of the SO$_2$($16_{0,16}-15_{1,15}$) line in VY\,CMa.  Shown in the panels is the integrated emission in a 6.4\,km\,s$^{-1}$ interval starting at the velocity indicated in the top left corner of each map. The beam size is shown with a crossed ellipse in the left bottom corner of each panel. The cross (+) marks the position of the continuum peak. See the electronic edition of the Journal for a color version of this figure.
\label{Fig-chann}}
\end{figure}

Shown in right panels of Fig.\,\ref{Fig-HST} is the spatio-kinematic structure of the nebula observed in CO(3--2) and SO$_2$($16_{0,16}-15_{1,15}$). The emission is divided into three velocity ranges, $-40:5$, $5:30$, and $30:90$\,\kms, or, respectively, blue-shifted, stationary, and red-shifted with respect to the systemic velocity. All three velocity components are present in the inner (unresolved) part of the nebula. The emission is red-shifted in the west and northwest (W\,Arc and NW\,Arc), and up to 1\arcsec\ north. In the southern part (S\,Arc, Arc\,2, and possibly Arc\,1), the expansion likely occurs in the plane of the sky, because radial motions in these parts are small. Additionally, our map of CO(3--2) suggests that the main nebula is surrounded in the north, west, and south by gas that is radially stationary, but it may appear so owing to a high  optical thickness of the line. Finally, the blue-shifted component dominates in the east/north-eastern part of the radio nebula. The kinematic structure is also readily discernible in more conventional channel maps of SO$_2$($16_{0,16}-15_{1,15}$), as shown in Fig.\,\ref{Fig-chann}.

The general picture of the spatio-kinematic structure of the nebula obtained here is consistent with constraints derived from spectroscopic observations in the optical \citep{hump-hst,smith_CO}. A more thorough discussion of the extended molecular emission around VY\,CMa will be presented in a separate paper.

\noindent {\bf Caption to Fig.\,\ref{Fig-HST}.---} Maps of the spatio-kinematic structure of the nebula of VY\,CMa observed in CO(3--2) and SO$_2$($16_{0,16} - 15_{1,15}$). {\bf Left:} The contours of CO (top) and SO$_2$ (middle) are overlaid on an image of the scattered light observed with HST filter F547M in 1999 \citep{smith-HST}. In the image at the bottom, the contours in white show the morphology of the continuum emission in its brightest parts. 
{\bf Right:} Images in the top and middle panels are maps of the kinematical structure of CO and SO$_2$. 
The different velocity components of the emission are color coded: 
red corresponds to  $-40 < V_{\rm{LSR}} < 5$\,km\,s$^{-1}$, 
blue to $30 < V_{\rm{LSR}} < 90$\,km\,s$^{-1}$, and 
green to $5 < V_{\rm{LSR}} < 30$\,km\,s$^{-1}$.
In regions where the different components overlap, the colors are mixtures of the three basic colors:
purple/pink/magenta for red- and blue-shifted emission; yellow/orange for central and red-shifted; purple for blue- and red-shifted; cyan for central and blue-shifted; and white for those regions where all three velocity components are present. 
Shown in the bottom is the continuum emission of the nebula observed with three different HST filters: F547W (coded as blue), F656N (green), and F1042M (red). The synthesized beam of the SMA is shown in the bottom left corner in the panels which show the molecular emission. 
See the electronic edition of the Journal for a color version of this figure.

\begin{deluxetable}{lccccccccl}
\tabletypesize\scriptsize
\tablecaption{Morphology of molecular emission in VY\,CMa\label{Tab-morph}}
\tablewidth{0pt}
\tablehead{
\multicolumn{4}{l}{}     & \multicolumn{4}{c}{Extended\tablenotemark{c}}  &  \multicolumn{2}{l}{}     \\
\cline{5-8}
\multicolumn{1}{l}{Molecule} &\multicolumn{1}{c}{Point-like\tablenotemark{a}}   &\multicolumn{1}{c}{Double\tablenotemark{b}}    & 
\multicolumn{1}{l}{} &\colhead{S~Arc}  & \colhead{Arc~2} & \colhead{NW~Arc}  & \colhead{NE}   & \colhead{Ring like\tablenotemark{d}} 
&\multicolumn{1}{l}{Map\tablenotemark{e}}  \\
}
\startdata
AlO  & $\surd$ & & & & & & & &  Fig.\,\ref{Fig-morph} \\
AlOH & $\surd$ & & & & & & & & \\
AlCl & $\surd$ & & & & & & & & \\
NaCl & $\surd$ & [$\surd$]\tablenotemark{f} & & & & & & & \citet{kami_tio}\\
TiO  & $\surd$ & & & & & & & & \citet{kami_tio}\\
TiO$_2$ & $\surd$ & & & & & & & & \citet{kami_tio}\\
H$_2$O & $\surd$ & & & & & & & & \citet{kami_tio}\\
SiO\,($v \not= 0$)& $\surd$ & & & & & & & & \\
SO~($v=1$) & $\surd$ & & & & & & & & \\
PO & $\surd$ & & & & & & & & \\
PN & $\surd$ & & & & & & & &  \citet{fu}\\[10pt]
SiS & & $\surd$ & & & & & & & \\
CS & & $\surd$ & & & & & & & Fig.\,\ref{Fig-morph}\\
H$_2$S &  & $\surd$ & & & & & & & Fig.\,\ref{Fig-morph} \\
NS &  & $ \surd$  &  &  &  &  &  &   &  \\[10pt]
SO$_2$ & & & & $\surd$ & $\surd$ & $\surd$ & $\surd$ & & Figs.\,\ref{Fig-morph},\ref{Fig-HST},\ref{Fig-chann} \\
SO~($v = 0$) & & & & $\surd$ & & $\surd$ & $\surd$ & & Fig.\,\ref{Fig-morph} \\
CO & & & & $\surd$ & $\surd$ & $\surd$ & $\surd$ & & Figs.\,\ref{Fig-morph},\ref{Fig-HST}\\
HCN & & & & $\surd$ & & $\surd$ &$\surd$ & & Fig.\,\ref{Fig-morph} \\
SiO\,($v = 0$) & & & & $\surd$ & [$\surd$]\tablenotemark{g} & $\surd$ & $\surd$ & & Fig.\,\ref{Fig-morph}\\[10pt]
CN & & & & & & & & $\surd$ & Fig.\,\ref{Fig-morph} \\  
\enddata
\tablenotetext{a}{``Point like'' refers to emission with a single spatial component which appears featureless at our angular resolution.}
\tablenotetext{b}{``Double'' refers to molecular emission from the central region, and emission that may coincide in position with the SW~Clump that is displaced from the star by 0\farcs8--1\farcs2.}
\tablenotetext{c}{The naming convention of the individual features of extend emission follows that of \citet{hump-hst}. See also Sect.\,\ref{morph}.}
\tablenotetext{d}{``Ring like", refers to hollow emission that appears in a ring or an incomplete ring.}
\tablenotetext{e}{See figures in this work or reference cited for map.}
\tablenotetext{f}{SW~Clump is only observed in the most intense line of NaCl.}
\tablenotetext{g}{Arc~2 is only observed in the most intense lines of  SiO ($v = 0$).}
\end{deluxetable}

\subsection{Continuum emission\label{Sect-conti}}  

Preliminary results of the continuum measurements were given in \citet{kami_tio}. Here we present a more comprehensive description of the continuum emission of VY\,CMa. The line-free continuum was measured in 36 subbands, each 2\,GHz wide (see Sect.\,\ref{obs}). Four subbands between 319 and 327\,GHz were ignored, because the S/N was low owing to telluric absorption at these frequencies. The average flux density between 279 and 355\,GHz is 0.67\,Jy, with a standard deviation of 0.11\,Jy determined from the scatter in the measurements in the different subbands. The continuum shows a slope with the best-fit value of the spectral index of $\alpha=1.5\pm0.3$ 
(where $F_{\nu}\propto \nu^{\alpha}$). Because of the relatively narrow frequency range covered here, this value is very approximate.  Combining these measurements with earlier ones also obtained with the SMA \citep{fu,shinnaga} and with other measurements in the literature, the spectral energy distribution in the (sub-)millimeter range is better described by $\alpha\approx2.6$ \citep[see also][]{shinnaga}. At the observed frequencies, the continuum emission is dominated by thermal dust emission with negligible contribution of the stellar photosphere \citep[see e.g.][]{sed}. The total continuum flux in the survey is 3.9 times higher than the total flux of all lines in Table\,\ref{Tab-main} observed with a $1\arcsec \times 1\arcsec$ aperture. 

\begin{figure}[ht!]\centering
\includegraphics[angle=270,width=0.49\textwidth]{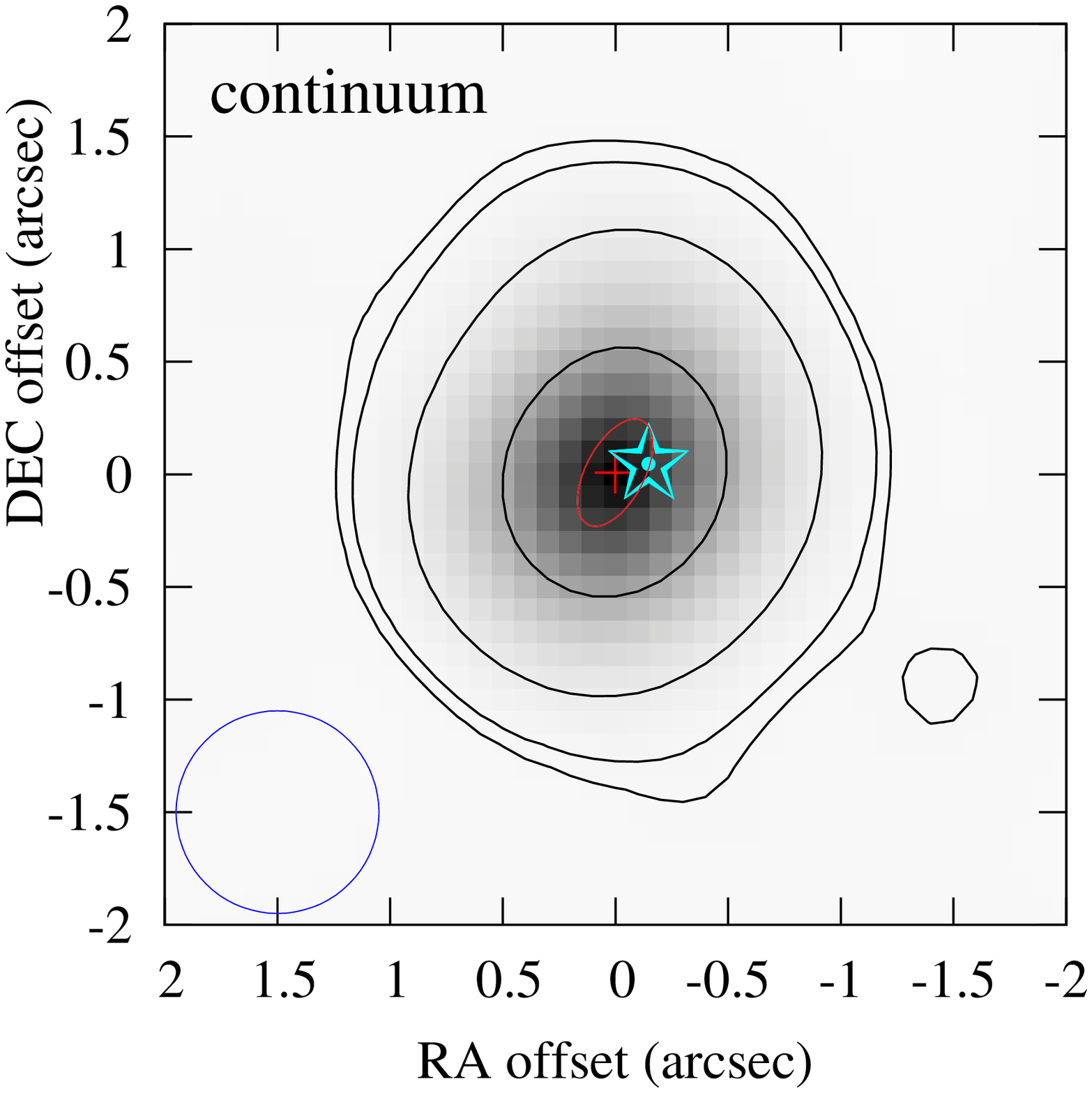}
\includegraphics[angle=270,width=0.49\textwidth]{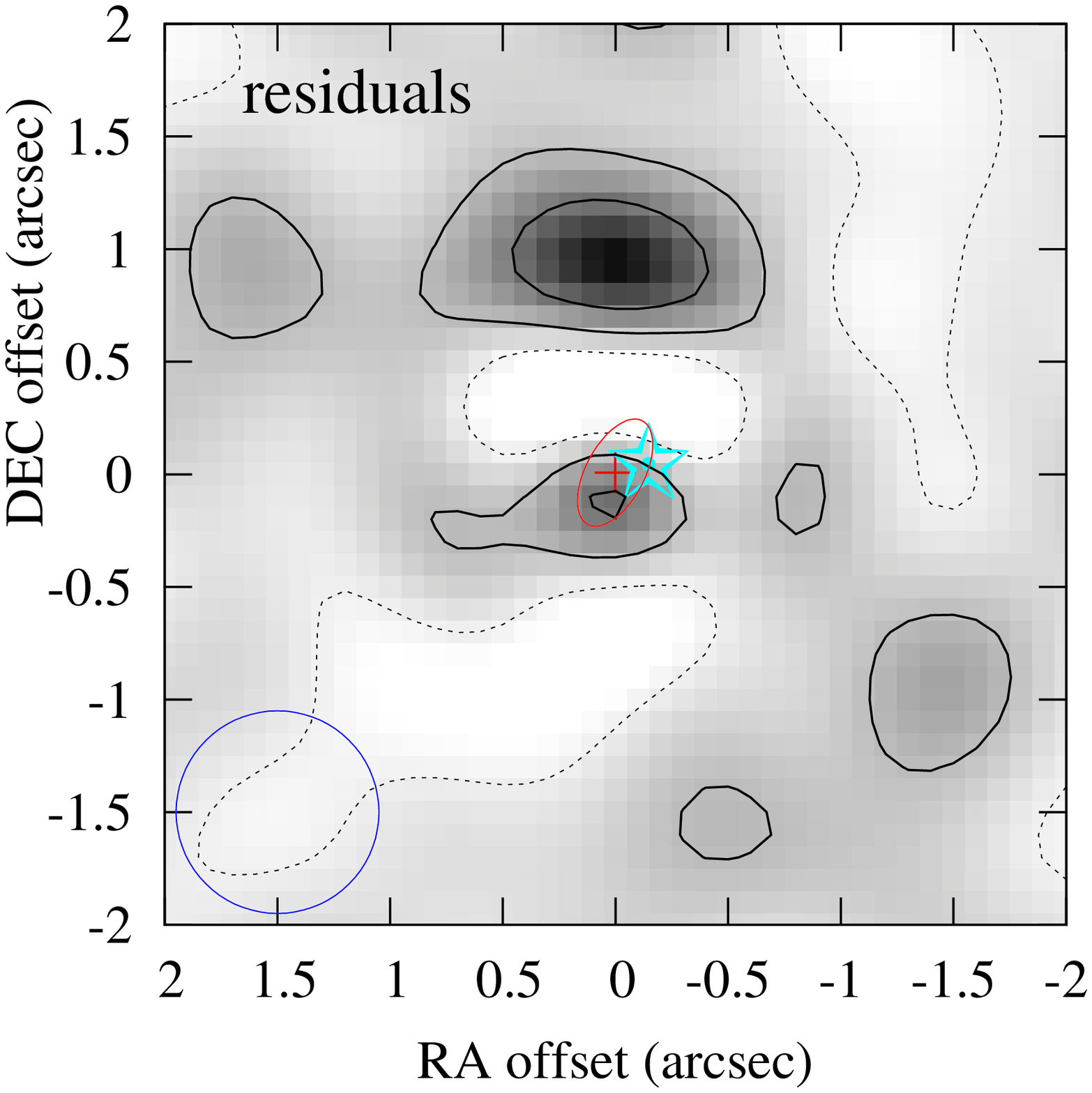}
\caption{{\bf Left:} A map of the average continuum emission of VY\,CMa in the 345\,GHz band. Contours are drawn at 5, 10, 50, and 300\,mJy/beam. {\bf Right:} Average residuals after removing a best-fit Gaussian component from the continuum map. Contours are drawn at --3 (dashed), 3 and 10\,mJy/beam (solid). In both panels, the stellar position (derived from maser measurements) is shown with the star symbol (cyan), while the average beam size is shown in bottom left corner (blue). The FWHMs of the main continuum source derived by deconvolution is represented by a red ellipse whose center 
is marked with red plus (+). See the electronic edition of the Journal for a color version of this figure.
\label{Fig-conti-decon}}
\end{figure}

The continuum source is resolved in our observations. An average deconvolved size of the continuum of $(522\pm3)  \times (257\pm2)$ mas (FWHMs) with the position angle of the major axis of $-27\fdg8 \pm 0\fdg5$ was derived by fitting a two-dimensional Gaussian in each of the 2-GHz wide bands. 
The average map of the continuum emission is shown in Fig.\,\ref{Fig-conti-decon}, where an ellipse represents the FWHMs of the best-fit Gaussian component. The residuals of the fit, included in Fig.\,\ref{Fig-conti-decon} in a separate panel, show an extra emission component 1\arcsec\ north from the main continuum source and 40 times weaker than the main component. Observations at higher angular resolution are needed to determine the actual structure of the submillimeter continuum source, but already our simple analysis indicates that the source is asymmetric at the 1\arcsec\ scale. 

\begin{figure}[ht!]\centering
\includegraphics[trim=0 0 0 120,angle=270,width=0.4\textwidth]{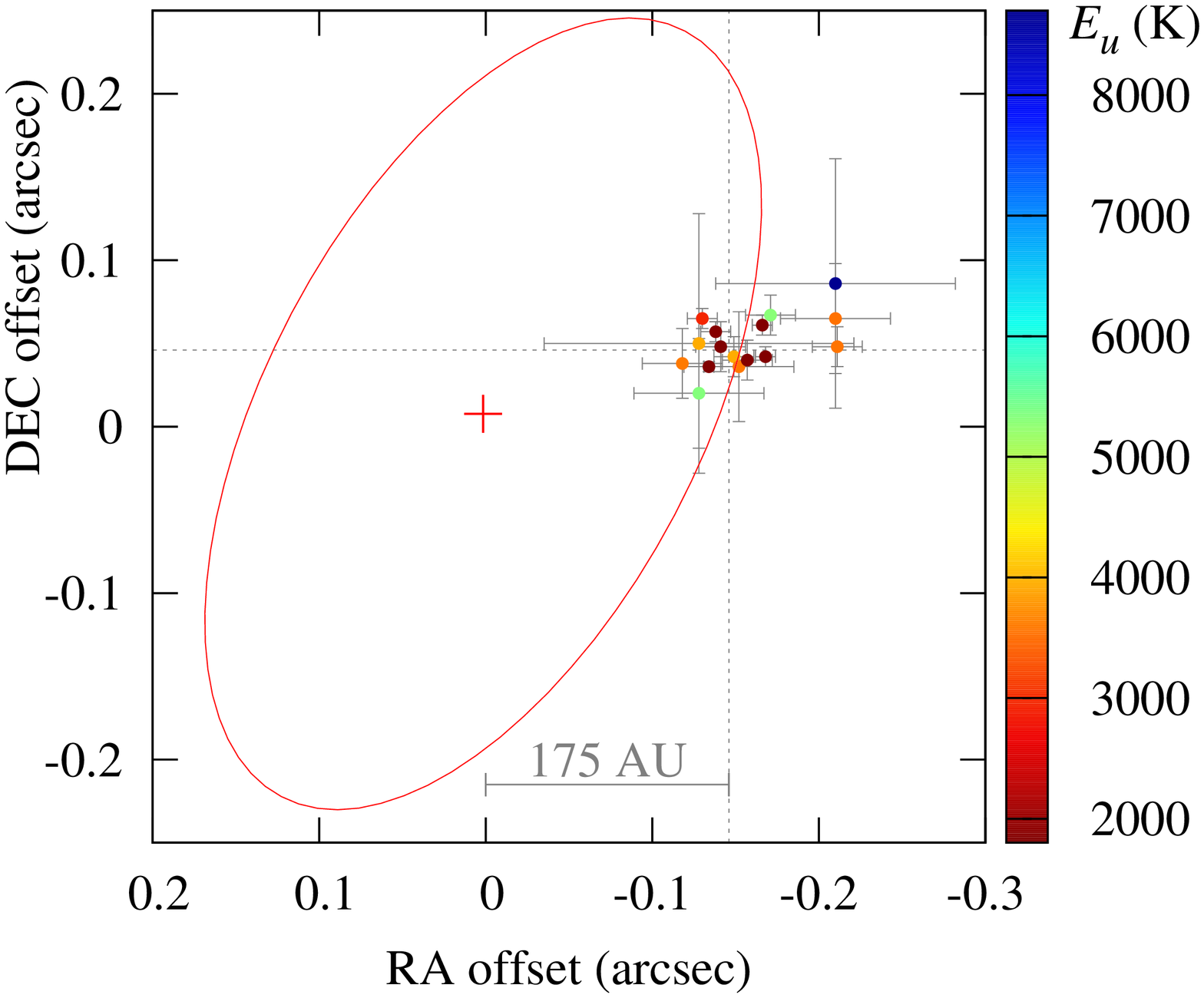}
\includegraphics[trim=0 0 0 120,angle=270,width=0.4\textwidth]{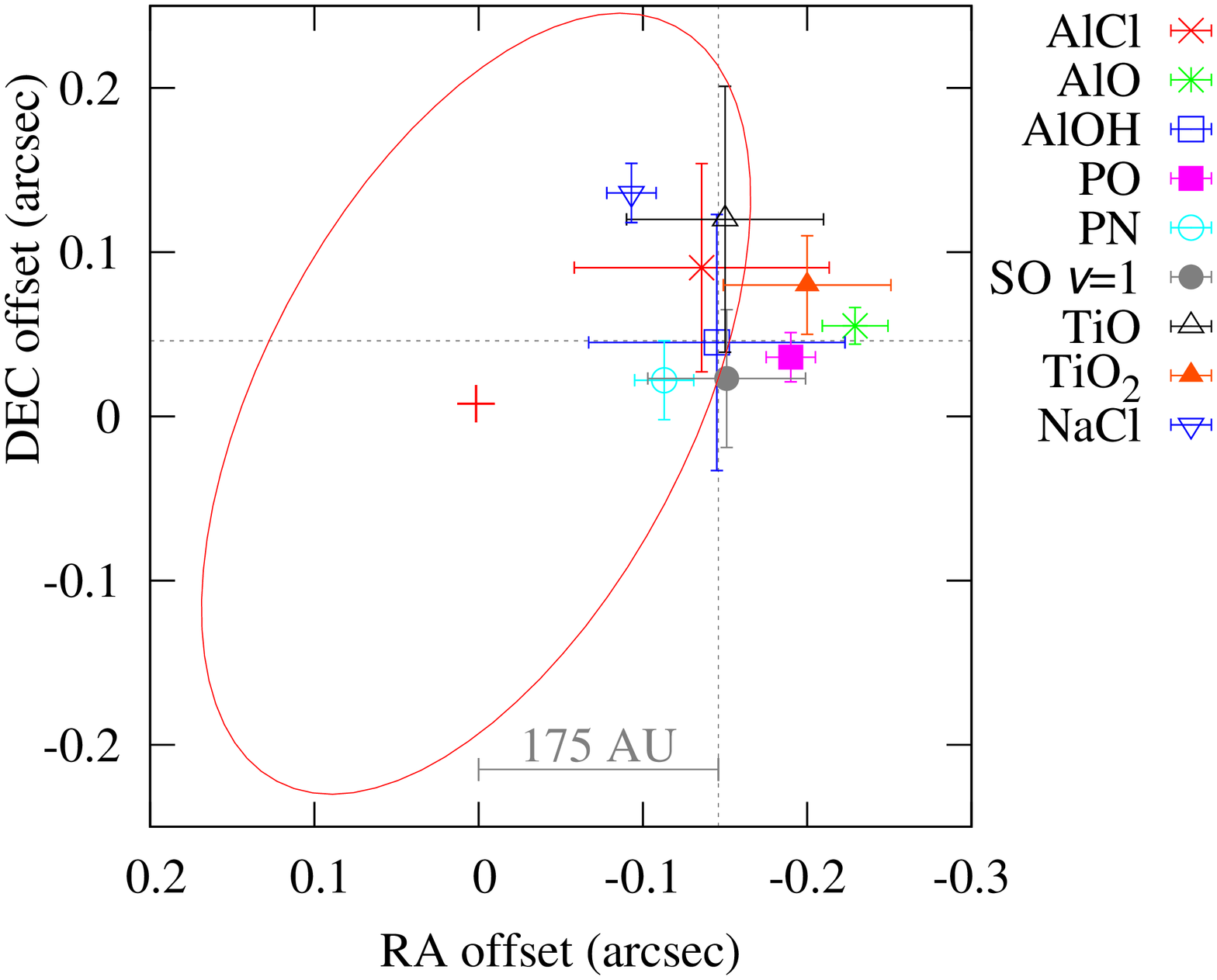}
\caption{The position of molecular emission with respect to continuum. The red ellipse represents the  size of the continuum source (the same as in Fig.\,\ref{Fig-conti-decon}). {\bf Left:} Points show central positions of emission of SiO and H$_2$O at excited vibrational states. The color of each symbol corresponds to the value of $E_u$ of the given transition at the scale indicated in the color bar. {\bf Right:} Central positions of emission of species that have single spatial component at our angular resolution. The color of the symbols is chosen arbitrarily. On both panels, the dashed grey lines cross at the average position of the vibrationally excited emission of SiO and H$_2$O, which, as we argue, is the stellar position. The errorbars correspond to 3$\sigma$ uncertainties. See the electronic edition of the Journal for a color version of this figure.
\label{Fig-conti-lines}}
\end{figure}

As noted in earlier studies \citep{muller,shinnaga}, the peak of the (sub-)millimeter continuum in VY\,CMa is shifted with respect to centers of molecular emission. Shown in Fig.\,\ref{Fig-conti-lines} are central positions of molecular emission observed in our survey. We limit the analysis here to species observed as single sources (i.e., those classified as ``point-like'' in Table\,\ref{Tab-morph}). The central positions were determined from Gaussian fits to maps of integrated emission. The left panel of Fig.\,\ref{Fig-conti-lines} shows spatial centers of SiO and H$_2$O rotational lines in vibrationally excited levels with energies above the ground of $E_u \gtrsim 2000$~K. The points are closely clustered to the west of the continuum peak. Their mean offset is $(-146, 46)$\,mas with a standard deviation of only 1\,mas in both axes. Because these high-excitation lines require a nearby excitation source, we interpret this position as the most likely location of the star. In support of such interpretation are observations of the SiO (and H$_2$O) masers observed with VLBI techniques which show that the maser spots are coincident with the stellar photosphere to within $\sim$20\,mas \citep[e.g.,][]{bo}. Shown in the right panel of Fig.\,\ref{Fig-conti-decon} are representative central positions of (thermal) emission of other species. These are located close to the centroid of SiO and H$_2$O emission, although the scatter in point positions is larger in this case. In fact, the emission of NaCl, AlO, PO, and PN likely peaks a few tens of mas  from the star/maser position, in different directions for different species, but this needs to be verified by measurements with a higher angular resolution. A more thorough discussion of the implications of the above findings and comparison to observations at other wavelengths will be presented in a separate paper.
  

\section{Summary}

Two hundred twenty three features between 279 and 355\,GHz were observed in an interferometric spectral line survey of VY\,CMa with the SMA.
Nineteen molecules were identified (nearly all with multiple transitions) including: TiO$_2$, which had not been observed in space prior to this work; TiO, whose spectrum in the visible is prominent in late-type stars, but had not been observed in the radio band; and AlCl, which although present in the circumstellar envelope of the carbon-rich star IRC$+10\degr216$, had not been observed in VY\,CMa. 
These two new species plus the 21 molecules observed previously at radio and (sub-)millimeter wavelengths, and three others observed exclusively in the optical band (VO, ScO, and YO), yields 26 molecules in all. The molecules with the highest number of lines (including all isotopic species and excited vibrational levels) are SO$_2$ (53), TiO$_2$ ($\ge 33$), NaCl  (33), SiO (29), and SO (26). 

Integrated intensity maps were made for all the lines in the survey at a typical resolution of 0\farcs9. Prior to this work, only a few molecular lines had been observed in this source with interferometers. 
The maps obtained here provide fundamental information needed for chemical modeling, and the development of more sophisticated physical models of this complex source.  
The emission from several molecules consists of nearly featureless unresolved point-like emission from the central spatial component.
The molecular emission from four sulfur-bearing molecules is from the central region, and a second source displaced to the SW by about $1\arcsec$ from the star. The emission in five molecules is extended by several arcsec, while that of CN is ring-like. The center of molecular emission is typically offset by 0\farcs15 from the continuum peak, and is most likely coincident with the location of the star.  The maps reveal the kinematic structure and the apparent correlation of the molecular distribution with some arcs and knots/clumps of the circumstellar nebula present in the HST images. 


The observed distributions and derived molecular abundances in the circumstellar nebula of VY\,CMa, provide a road map for future interferometric spectral line studies. Measurements of the electronic spectra of small metal-bearing molecules by optical astronomers (e.g., TiO, VO, AlO, ScO) has helped elucidate the properties of the inner expansion zone of this complex source. Now that the rotational spectra of TiO and TiO$_2$ have been observed in the cooler portion of the inner expansion zone, astronomers  may begin to be develop a more comprehensive description of the chemical and physical properties of this source. Our observation of TiO and TiO$_2$ implies that other transition metal-bearing molecules such as VO, ScO, CrO, NiO, and FO might also be observed with interferometers.  The rotational spectra for many of these have been measured to high precision in the laboratory.  What is needed are observations of TiO, TiO$_2$, and other transition metal oxides at higher angular resolution and sensitivity with ALMA when it reaches full design specifications.

\newpage
\centerline{\bf APPENDIX}
\appendix
\section{Spectral-line atlas of VY\,CMa in the 345\,GHz band}\label{atlas}
Fig.\,\ref{Fig-Taco} presents spectra of VY\,CMa extracted from a $1\arcsec \times 1\arcsec$ region centered on the position of the continuum peak. Flux densities for the lines shown in green, blue, and red were scaled by dividing by 10, 40, and 500, respectively.  Each line is labeled with the molecular (isotopic) species and the index number in Table\,\ref{Tab-main}. Narrow ``emission'' spikes of instrumental origin are labeled as "Instr." (consult Table\,\ref{Tab-main}). 

\begin{figure*}\vskip -0.5cm
\includegraphics[width=0.95\textwidth]
{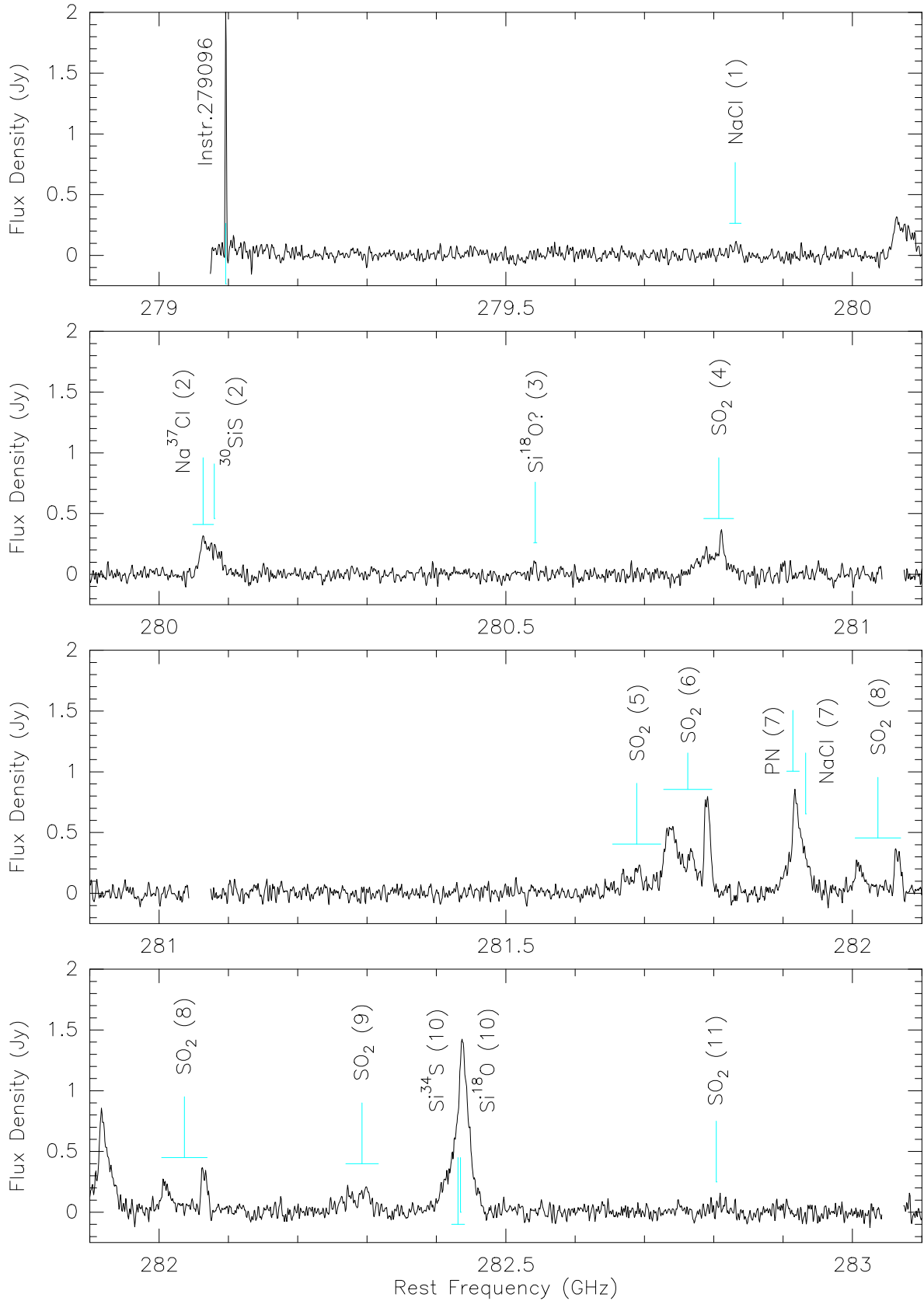}
\caption{The spectral atlas of VY\,CMa in the 345\,GHz band.\label{Fig-Taco}}
\end{figure*}
\clearpage

\begin{figure*}\vskip -0.5cm
\includegraphics[width=0.95\textwidth]
{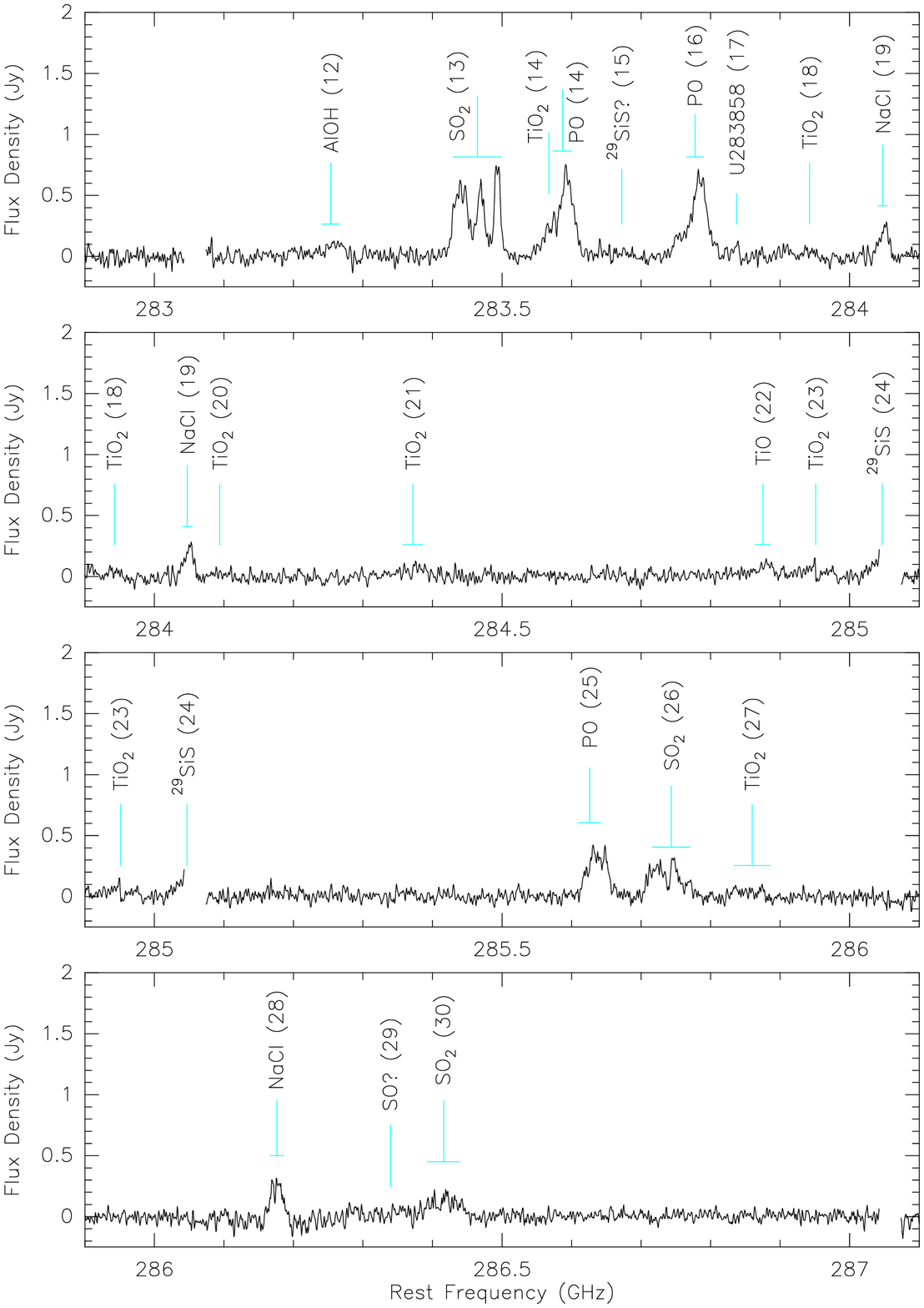}
\addtocounter{figure}{-1}
\caption{\it (Continued)}
\end{figure*}

\begin{figure*}\vskip -0.5cm
\includegraphics[width=0.95\textwidth]
{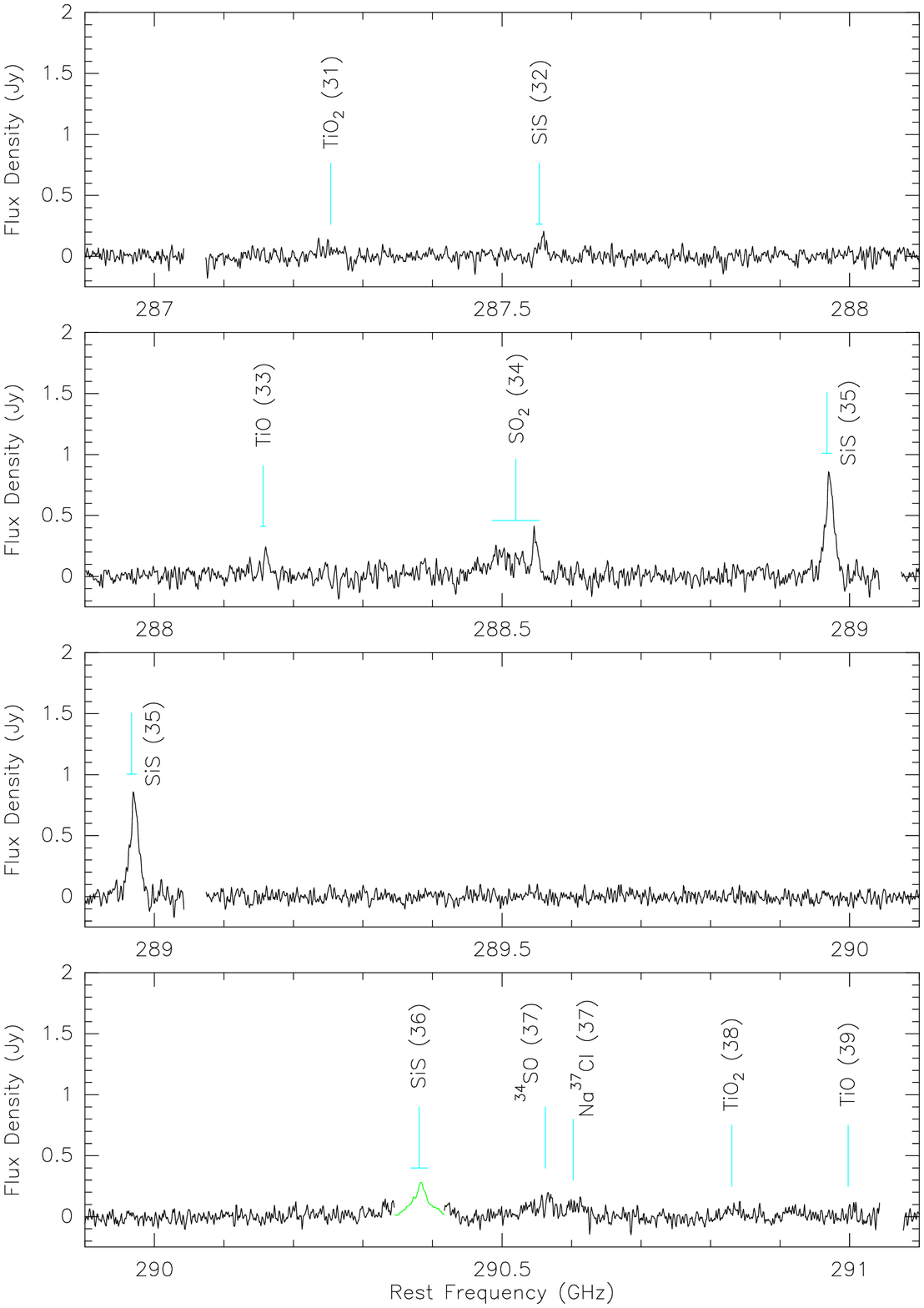}
\addtocounter{figure}{-1}
\caption{\it (Continued)}
\end{figure*}

\begin{figure*}\vskip -0.5cm
\includegraphics[width=0.95\textwidth]
{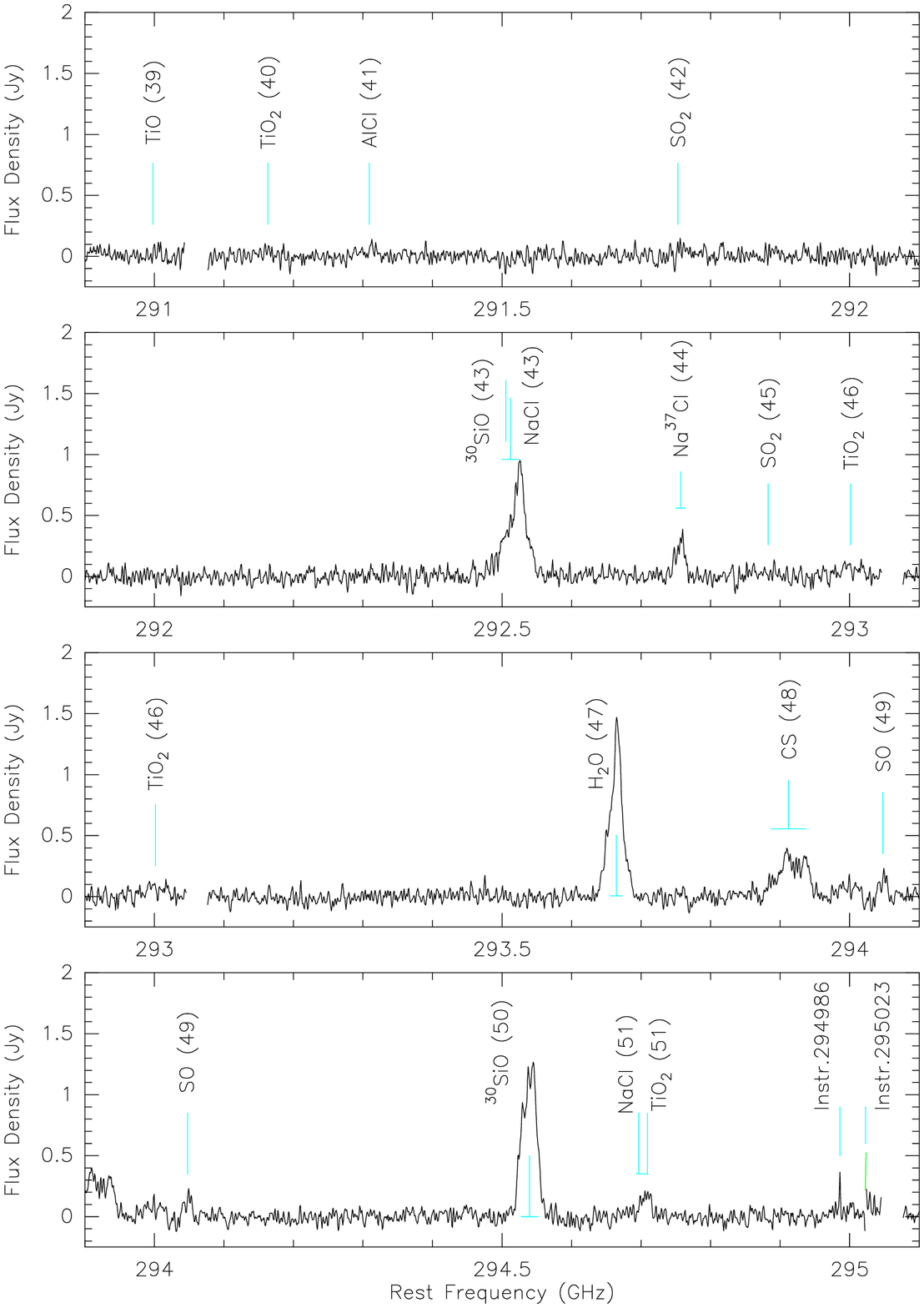}
\addtocounter{figure}{-1}
\caption{\it (Continued)}
\end{figure*}

\begin{figure*}\vskip -0.5cm
\includegraphics[width=0.95\textwidth]
{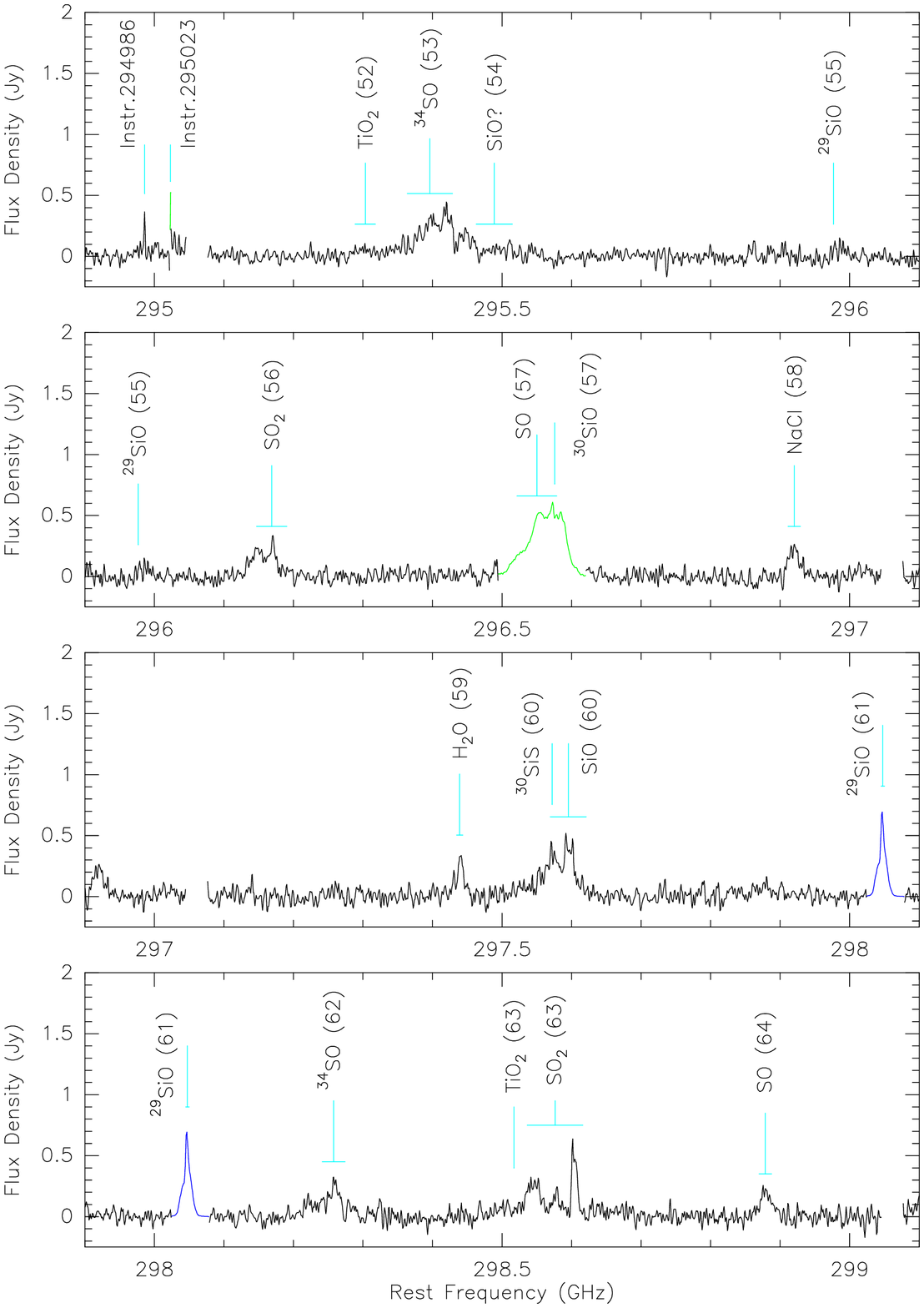}
\addtocounter{figure}{-1}
\caption{\it (Continued)}
\end{figure*}

\begin{figure*}\vskip -0.5cm
\includegraphics[width=0.95\textwidth]
{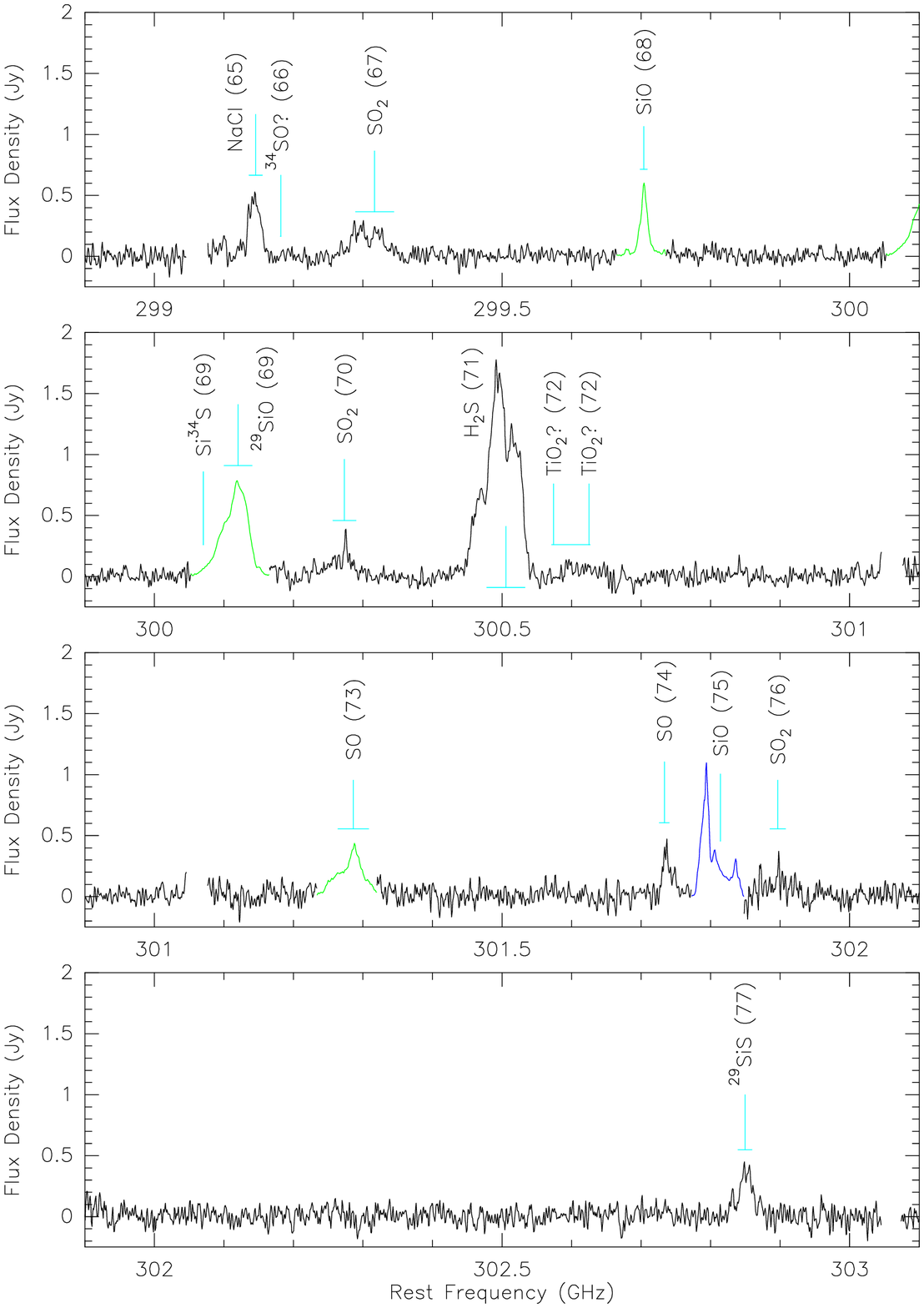}
\addtocounter{figure}{-1}
\caption{\it (Continued)}
\end{figure*}

\begin{figure*}\vskip -0.5cm
\includegraphics[width=0.95\textwidth]
{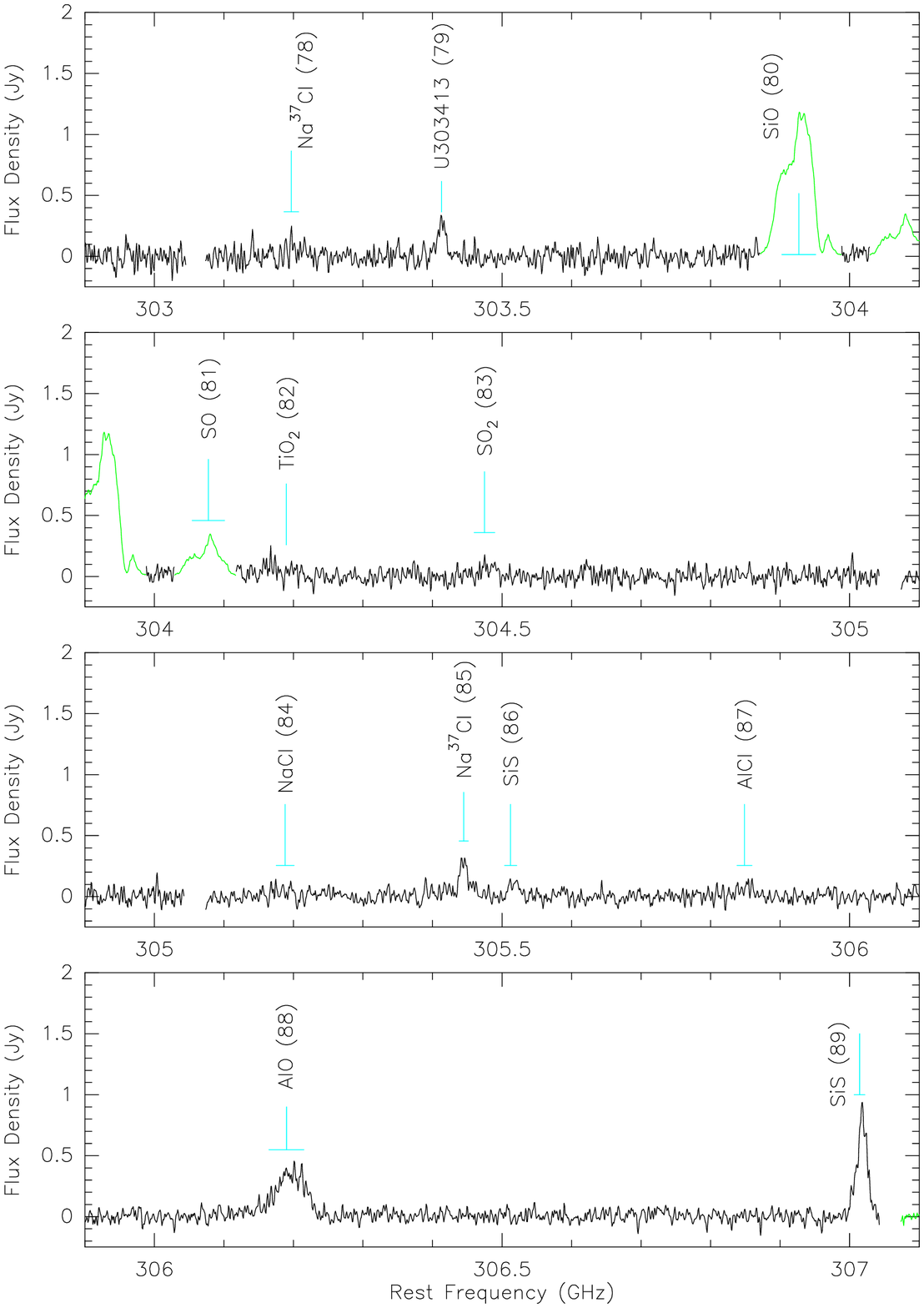}
\addtocounter{figure}{-1}
\caption{\it (Continued)}
\end{figure*}

\begin{figure*}\vskip -0.5cm
\includegraphics[width=0.95\textwidth]
{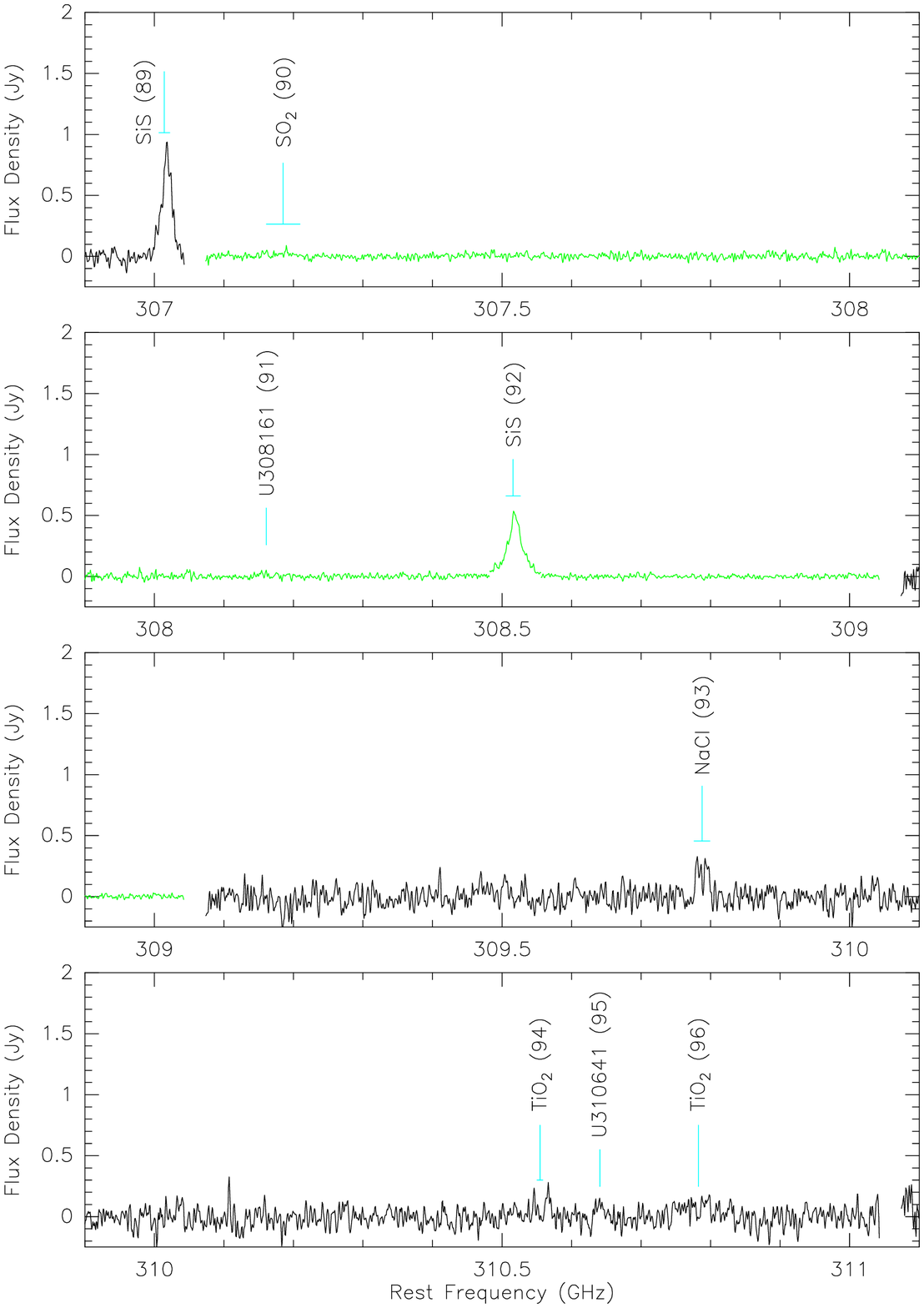}
\addtocounter{figure}{-1}
\caption{\it (Continued)}
\end{figure*}

\begin{figure*}\vskip -0.5cm
\includegraphics[width=0.95\textwidth]
{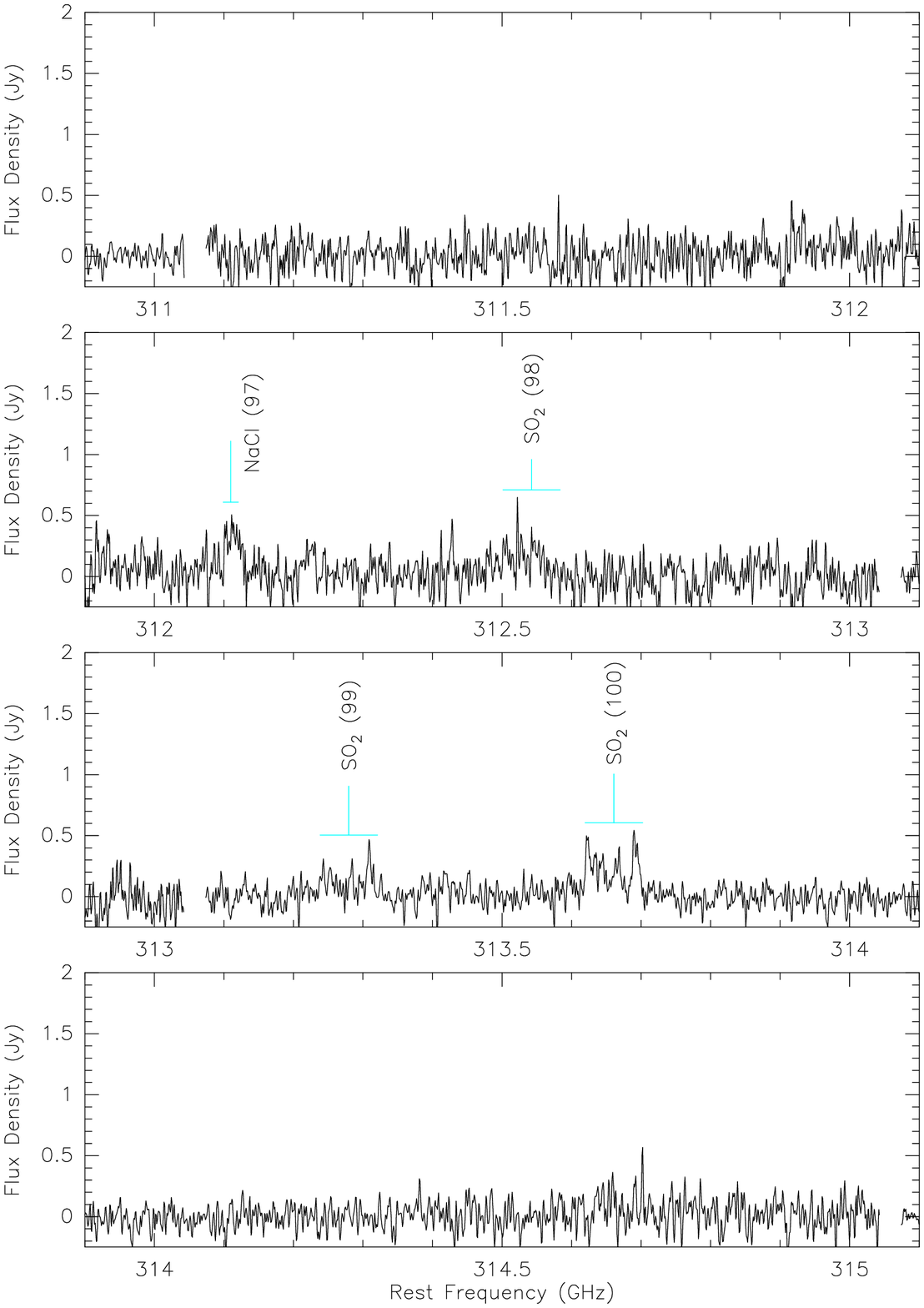}
\addtocounter{figure}{-1}
\caption{\it (Continued)}
\end{figure*}

\begin{figure*}\vskip -0.5cm
\includegraphics[width=0.95\textwidth]
{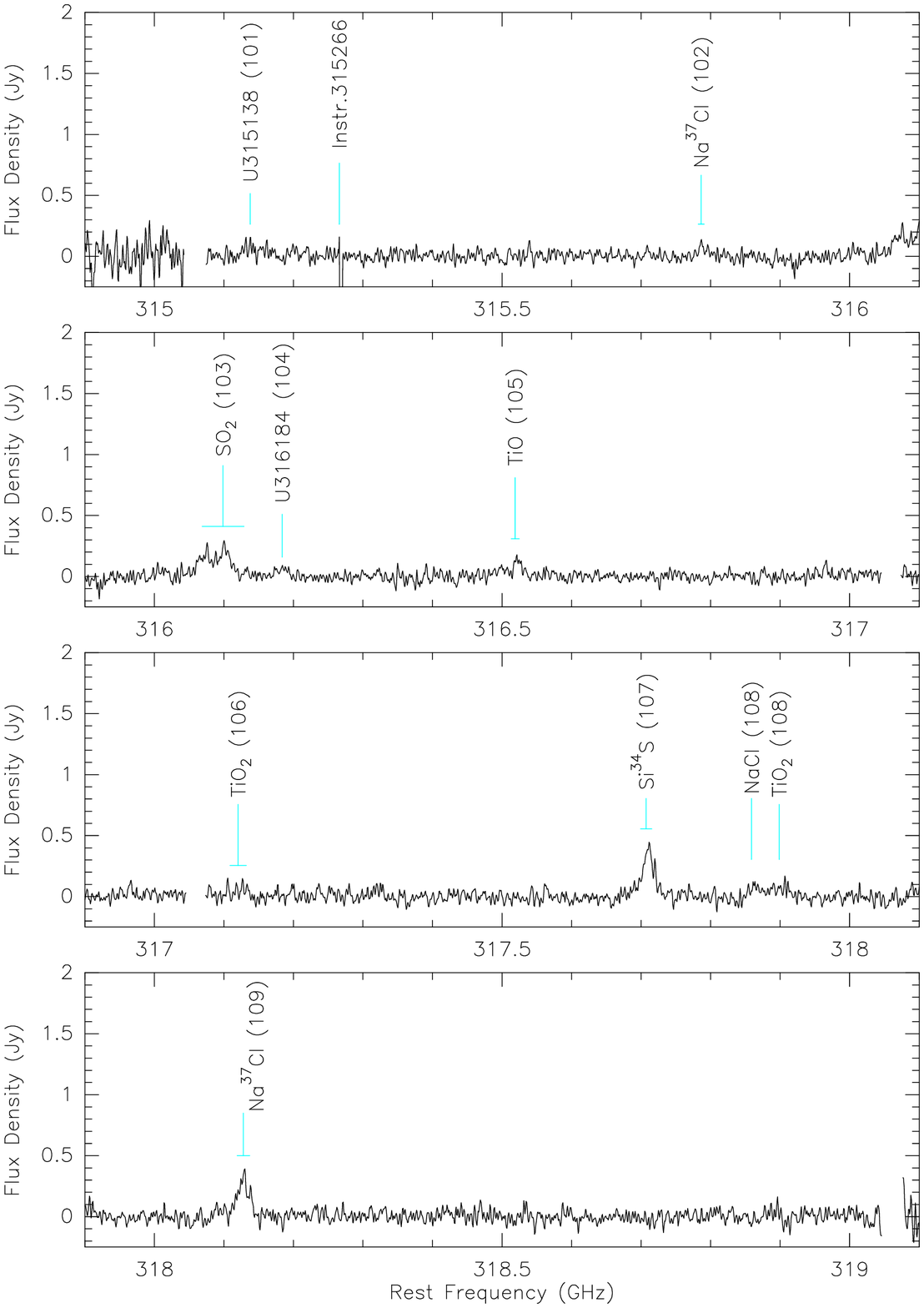}
\addtocounter{figure}{-1}
\caption{\it (Continued)}
\end{figure*}

\begin{figure*}\vskip -0.5cm
\includegraphics[width=0.95\textwidth]
{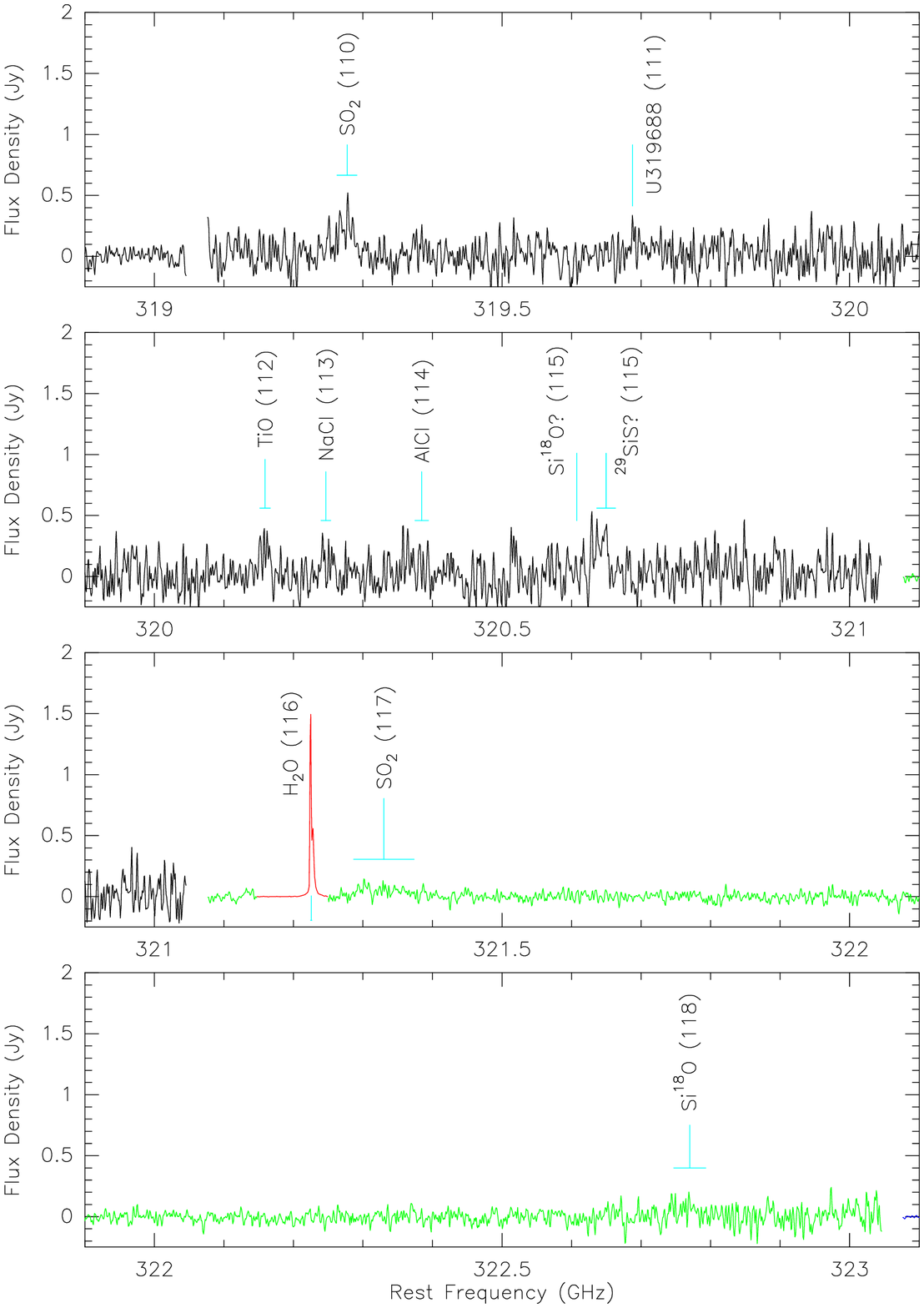}
\addtocounter{figure}{-1}
\caption{\it (Continued)}
\end{figure*}

\begin{figure*}\vskip -0.5cm
\includegraphics[width=0.95\textwidth]
{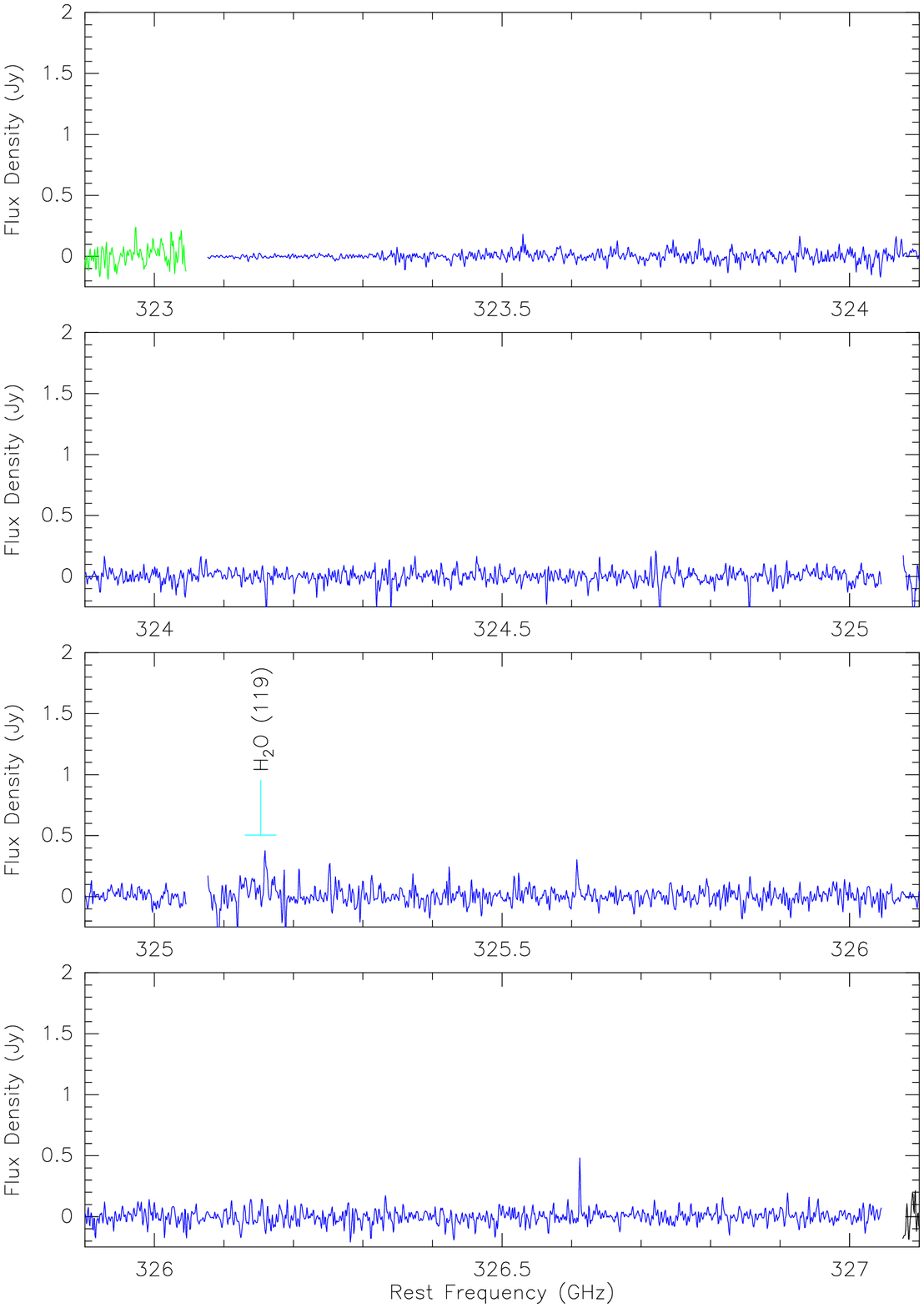}
\addtocounter{figure}{-1}
\caption{\it (Continued)}
\end{figure*}

\begin{figure*}\vskip -0.5cm
\includegraphics[width=0.95\textwidth]
{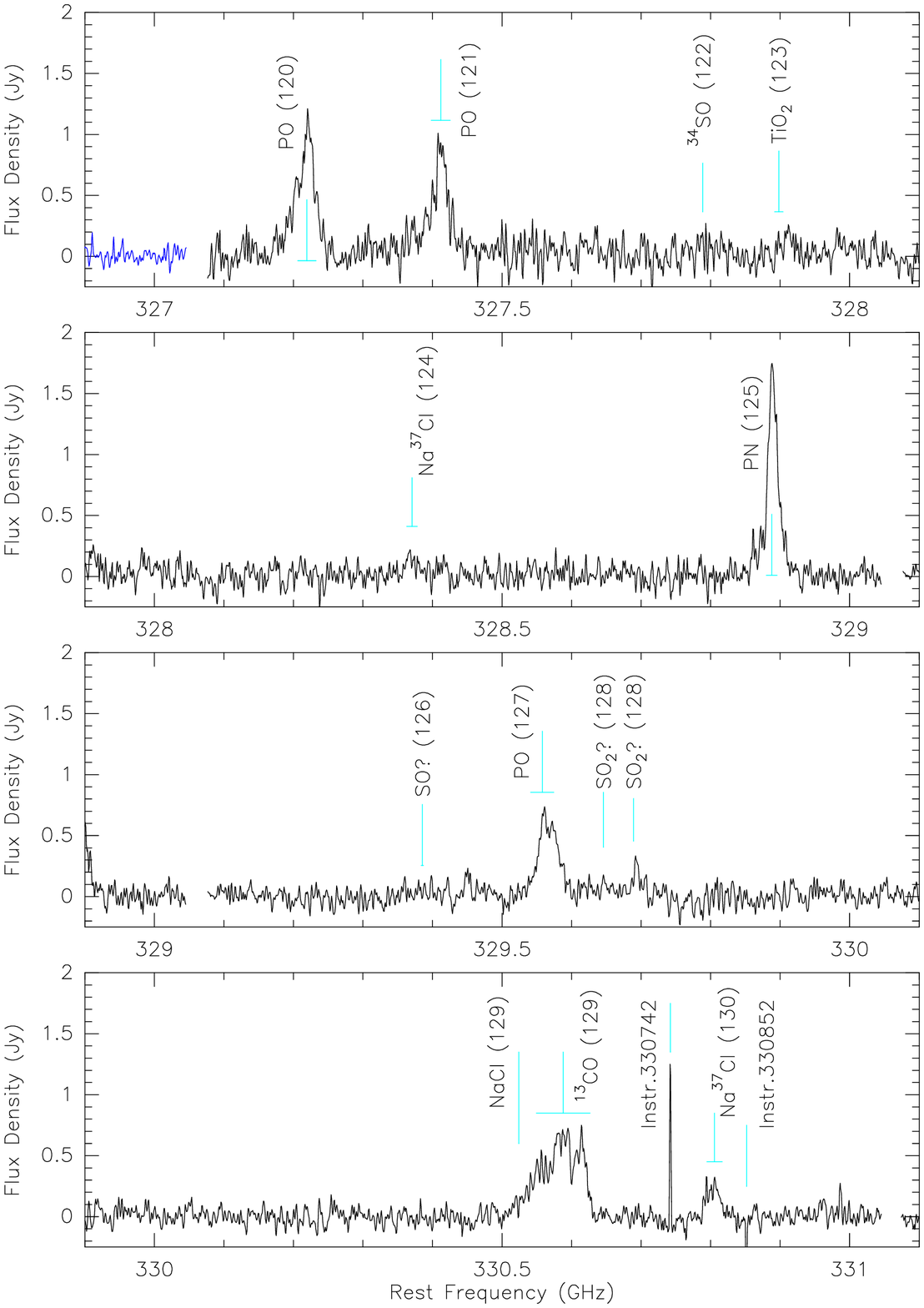}
\addtocounter{figure}{-1}
\caption{\it (Continued)}
\end{figure*}

\begin{figure*}\vskip -0.5cm
\includegraphics[width=0.95\textwidth]
{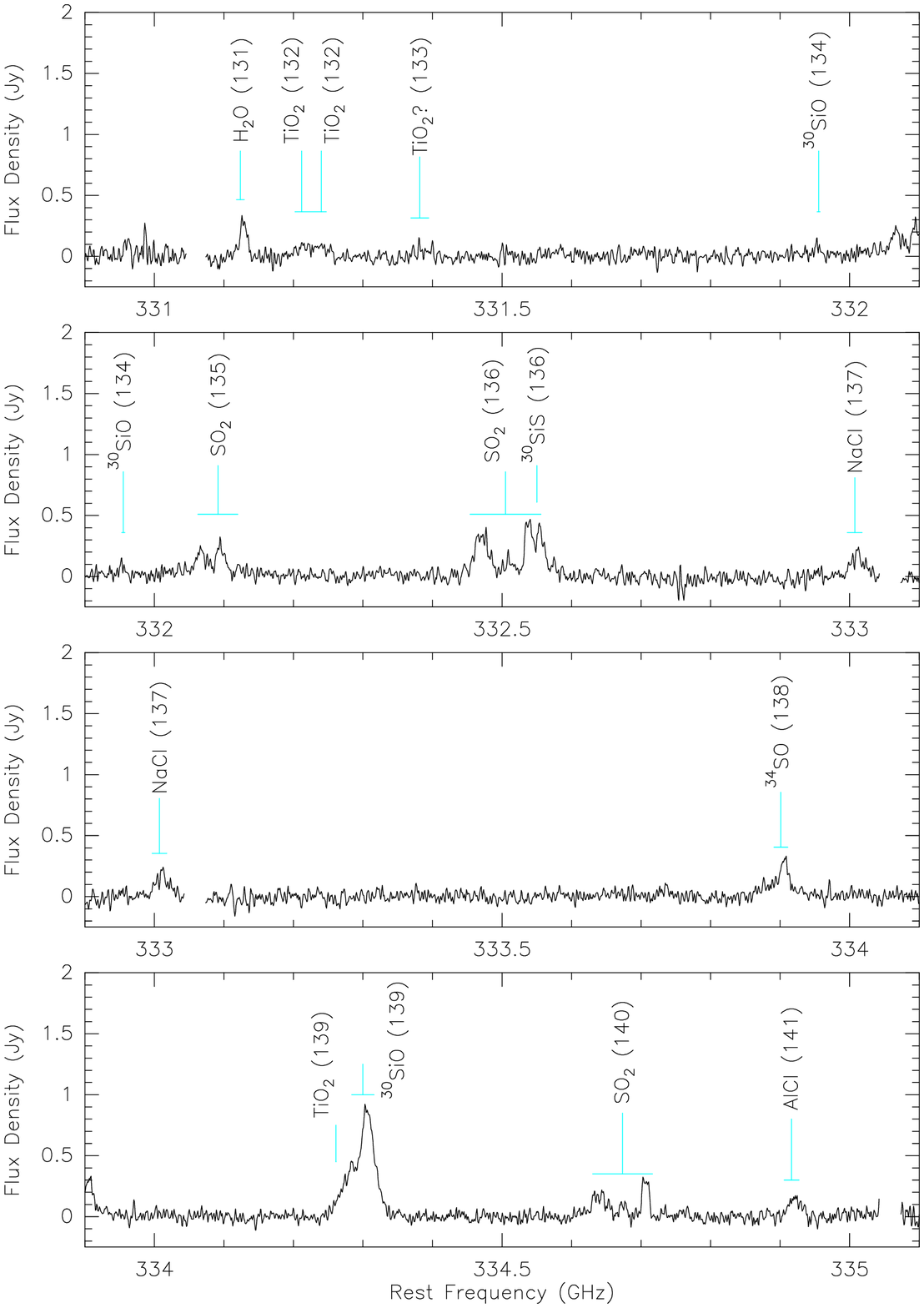}
\addtocounter{figure}{-1}
\caption{\it (Continued)}
\end{figure*}

\begin{figure*}\vskip -0.5cm
\includegraphics[width=0.95\textwidth]
{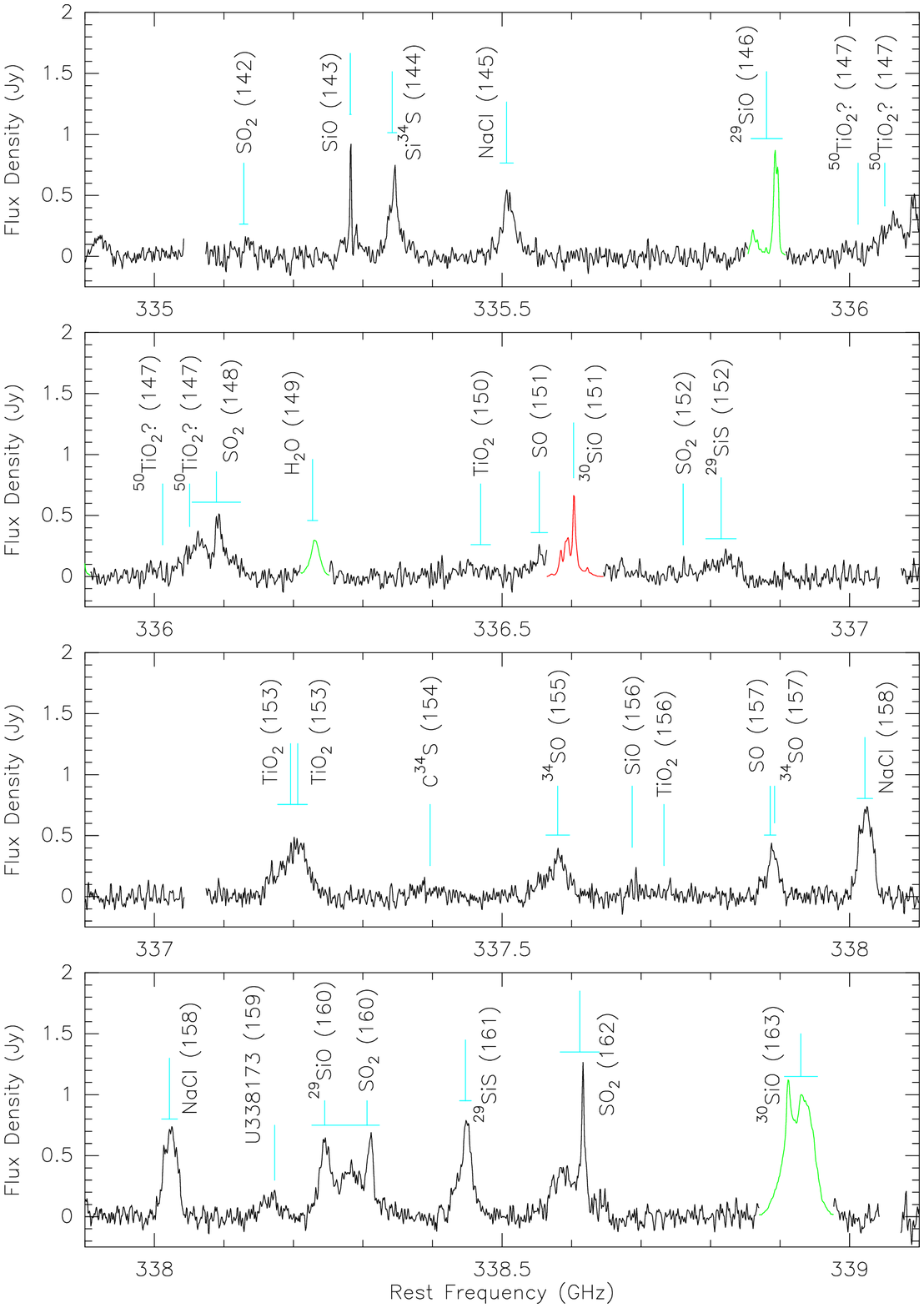}
\addtocounter{figure}{-1}
\caption{\it (Continued)}
\end{figure*}

\begin{figure*}\vskip -0.5cm
\includegraphics[width=0.95\textwidth]
{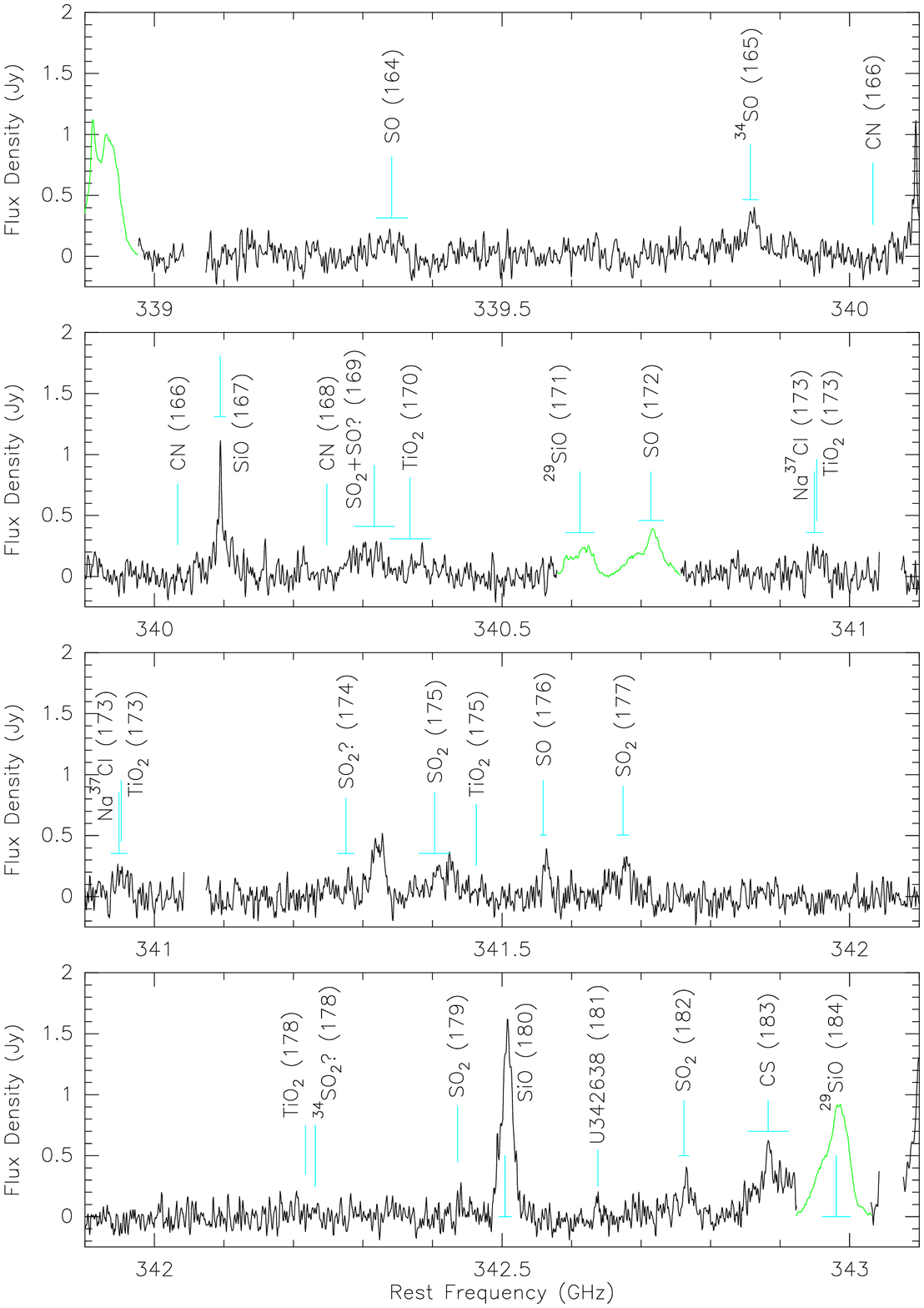}
\addtocounter{figure}{-1}
\caption{\it (Continued)}
\end{figure*}

\begin{figure*}\vskip -0.5cm
\includegraphics[width=0.95\textwidth]
{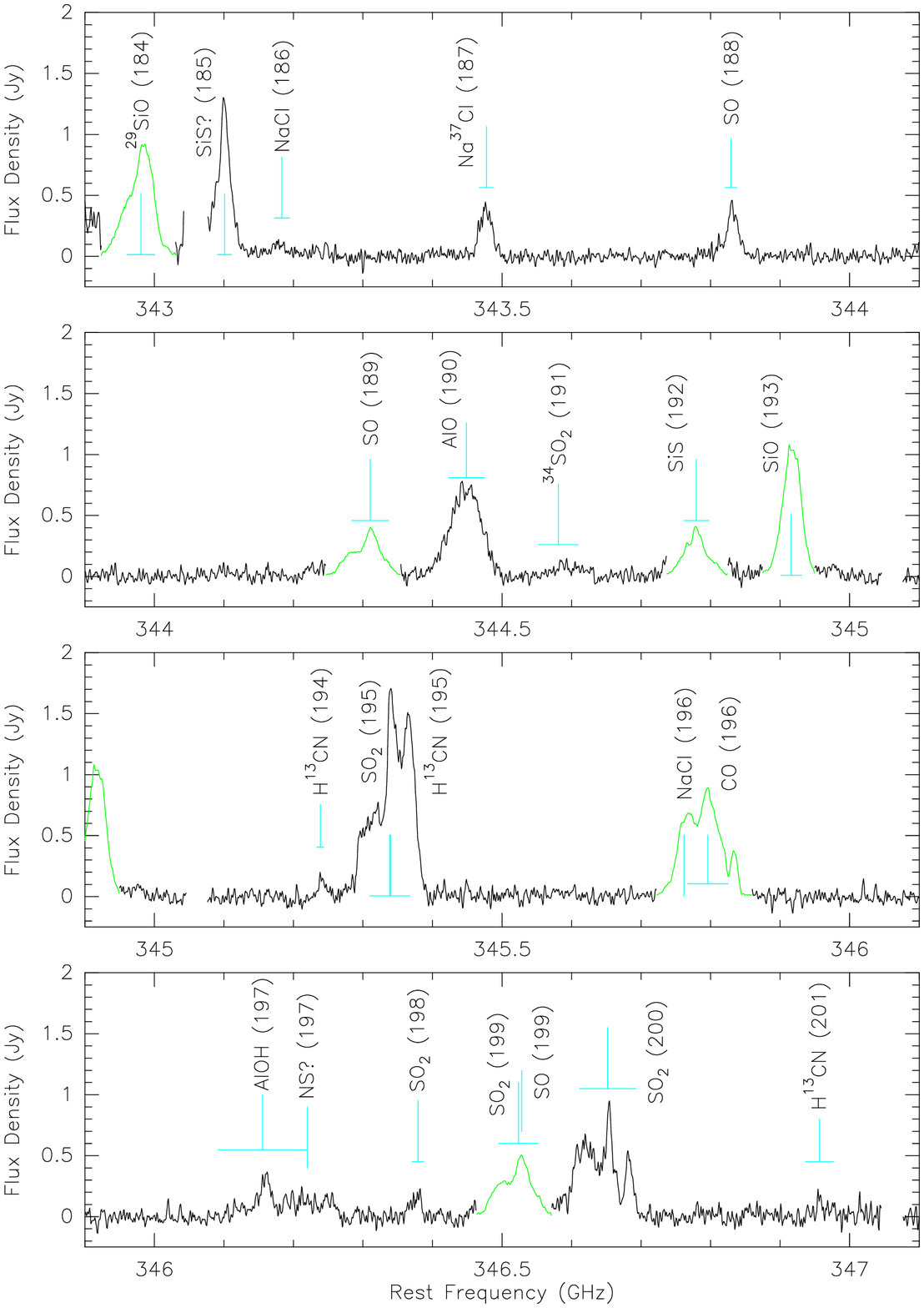}
\addtocounter{figure}{-1}
\caption{\it (Continued)}
\end{figure*}

\begin{figure*}\vskip -0.5cm
\includegraphics[width=0.95\textwidth]
{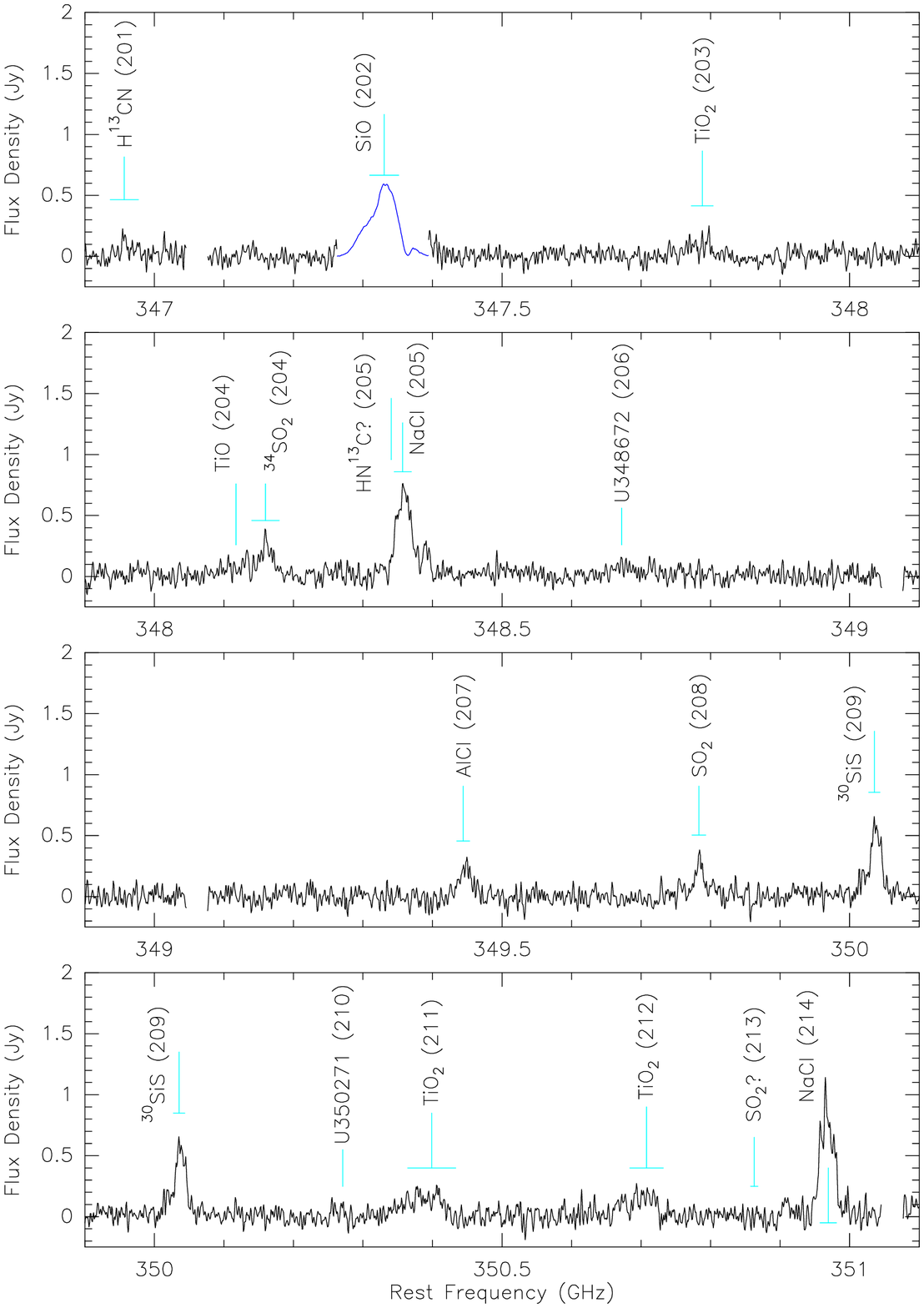}
\addtocounter{figure}{-1}
\caption{\it (Continued)}
\end{figure*}

\begin{figure*}\vskip -0.5cm
\includegraphics[width=0.95\textwidth]
{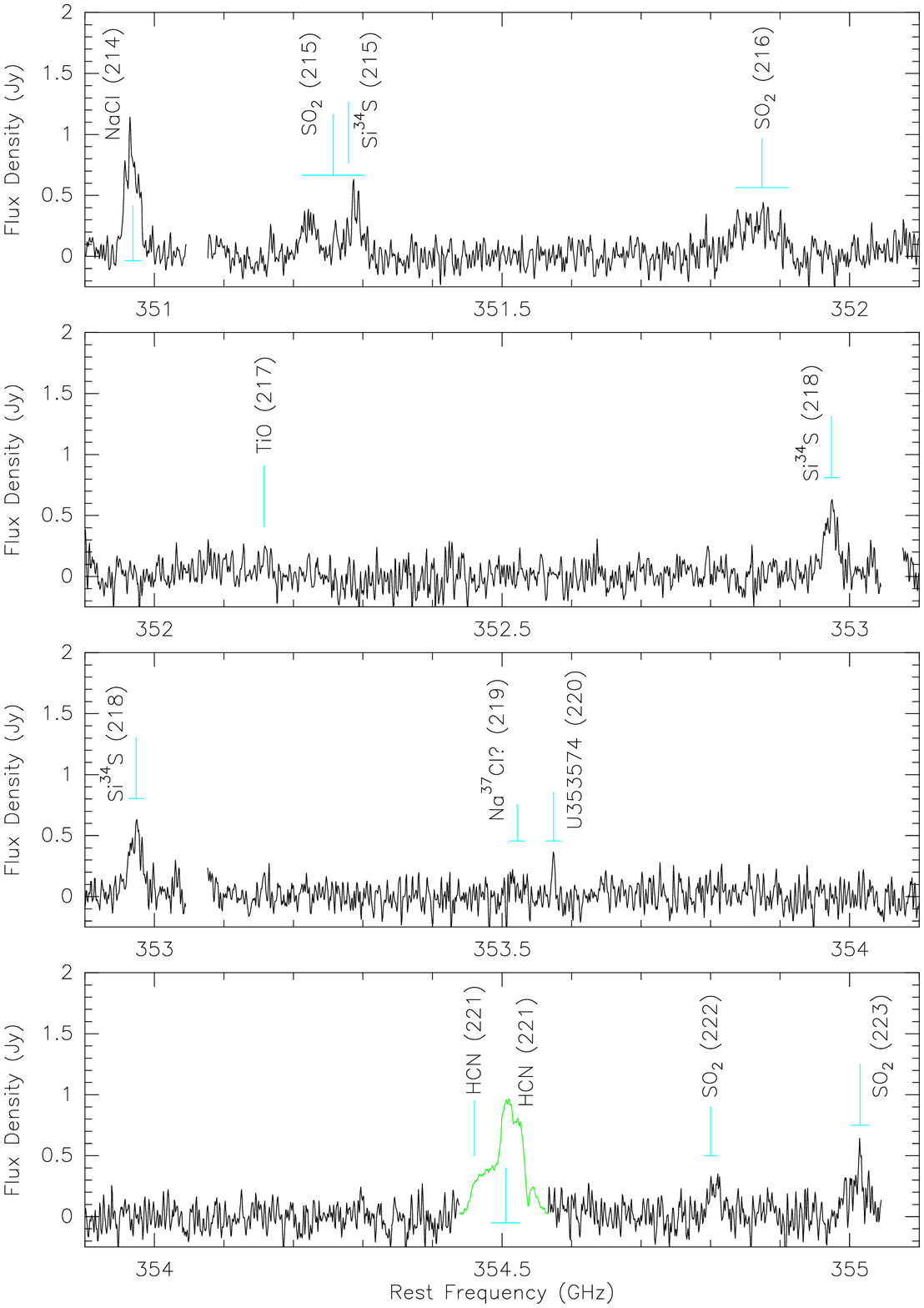}
\addtocounter{figure}{-1}
\caption{\it (Continued)}
\end{figure*}

\clearpage
\section{Analysis of population diagrams \label{sect-RDintro}}

We performed a population diagram analysis of majority of the molecules observed in the SMA survey. The primary purpose was to verify the identification of the more than 200 spectral features, but the excitation temperatures and column densities that we derive may also serve as a first step in more comprehensive studies in the future. Five sulfur-bearing molecules SO$_2$, SO, CS, H$_2$S, and SiS; two phosphorous-bearing molecules PO and PN; three aluminum-bearing molecules AlOH, AlCl, and AlO; and NaCl were analyzed. The excitation and abundances of TiO and TiO$_2$ was discussed elsewhere \citep{kami_tio}. 
Both H$_2$O and SiO were omitted in the analysis, because they are known to contain a strong maser component. Also omitted were CN and NS, because too few lines were observed or those observed were of insufficient quality. Following a general discussion of the analysis as it applies to the work here (Sect.\,\ref{sect-RDintro}), a brief description of the population diagrams for each molecule is presented. 


A population diagram is a graphical representation of the Boltzmann distribution \citep[e.g.,][]{goldsmith}
\begin{equation}
\ln \Biggl(\frac{N_u}{g_u}\Biggr) = 
\ln \Biggl(\frac{8 \pi k \nu^2 \int T_{\rm mb}\;dV}{B h c^3 A_{ul} g_u} \Biggr) =
\ln \Biggl(\frac{N_{\rm tot}}{Q}\Biggr)+\ln\biggl(e^{-E_u/k/T_{\rm exc}}\biggr)
\label{Eq-RD}
\end{equation}
as a function of $E_u/k$. Here, $k$ and $h$ are the Boltzmann and Planck constants; $c$ is the speed of light; $N_u$ and $N_{\rm tot}$ are the column density of molecules in the upper level of a transition and the total column density of the species of interest; $A_{ul}$, $g_u$, and $E_u$ are the Einstein coefficient for spontaneous emission, and the statistical weight  and the energy of the upper level; $T_{\rm exc}$ is the excitation temperature and $Q=Q(T_{\rm exc})$ is the partition function for that temperature. 
The line intensities measured with the SMA are in flux units (Jy/beam).
From these, the line intensity integrated over velocity in temperature units ($\int T_{\rm mb}\;dV$) was calculated in the Rayleigh-Jeans limit from the velocity-integrated fluxes ($I_F=\int F_{\nu}\;dV$), where
\begin{equation}
\int T_{\rm mb}\;dV=\frac{c^2 I_F}{2 k \nu^2 \Omega_{\rm b}},
\end{equation} 
and $\Omega_{\rm b}$ is the beam solid angle.  
For a Gaussian beam with major and minor axes of $\theta_a$ and $\theta_b$ (FWHM), 
$\Omega_{\rm b}=\pi/(4 \ln2)\theta_a \theta_b$. 
If  we had used the Planck expression instead of the Rayleigh-Jeans approximation, the intensities expressed in K would typically be only a few percent higher at the observed frequencies and temperatures, i.e., much less than the uncertainties in the measurements.

The beam filling factor ($B$ in Eq.\,\ref{Eq-RD}), is especially important for extended emission and when measurements with different telescopes are analyzed simultaneously. 
For flux integrated within the solid angle $\Omega_{\rm ap}$  and a source size of $\Omega_s$, the beam filling factor is $B=\Omega_s/(\Omega_s+\Omega_{\rm ap})$. 
In the analysis here, $\Omega_s$ is obtained from the angular distribution of the resolved emission or,  if necessary, by deconvolution, although  in many cases the source size was assumed.\footnote{Listed in Table\,\ref{Tab-RD-results} is $\Omega_s$ used in the analysis, so that the column densities may be  recalculated if better constraints on the source size become available.} 
For unresolved sources, we usually used the beam-averaged fluxes which for the typical beam size of our survey ($\theta \approx 0\farcs9$) is $\Omega_{\rm ap}$=$\Omega_{\rm b}$=0.92\,arcsec$^2$. 
For CS and SO$_2$, the population diagrams were constructed for a few spatial and/or kinematical components and the solid angles are specified for these components accordingly.

The Boltzmannian distribution of states applies to an isothermal gas in (local) thermodynamic equilibrium (LTE). Undoubtedly, LTE does not apply to unresolved emission arising close to the strong source of infrared and optical radiation which can radiatively excite molecules. 
Also, the molecular envelope is not isothermal, but rather there is a temperature gradient that was not taken into account here. Not surprisingly, many of the rotational temperature diagrams cannot be reproduced by a single linear fit, particularly when rotational transitions from excited vibrational levels  as well as those in the  ground state are observed (NaCl, SiS, and SO).
In that case, rotational lines from high lying and low lying levels were fit separately.
An excitation temperature derived from transitions in the ground vibrational state is referred to as a rotational temperature ($T_{\rm rot}$), while the temperature derived from transitions including those in vibrationally excited levels is interpreted as a vibrational temperature ($T_{\rm vib}$).    

The parameters characterizing a rotational transition ($A_{ul}g_u$ or $\mu^2S_{ul}$, and $E_u$) are from CDMS or JPL \citep{cdms1,cdms2,jpl}. In Table\,\ref{Tab-RD-results}, we list the values of the electric dipole moments ($\mu$) from which the $A_{ul}$ coefficients were calculated in catalog entries that we used in the analysis. The partition functions were interpolated (or in a few cases extrapolated) for the given excitation temperature using the tabular data in CDMS or, if absent in CDMS, from JPL (e.g., partition functions for all the aluminum-bearing molecules are from JPL). In all cases where excitation of excited vibrational levels is important, the partition function includes the vibrational contribution.   

If  the rotational transitions of a molecule did not span a wide enough range of excitation energies to allow the determination of the rotational temperature, measurements in other frequency bands (preferably with interferometers, or otherwise with single antennas) were included in the analysis. 
Specifically, these included measurements with the 
SMA  \citep[Fu+;][]{fu}  and Plateau de Bure interferometers \citep[PdBI 2009;][]{ver_phd},\citep[PdBI 2012;][]{kami_tio}; and the ARO 10\,m  \citep[ARO;][]{tenen_survey}, {\it Herschel} 3.5\,m  \citep[{\it Herschel};][]{herschel}, and IRAM 30\,m \citep[IRAM;][]{sandra_tio2} single antennas.
(Shown in the parentheses are the symbols used to designate the interferometers and antennas referred to in Fig.\,\ref{Fig-RDs1}.) 
We also extracted spectra from the APEX archives (to be described elsewhere). 
When measurements with single antennas are combined with those from interferometers, part of  the fluxes derived with interferometers may be missing, especially if the emission is extended (see Sect.\,\ref{obs}). In general, we only used those lines in the analysis that are not blended  with other features.     

The excitation temperatures and total column densities were derived from a linear fit to points on the population diagram of each molecule \citep[see e.g.,][]{snyder}. A summary of the rotational and vibrational temperatures, column densities, aperture sizes, and velocity ranges are given in Table\,\ref{Tab-RD-results}. The indicated uncertainties are only approximate and were estimated from the noise in spectra, in particular they do not take into account calibration uncertainties. All the analyzed  diagrams are shown in Fig.\,\ref{Fig-RDs1}. 

A linear trend in the population diagrams is expected for lines that are optically thin \citep{goldsmith}.  
The optical depth at line center, $\tau$, was estimated from
\begin{equation}\label{eq_tau}
\tau=
\frac{A_{ul} g_u c^3 N_{\rm tot}}
     {8 \pi \nu^3 Q \Delta\!V} 
e^{-E_u/k/T_{\rm exc}}
\bigl(e^{h \nu/k/T_{\rm exc}}-1\bigr),
\end{equation}  
where $\Delta\!V$ is the FWHM of the line of interest, and $T_{\rm exc}$ and $N_{\rm tot}$ were derived from the linear fits. It appears that all lines analyzed here are optically thin (i.e., $\tau \ll 1$), including the most intense lines of SO in $v=0$ whose optical depths are higher than all other lines analyzed here ($\tau \sim 0.06$; see Sect.\,\ref{sect-RT-SO}).
We did not correct the population diagrams for optical depth effects, because the correction $C_{\tau}=\tau/(1-e^{-\tau})$ \citep[see e.g.,][]{goldsmith,snyder} is at most 1.03 for SO, and is
typically much less than this for all other molecules --- we did not analyze CO and HCN, because they are optically thick (see Sects.\,\ref{Sect-CO} and \ref{Sect-HCN}).

A more sophisticated treatment of the excitation conditions of the molecular gas around VY\,CMa, which includes non-LTE effects and a realistic thermal structure of the source, will be presented elsewhere.

\begin{table}
\begin{center}
\caption{Excitation temperatures and column densities derived from a population diagram analysis\label{Tab-RD-results}}
\begin{scriptsize}
\begin{tabular}{lcccccc}
\tableline \tableline
Species &$\mu$&$T_{\rm exc}$&$N_{\rm tot}$&$\Omega_s$&Vel.\,range&Spatial\\
               & (D) & (K) & (cm$^{-2}$) & (arcsec$^2$) & (\kms) & component\\ 
\tableline \tableline
SO$_2$ $v=0$       &1.63& $59\pm6$ & $(4.0\pm1.2)\times10^{15}$& 3.8 &[$-20:5$] (neg.)&central\\
$^{34}$SO$_2$ $v=0$&1.63& 59\tbm{a}& $(6.3\pm1.4)\times10^{14}$& 3.8 &[$-20:5$] (neg.)&central\\
SO$_2$ $v=0$       &1.63& $407\pm58$&$(1.1\pm0.1)\times10^{16}$& 3.8 & [$5:30$] (cent.)&central \\
$^{34}$SO$_2$ $v=0$&1.63& 407\tbm{a}&$(1.1\pm0.2)\times10^{15}$& 3.8 & [$5:30$] (cent.)&central\\
SO$_2$ $v=0$& 1.63 & $160\pm11$& $(1.1\pm0.1)\times10^{15}$& 3.8 & [$30:65$] (pos.)&central\\
SO$_2$ $v=0$& 1.63 & $62\pm6$ & $(4.1\pm1.1)\times10^{15}$& 3.8 &[$-20:5$] (neg.)&NE\,source\\
SO$_2$ $v=0$& 1.63 & $55\pm9$ & $(3.7\pm1.5)\times10^{15}$& 2.9 &[$30:65$] (pos.)&NW\,Arc\\[4pt]
%
SO $v=0$       & 1.54 &  $67\pm13$& $(1.4\pm0.3)\times10^{16}$& 0.56 &[$ 10:30$] (cent.)& central\\
SO $v \leq 1$  & 1.54 &$ 590\pm36$& $(6.2\pm0.8)\times10^{16}$& 0.56 &[$ 10:30$] (cent.)& central\\
SO $v \leq 1$  & 1.54 & $512\pm23$& $(2.1\pm0.2)\times10^{17}$& 0.56 &[$-20:70$] (full) & central\\
$^{34}$SO $v=0$& 1.54&  590\tbm{a}& $(4.5\pm0.3)\times10^{15}$& 0.56 &[$ 10:30$] (full)& central\\
$^{34}$SO $v=0$& 1.54 &  $69\pm32$& $(9.6\pm4.3)\times10^{15}$& 0.73 &[$ 10:30$] (full)& central \\[4pt]
SiS $v=0$        & 1.74 & $157\pm34$& $(3.6\pm0.7)\times10^{16}$& 1.0 &[$-30:75$] (full)& central \\
SiS $v \leq 2$   & 1.74 & $576\pm39$& $(9.0\pm1.8)\times10^{16}$& 1.0 &[$-30:75$] (full)& central \\
Si$^{34}$S $v=0$ & 1.74 & $436\pm10$& $(5.1\pm0.3)\times10^{15}$& 1.0 &[$-30:75$] (full)& central \\
$^{29}$SiS $v=0$ & 1.74 & 576\tbm{a}& $(9.1\pm0.6)\times10^{15}$& 1.0 &[$-30:75$] (full)& central \\[4pt]
CS $v=0$&1.96& $39\pm1$& $(1.4\pm0.1)\times10^{15}$& 0.79 & [$-15:60$] (full)& central\\
CS $v=0$&1.96& $28\pm4$& $(1.6\pm0.4)\times10^{15}$& 0.79 & [$-15:60$] (full)& SW\,Arc\\
CS $v=0$&1.96& $16\pm4$& $(1.1\pm0.8)\times10^{15}$& 6.88 & [$-15:60$] (full)& full\\
C$^{34}$S $v=0$&1.96&39\tbm{a}&$\approx1.2\times10^{14}$& 0.79& [$-15:60$] (full)& central\\[4pt]
H$_2$S $v=0$& 0.98 & $192\pm17$& $(1.9\pm0.2)\times10^{17}$& 0.94 &[$-25:85$] (full) &central\\[4pt]
%
PO &1.88&$244\pm29$&$(6.5\pm1.2)\times10^{16}$& 0.18 &[$-20:70$] (full)&central\\
%
PN $v=0$ & 2.75 & $130\pm10$& $(3.7\pm0.3)\times10^{15}$& 0.28 &[$-5:40$] (full) & central \\[4pt]
AlOH $v=0$& 1.04 &$>50$& $\sim 10^{16}$& 0.10 & [$0:40$] (full) & central \\
AlCl $v=0$& 1.00 &$290\pm150$&$(1.0\pm0.3)\times$10$^{16}$& 0.28 &[$-5:40$] (full) & central\\
AlO $v=0$ & 4.60 &$1055\pm191$& $(1.9\pm0.1)\times10^{16}$& 0.10 &[$-20:70$]\tbm{b}(full)& central\\[4pt]
%
NaCl $v=0$    & 9.00     & $210\pm25$ &$(1.3\pm0.1)\times10^{15}$& 0.10 & [$0:45$] (full)& central\\
NaCl $v\leq 3$&9.1\tbm{c}& $591\pm61$ &$(3.6\pm0.2)\times10^{15}$& 0.10 & [$0:45$] (full)& central\\
NaCl $v\leq 3$\tbm{d}&9.1\tbm{c}&$959\pm123$&$(4.8\pm0.3)\times10^{15}$& 0.10 & [$0:45$] (full)& central\\
Na$^{37}$Cl $v\leq1$&9.00& $562\pm139$&$(1.2\pm0.1)\times10^{15}$& 0.10 & [$0:45$] (full)& central\\
\tableline
\end{tabular}
\end{scriptsize}
 
\tablenotetext{a}{The excitation temperature is assumed to be the same as that of the main species.}
\tablenotetext{b}{The velocity is given with respect to the central hyperfine component (see Sect.\,\ref{sect-al}).}
\tablenotetext{c}{The dipole moments for $v=1$,\,2,\,3 are 9.06, 9.12, and 9.18\,D.}
\tablenotetext{d}{Fit to SMA survey data only.}
\end{center}
\end{table}

\subsection{SO$_2$}\label{sect-RT-SO2}

Owing to the complex line shape and extended emission of SO$_2$, the flux was derived from integrated intensity maps in three LSR velocity ranges: --20:5\,\kms\ (negative); 5:30\,\kms\ (central); and 30:65\,\kms\ (positive) which correspond to the velocity ranges of the spatial and kinematical components A, D, and B in \citet{fu}. For the central source (D), we used a circular aperture with a radius of 1\farcs1 that was offset 0\farcs5 west and 0\farcs2 north of the  continuum source. For the NE source (A), dominated by emission at negative velocities, a circular aperture of the same size as that of the central source was used. In order to provide better coverage of source A, the aperture was displaced 0\farcs5 east and 0\farcs32 north from the continuum center. For the NW\,Arc (source B), an elliptical aperture enclosing most of the signal for the ``positive'' velocity component was used. This particular aperture does not include the region associated with the southern arcs --- i.e., Arcs\,1 and 2, the SW\,Arc, and S\,Arc were omitted, and we did not analyze source C of Fu et al. The ellipse is $13\arcsec \times 7\arcsec$ with the center located 2\farcs6 west and 0\farcs1 north of the continuum peak, and the major axis inclined by 55\degr\ to the meridian (counting west from north).

Typically 10--20 transitions, spanning excitation energies ($E_u$) as high as 200--1000~K, were observed in the survey. As a result, $T_{\rm{rot}}$ was well determined with uncertainties of about 10\%, except for the NW\,Arc where the uncertainty in $T_{\rm{rot}}$ is about 20\%. As expected, the column densities are also well determined. The rotational temperatures of the different components are within a broad range from 55~K in the NW\,Arc to 407~K for the central velocity component in the central source.
On the basis of the derived parameters from the rotational-diagram analysis, we conclude that the lines of SO$_2$ are optically thin with $\tau \sim 0.001$ or less.

Although \citet{fu} measured only four transitions in SO$_2$ from levels over a limited range in energy above ground of $E_u=19-131$~K, their derived rotational temperatures and column densities are consistent with those derived from the much more extensive measurements here. We find that our derived rotational temperatures and column densities agree with the radiative transfer analysis by Fu et al. 

There are only two unblended lines of $^{34}$SO$_2$ in the survey. The line at 344.6\,GHz ($19_{1,19}-18_{0,18}$) is barely discernible in the spectrum obtained with an aperture of $1\arcsec \times 1\arcsec$, and the second line ($57_{6,52}-56_{7,49}$ at 327.8\,GHz) is in a noisy part of the survey and may be blended with $^{34}$SO. Only the line of $^{34}$SO$_2$ at 344.6\,GHz was analyzed. The negative and central velocity components are both present in this line, but the positive velocity component is absent. The integrated intensity maps were made with the same velocity intervals as that for SO$_2$. On the assumption that the excitation of $^{34}$SO$_2$ is the same as that for the main isotopic species, the $N$(SO$_2$)/$N(^{34}$SO$_2$) ratio is $6\pm3$ for the negative velocity component and $10\pm3$ for the central velocity component --- i.e., for both velocity components, the S/$^{34}$S ratio in SO$_2$ is somewhat less than that in SO (Sect.\,\ref{sect-RT-SO}). This ratio is even lower for the positive velocity component, because it was not detected in $^{34}$SO$_2$.  The derived S/$^{34}$S ratios, which are significantly smaller than the solar elemental ratio \citep[22.6;][]{lodders}, are not attributable to optical depth effects.

\subsection{SO and $^{34}$SO}\label{sect-RT-SO}

Owing to the broad range of $E_u$, the vibrational temperature of SO was well determined from the SMA measurements. Two velocity ranges were considered in the analysis: a wide range ($-20$ to 70\,\kms) which encompasses the entire emission of the main species of SO in the ground vibrational state in the central source; and a narrow range (10 to 30\,\kms) appropriate for $^{34}$SO and the $v = 1$ excited vibrational level of SO. No attempt was made to analyze other spatial components, because they only are detected in the $v=0$ lines of SO.

Shown in Fig.\,\ref{Fig-RDs1} are population diagrams for the narrow velocity range (results for the broader range are included in Table\,\ref{Tab-RD-results}), and a circular aperture of 0\farcs7 diameter that encloses the central source only. For SO, independent fits to all the measurements in $v=0$ and 1 and to those in $v=0$ alone, yield vibrational and rotational temperatures which differ by one order of magnitude ($T_{\rm vib}=590$~K vs. $T_{\rm rot}=67$~K). Considering the high S/N of the analyzed data, this effect is real and indicates excitation conditions violating LTE (or the assumption of an isothermal source). Because $^{34}$SO was only detected in $v=0$, the column density was derived on the assumption that the vibrational temperature is the same as that of the main species, thereby allowing us to derive the ratio of the column densities of the main and sulfur--34 species. The derived 
SO/$^{34}$SO ratio of 14$\pm$3 is higher than that found for SO$_2$, but is much smaller than the solar elemental ratio of the two isotopes 
\citep[S/$^{34}$S=22.6;][]{lodders}.    

An independent estimate on the rotational temperature of $^{34}$SO was obtained by combining the SMA measurements with single-antenna measurements obtained elsewhere.  These included a line observed with APEX ($12_{11}-11_{10}$ at 465.3\,GHz), and two lines observed with the ARO ($6_5-5_4$ and $6_6-5_5$). Because single-antenna measurements correspond  to the total spatial emission, the SMA lines were measured for the entire source (i.e., they were not restricted by aperture as described above for the SMA measurements alone). The rotational temperature diagram for the combined analysis is shown in a separate panel of Fig.\,\ref{Fig-RDs1}. The rotational temperature for $^{34}$SO of 69~K is in very good agreement with that derived for the main species ($T_{\rm rot}=$67~K).

As mentioned in Sect.\,\ref{sect-RDintro}, we estimate that the lines of SO at $v=0$ have the highest opacity among all lines analyzed, nevertheless they are optically thin with $\tau \lesssim 0.1$. Therefore, optical depth effects appear to not be the cause of the anomalous S/$^{34}$S ratio derived here. Additional evidence in support of this interpretation is the agreement of $T_{\rm rot}$  for SO and $^{34}$SO.


\subsection{SiS and Si$^{34}$S (and $^{29}$SiS)}


In SiS, the source consists of two spatial components (see Fig.\,\ref{Fig-morph}). In principle, the two regions which contribute to overall flux may have different temperatures. We find that the second source (SW\,Arc) is only observed in rotational transitions in $v=0$ and the flux is much less than that of the main (central) source. Therefore, vibrational-temperature diagrams were constructed for SiS and Si$^{34}$S from lines extracted with an aperture which encompasses the central source, but does not include emission from the weaker source in the SW\,Arc. The vibrational temperature derived for SiS of  $T_{\rm{vib}}=580 \pm 40$~K agrees to within the $3\sigma$ uncertainties with that derived for Si$^{34}$S, $T_{\rm{vib}}=436 \pm 10$~K. The ratio of column densities $N({\rm{SiS}})/N({\rm{Si}}^{34}{\rm{S}})$ of $18 \pm 5$ is close to the solar S/$^{34}$S ratio \citep[22.5;][]{lodders}. 

We also measured lines of the $^{29}$SiS molecule in $v=0$. The points in the rotational temperature diagram corresponding to $^{29}$SiS lie in the same part of the diagram as those of Si$^{34}$S, but are not shown in Fig.\,\ref{Fig-RDs1}. On the assumption that $T_{\rm{vib}}$  for $^{29}$SiS and SiS are the same, we estimate that the column density of $^{29}$SiS is comparable to that of Si$^{34}$S. Our estimate of $N$($^{29}$SiS) yields an elemental abundance of Si/$^{29}$Si$=10\pm1$ that is two times less than the solar value \citep[19.7;][]{lodders}. 

 The rotational temperature of SiS could not be determined from the survey data alone, because only four rotational transitions were observed in each vibrational level at most. Instead, we attempted to derive $T_{\rm{rot}}$ by combining our measurements with lines of SiS observed with the SMA by \citet{fu}, the $J=5-4$ transition observed with the PdBI in 2009 \citep{ver_phd}, ten transitions measured with ARO \citep{tenen_survey}, and two measurements with APEX at 361.1 and 362.9\,GHz. Although we were unable to determine  $T_{\rm{rot}}$ in the combined analysis owing to uncertainties in the measurements, it appears that  $T_{\rm{rot}}$ is about 160~K (see the gray line in the appropriate panel of Fig.\,\ref{Fig-RDs1}).  Lines of SiS in $v=1$ and those of Si$^{34}$S in $v=0$ show considerable scatter around the ``vibrational fit'', suggesting that $T_{\rm{rot}}$ is much smaller than $T_{\rm{vib}}$.  

  
\subsection{CS and C$^{34}$S}

Two lines of CS were observed in the survey ($J=6-5$ and $7-6$). Combining these with the $5-4$ line at 244\,GHz observed by \citet{fu}, a rotational temperature diagram was obtained from maps of integrated intensity with a circular aperture centered on the central source and on the SW\,Arc. We find that the two spatial components have similar temperatures and column densities. We therefore obtained a rotational temperature diagram for the full emission region, which however indicates that the average temperature is lower than in the two ``discrete'' sources.   In addition, an approximate estimate of the column density of C$^{34}$S in the central source was obtained from a single weak line in the survey ($J=7-6$ at 337.4\,GHz). The derived CS/C$^{34}$S ratio of 13.5 is in good agreement with the S/$^{34}$S ratio inferred from SO and SO$_2$ (see Sects.\,\ref{sect-RT-SO2} and \ref{sect-RT-SO}).

\subsection{H$_2$S }

Only one intense line of H$_2$S was observed in the survey ($3_{3,0}-3_{2,1}$ at 300.5\,GHz), because the rotational spectrum of H$_2$S is fairly sparse in the submillimeter band. In order to do a rotational analysis, the SMA measurement was combined with the $2_{2,0}-2_{1,1}$ line (216.7\,GHz) in the ARO survey \citep{tenen_survey}, and the $7_{6,1}-7_{5,2}$ line (626.5\,GHz) observed in a deep integration with APEX. From these, we derived a $T_{\rm{rot}}$ that is slightly less than 200\,K, and a column density that is two times higher than SO and is much higher than any other sulfur-bearing species we observed, implying that H$_2$S is an important molecular carrier of sulfur in VY\,CMa. Because the lines observed in the SMA survey and with APEX are transitions in {\it ortho}-H$_2$S and the ARO line is a {\it para} line, we used the values of partition function from CDMS where the two states are combined. (Although three  transitions were covered with {\it Herschel} in \citet{herschel}, the detections were tentative and the lines are strongly blended with those of other species.)

\subsection{PO and PN}

A population diagram for PO was constructed from the integrated fluxes of the six spectrally-resolved lines (two in  $^2\Pi_{1/2}$ and one in $^2\Pi_{3/2}$ for each rotational transition). For three of the lines of PO, small contributions to the observed emission from other species were considered:
TiO$_2$ at 283.6\,GHz, SO$_2$ at 327.2\,GHz, and possibly TiO$_2$ at 327.4\,GHz (see Table\,\ref{Tab-main} in main text).  However, a rotational analysis of TiO$_2$ \citep{kami_tio} and SO$_2$ (see Sect.\,\ref{sect-RT-SO2}) has shown that for a 1\,arcsec$^2$ aperture, the observed flux in the two lambda components in the $^2\Pi_{1/2}$ ladder of PO are not affected by blends with either species.  

Combining the measurements here with an earlier published observation with the SMA \citep{fu}, those from ARO \citep{tenen_survey}, and {\it Herschel} \citep{herschel}\footnote{The {\it Herschel} measurements were included with uncertainties of 10\% the total line intensity.}, we determined that the temperature describing the relative population of the two spin components ($^2\Pi_{1/2}$ and $^2\Pi_{3/2}$) is $T_{\rm{rot}} = 244 \pm 30$~K. The PO radical was first observed in VY\,CMa by \citet{tenen_PO}, but they only measured two rotational transitions in the lowest spin component with similar excitation energies, and the S/N of their lines were much lower than that here. As a result, Tenenbaum~et~al. obtained a $T_{\rm{rot}}$ (about 45\,K) that is not well determined, nevertheless the column density they derived on the assumption that the diameter of the source is 1\arcsec\ is essentially the  same as that derived here.


In order to construct a diagram of PN in the ground vibrational state, the two lines in survey were combined with the $J=5-4$ line at lower frequency observed by \citet{fu}, three lines in ARO  \citep[$J=3-2$, $5-4$, and $6-5$;][]{tenen_survey}, and two from {\it Herschel} spectra 
\citep[$J=16-15$ and $13-12$;][]{herschel}. Two of the three lines observed with the ARO were also observed with the SMA.  The $T_{\rm{rot}}$ for PN that we derive is about two times lower than that of PO and  the column density of PN is 18 times lower than that of PO (if the emission sources have relative sizes as in Table\,\ref{Tab-RD-results}).

\subsection{ AlOH, AlCl, and AlO}

Lower limits of $T_{\rm{rot}}$ of AlOH and AlCl were estimated from the population diagrams. By combining the SMA measurements with those obtained with single antennas \citep{TZ09,tenen_survey} we found $T_{\rm rot}$(AlOH)\,$> 50$\,K and $T_{\rm rot}$(AlCl)\,$\gtrsim 120$\,K (for the latter molecule the best fit shown in Fig.\,\ref{Fig-RDs2} has a very low statistical significance). For AlO, in addition to the ARO single antenna
measurements, one line was measured with {\it Herschel} \citep[$J=20-19$;][]{herschel}, allowing a good determination of $T_{\rm{rot}}$.
Combining the single antenna measurements with the SMA observations, we obtain $T_{\rm{rot}}\approx1060$~K, which is consistent with an analysis of optical bands of AlO bands in VY\,CMa \citep[$T_{\rm rot}\approx700$~K;][]{kami_alo}. This is the highest rotational temperature we found for all molecules analyzed here, but it is only slightly higher than the \emph{vibrational} temperature derived for NaCl ($T_{\rm vib}\approx 960$~K, Sect.\,\ref{sect-RT-NaCl}).
 

\subsection{NaCl and Na$^{37}$Cl \label{sect-RT-NaCl}}

Among the less abundant species in the inner outflow of VY\,CMa, the emission from NaCl is fairly intense in the ground and excited vibrational levels with $v\leq3$. As many as five rotational transitions in each vibrational level were accessible in the survey (a sixth at 325.1\,GHz is too close in frequency to a telluric water line to observe). We were unable to constrain $T_{\rm{rot}}$ from lines of NaCl in the survey, therefore measurements in other frequency bands were included in the analysis. Specifically, these were: (1) $J=18-17$ and $19-18$ in $v=0$ (234.3 and 247.2\,GHz) and $v=1$ (232.5 and 245.4\,GHz) with the SMA \citep{fu}; (2) $J=17-16$ (221.3\,GHz) with the PdBI \citep{kami_tio}; (3) $J=7-6$ (91.2\,GHz) and $8-7$ (104.2\,GHz), also with the PdBI \citep{ver_phd}; and (4) all five transitions between 221 and 273\,GHz in $v=0$ and one at 143.2\,GHz with the ARO \citep{tenen_survey,milam_nacl}; (5) $J=12-11$ in $v=0$ (153.9\,GHz) with IRAM 30\,m  \citep{sandra_tio2}; and (6)  $J=36-35$ and $37-36$ in $v=0$ with APEX.

In addition to analyzing the SMA survey measurements alone, we also made a global fit to all the available data. While the complementary data were necessary to derive $T_{\rm{rot}}$,  the survey data alone yield a better estimate of $T_{\rm vib}$ than that obtained in the global analysis, because lines in different vibrational levels in the survey are from similar rotational levels and the rotational excitation does not affect the fit. 
From the combined analysis, it appears that $T_{\rm{rot}}$ is a few times lower than $T_{\rm{vib}}$ (see Fig.\,\ref{Fig-RDs2}).

A population diagram for Na$^{37}$Cl is shown in a separate panel of Fig.\,\ref{Fig-RDs2}. A vibrational temperature derived from the $v=0$ and 1 data of $T_{\rm{vib}}=562\pm139$~K is consistent with that obtained for the main species. The ratio of the NaCl/Na$^{37}$Cl column density of $4\pm1$ is consistent with the solar elemental abundance ratio Cl/$^{37}$Cl=3.1 \citep{lodders}. 


\begin{figure*}\centering
\includegraphics[angle=270,width=0.45\textwidth]{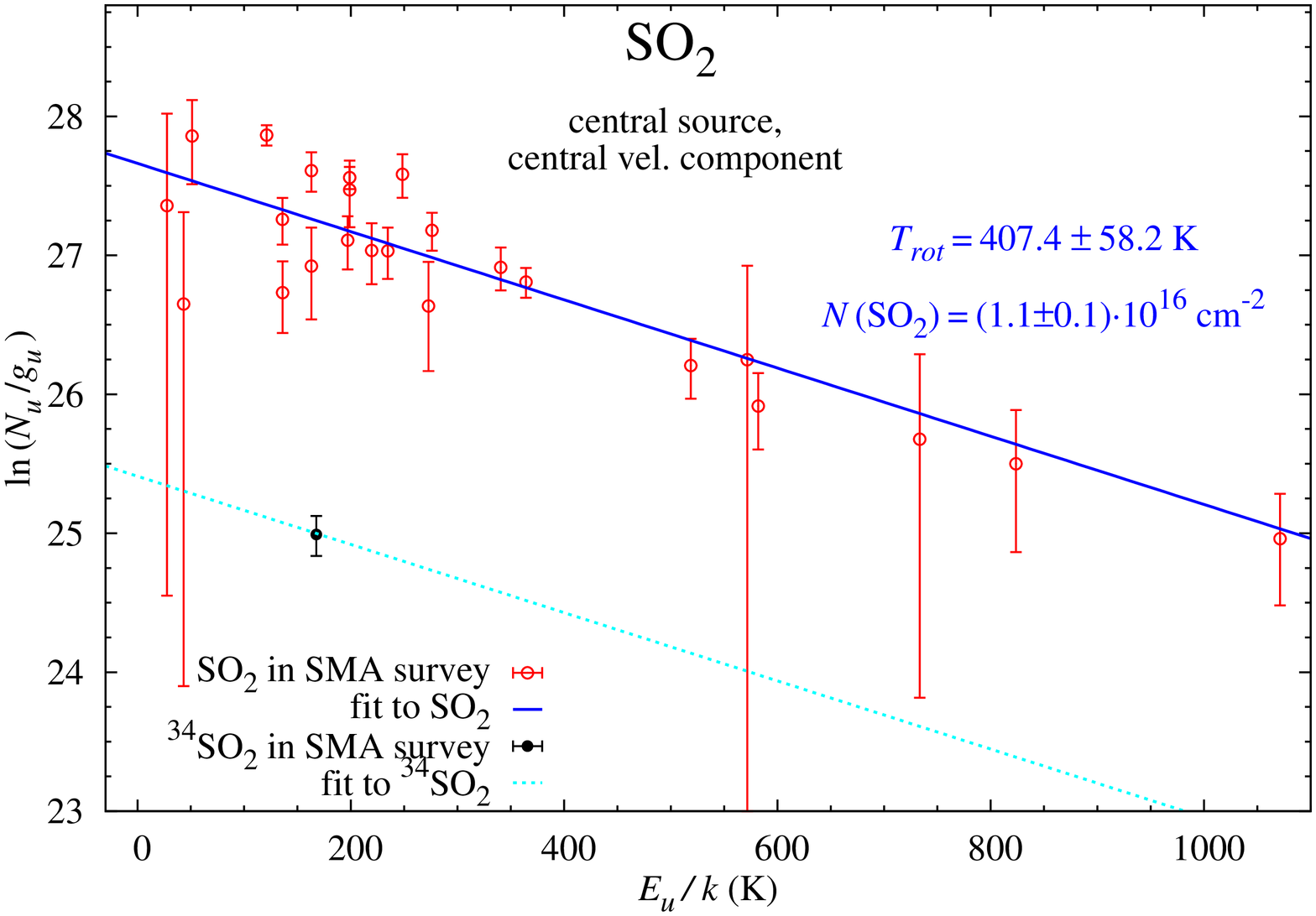}
\includegraphics[angle=270,width=0.45\textwidth]{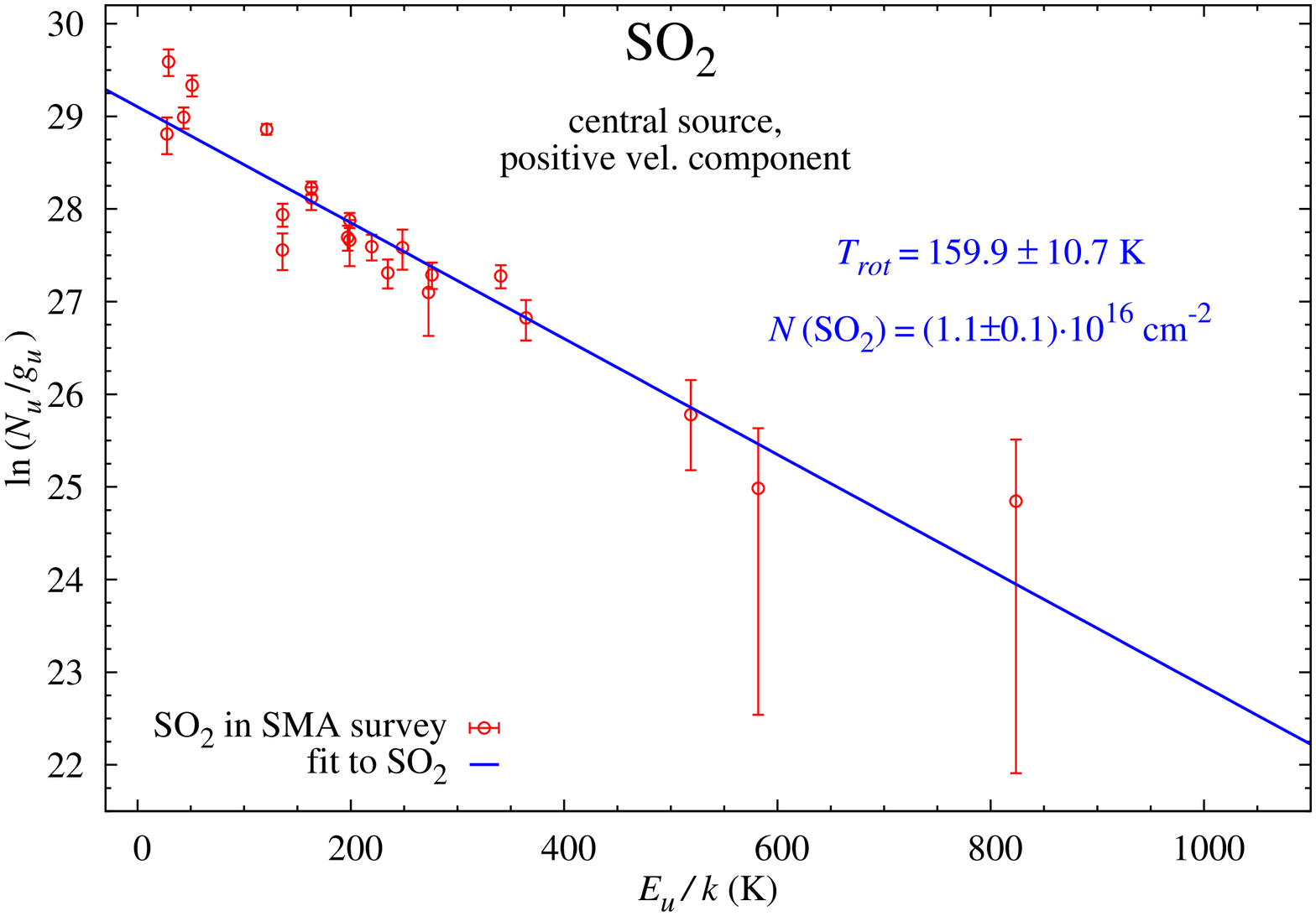}
\includegraphics[angle=270,width=0.45\textwidth]{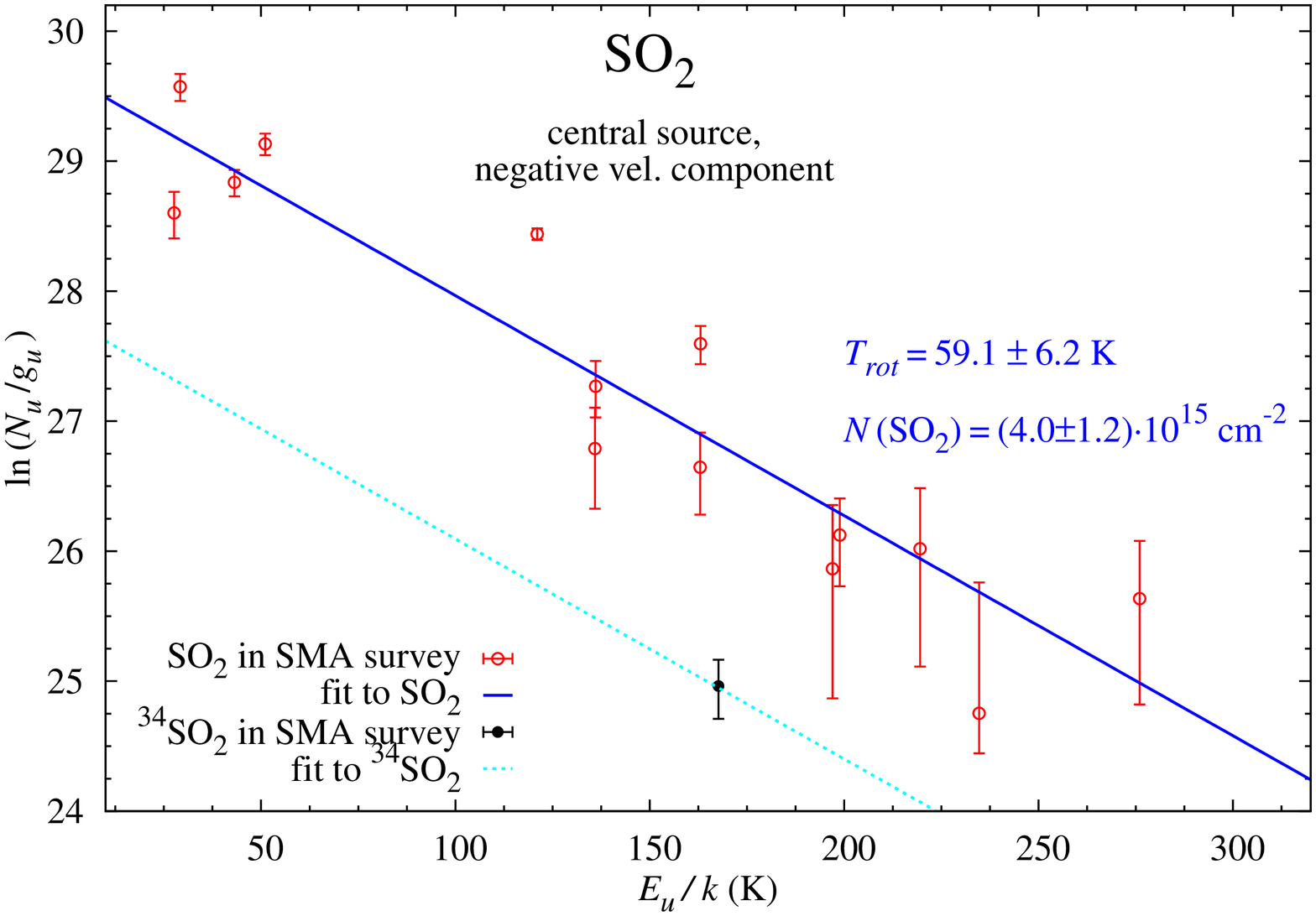}
\includegraphics[angle=270,width=0.45\textwidth]{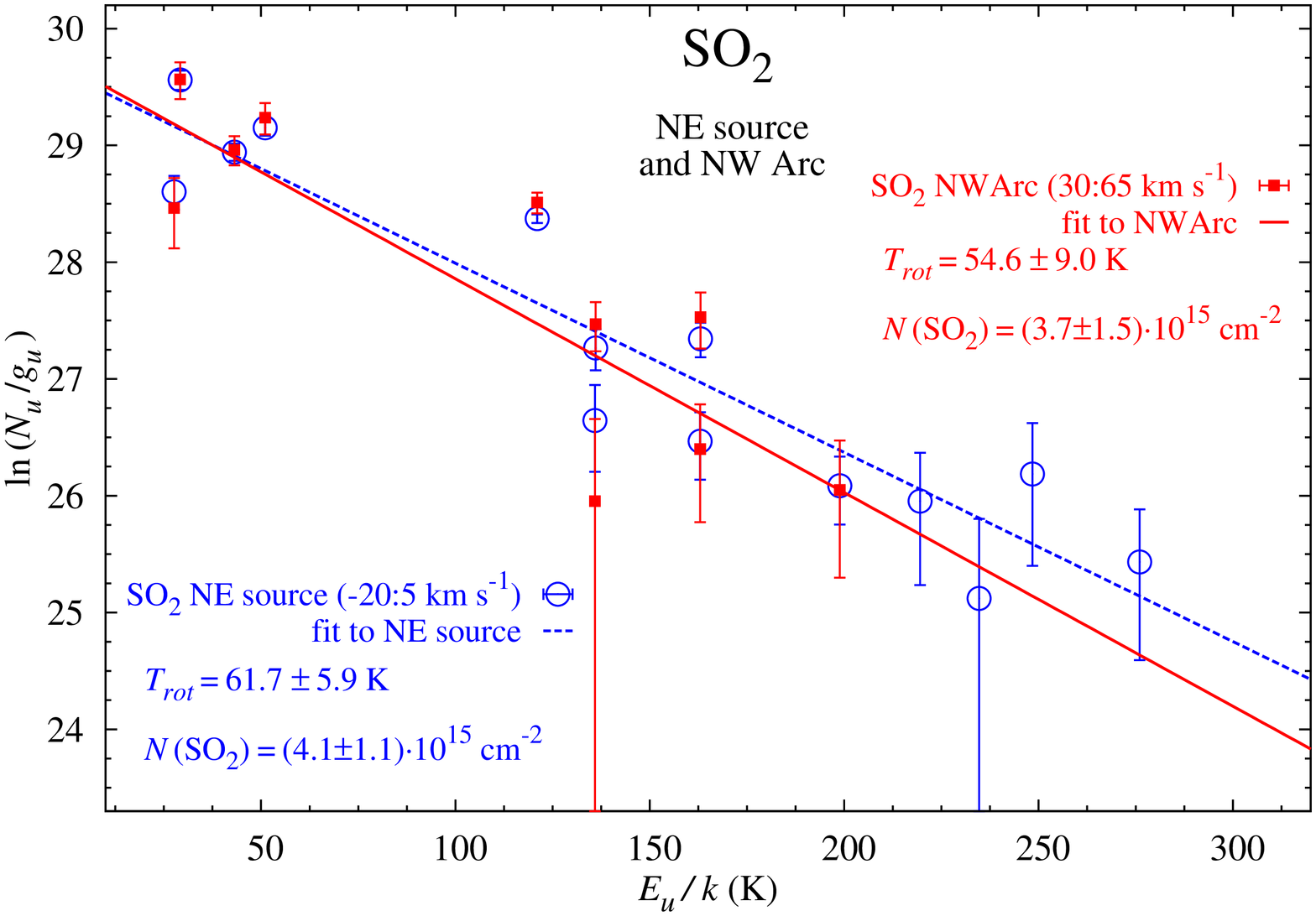}
\includegraphics[angle=270,width=0.45\textwidth]{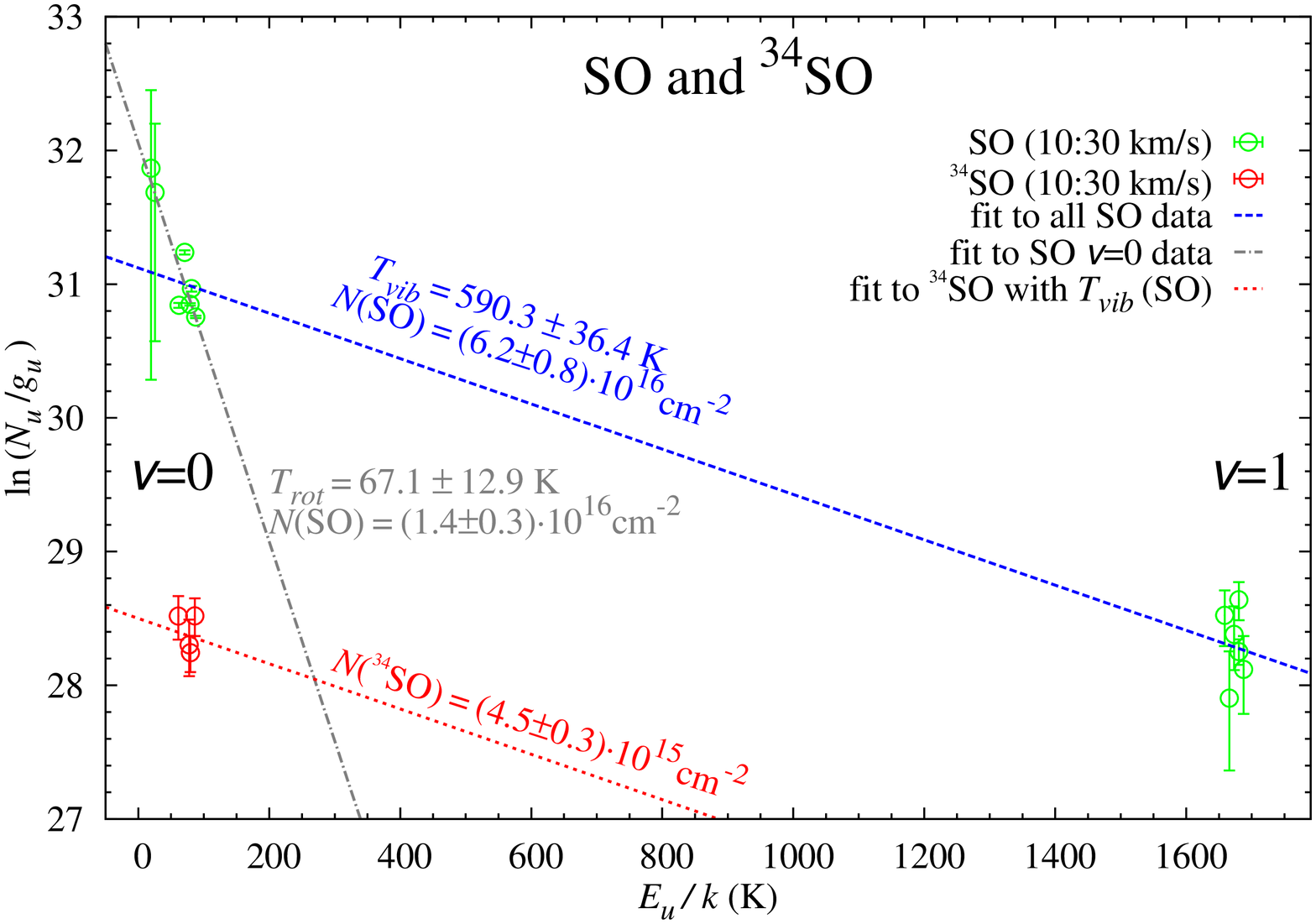}
\includegraphics[angle=270,width=0.45\textwidth]{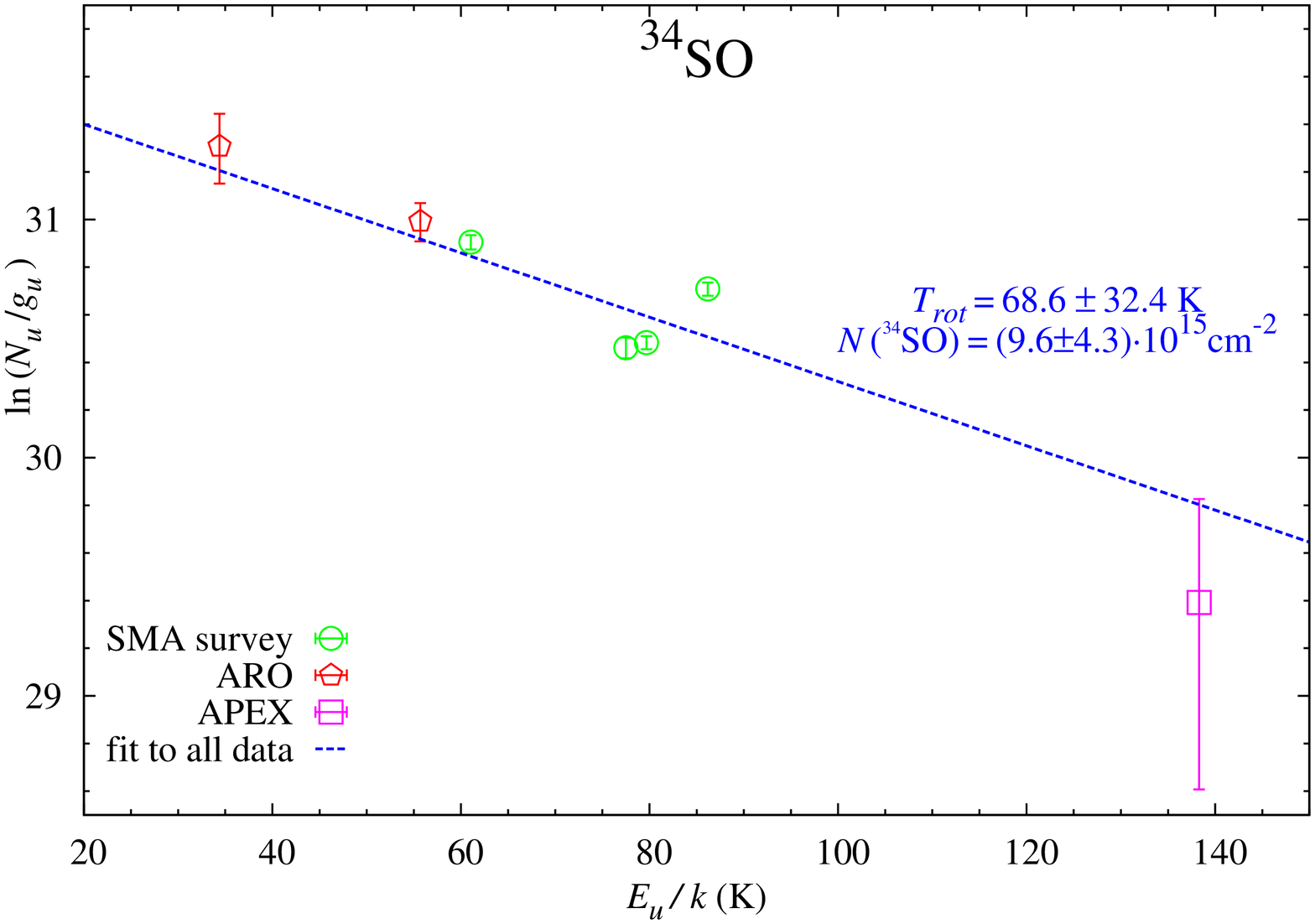}
\includegraphics[angle=270,width=0.45\textwidth]{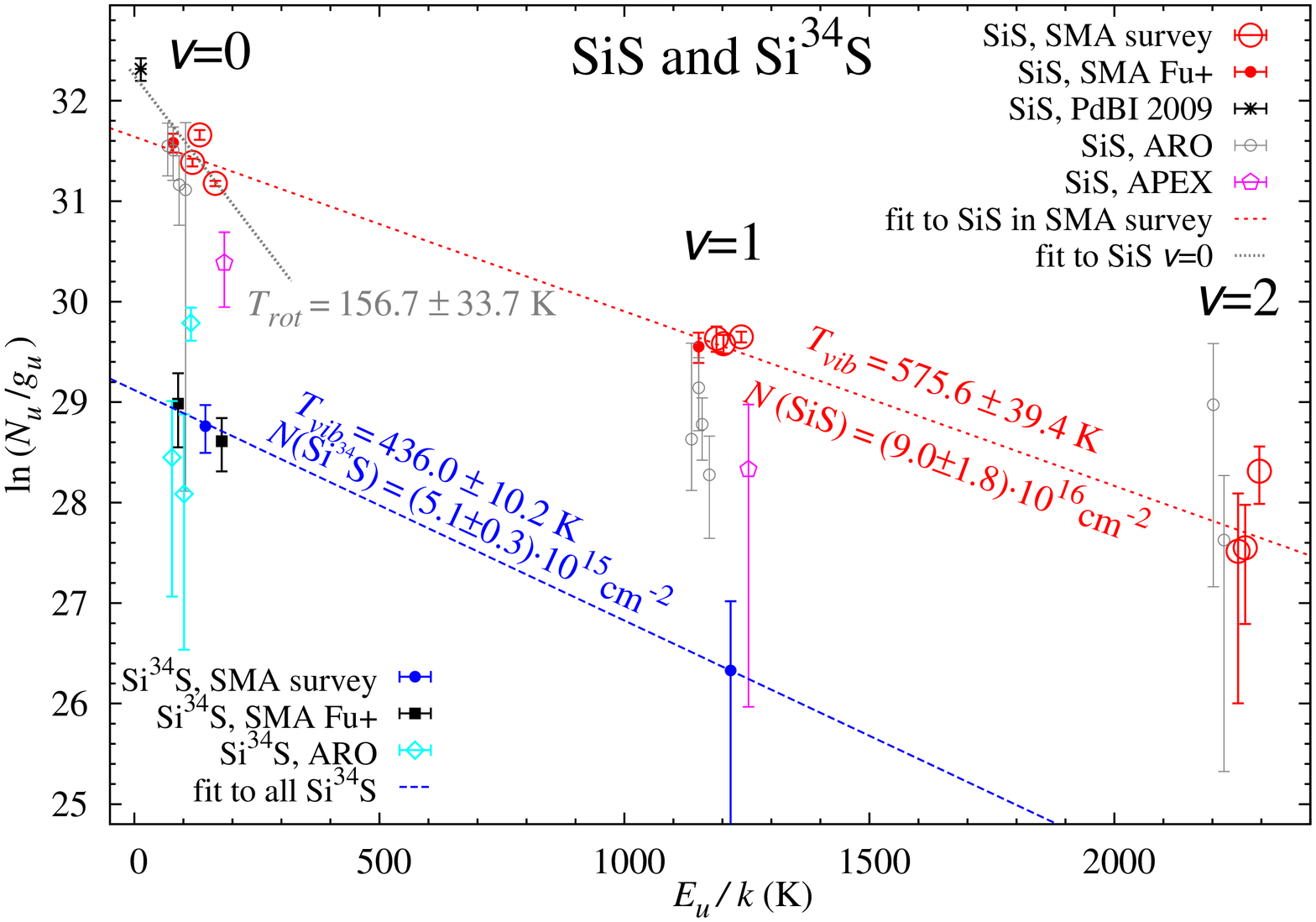}
\includegraphics[angle=270,width=0.45\textwidth]{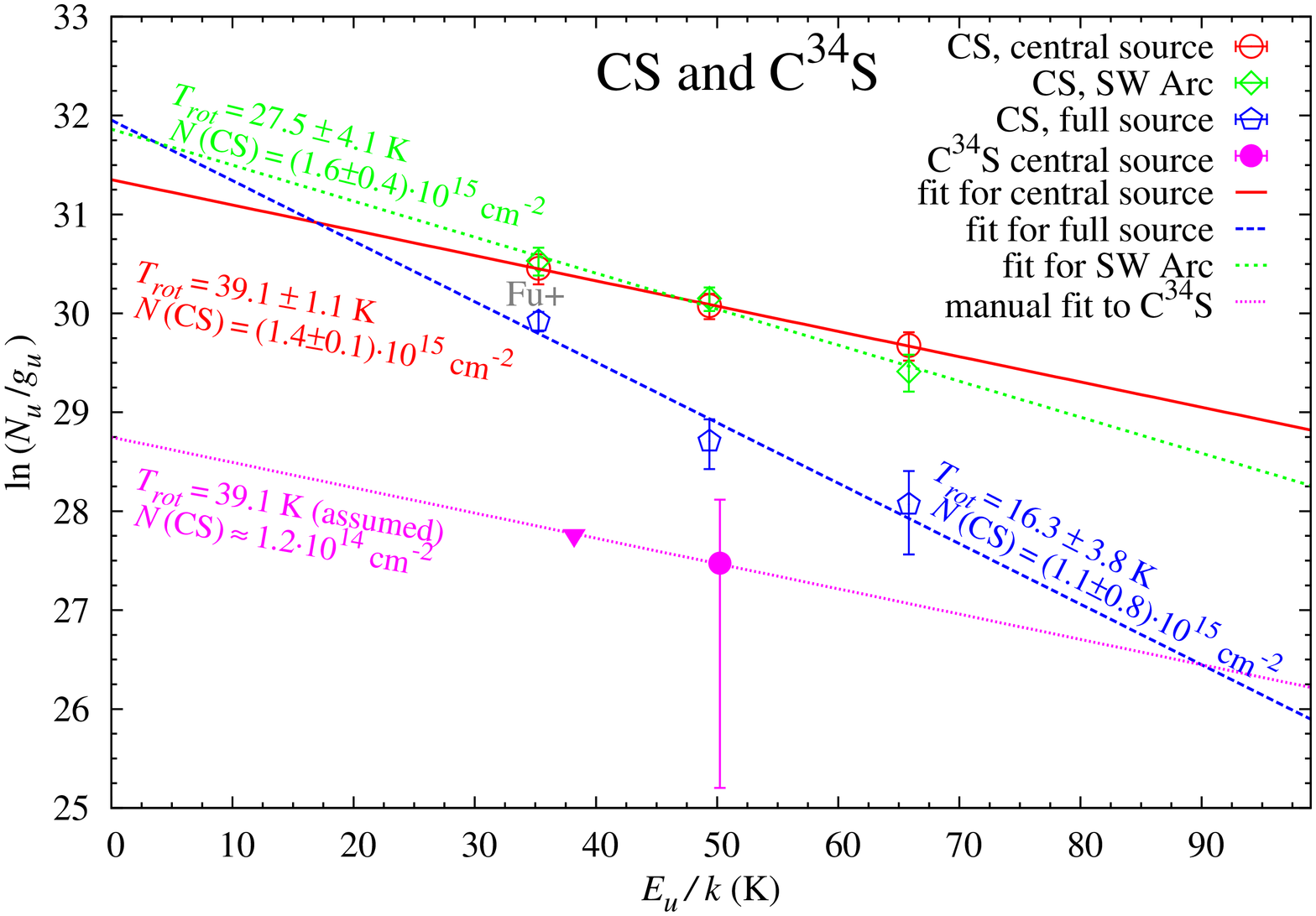}
\caption{Population diagrams. All errorbars are 3$\sigma$, except those from {\it Herschel}.}
\label{Fig-RDs1}
\end{figure*}

\begin{figure*}\centering
\includegraphics[angle=270,width=0.45\textwidth]{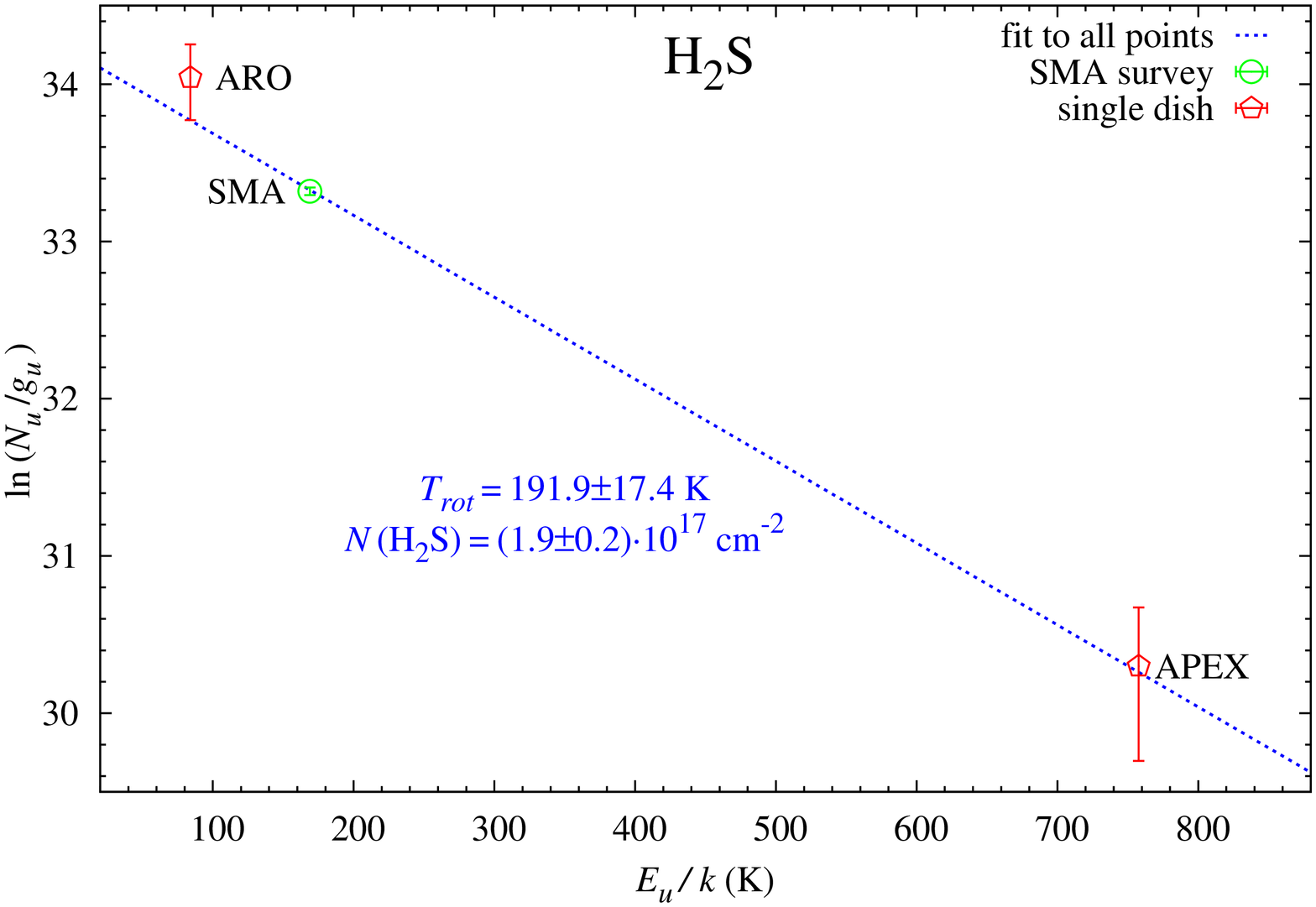}
\includegraphics[angle=270,width=0.45\textwidth]{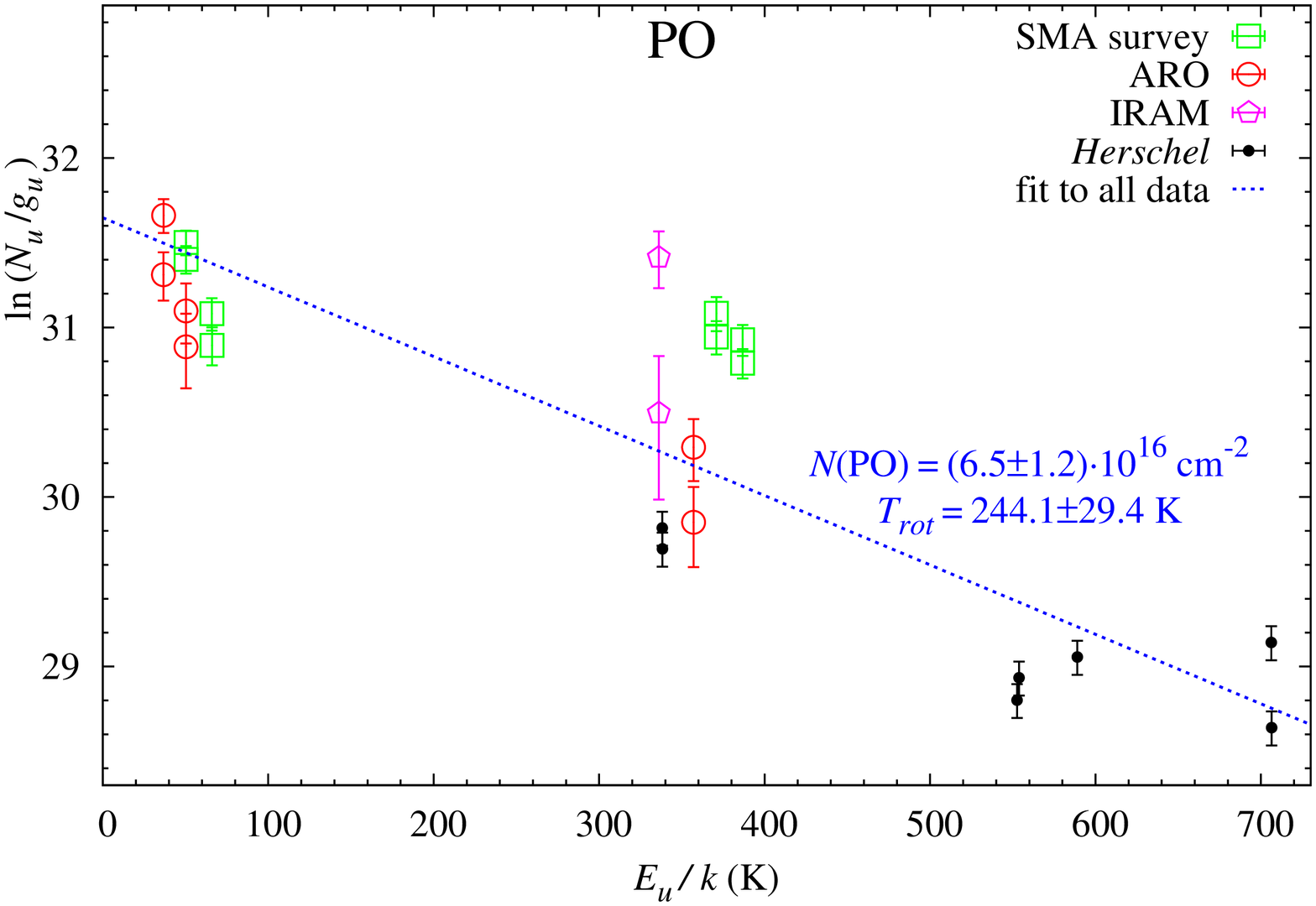}
\includegraphics[angle=270,width=0.45\textwidth]{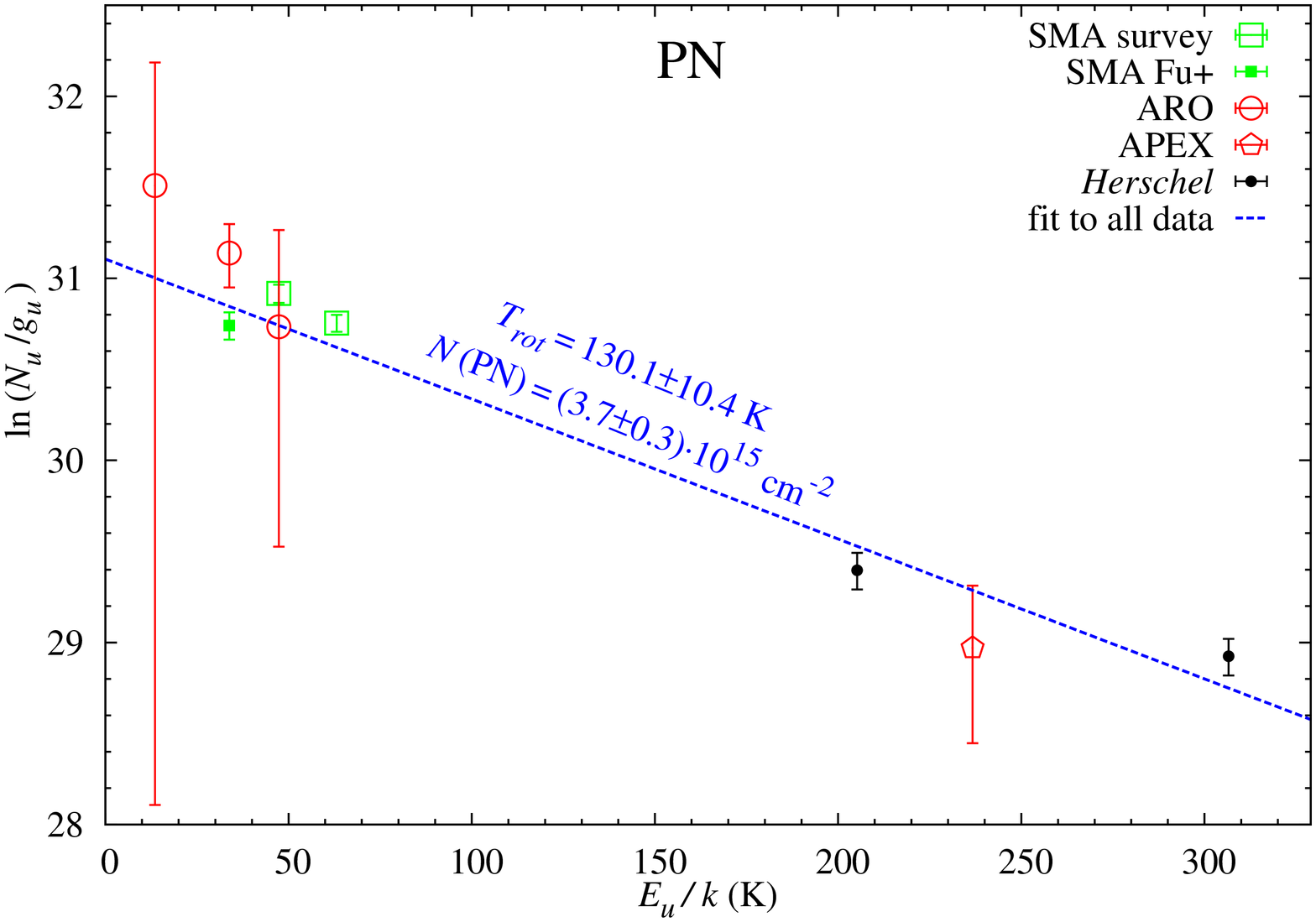}
\includegraphics[angle=270,width=0.45\textwidth]{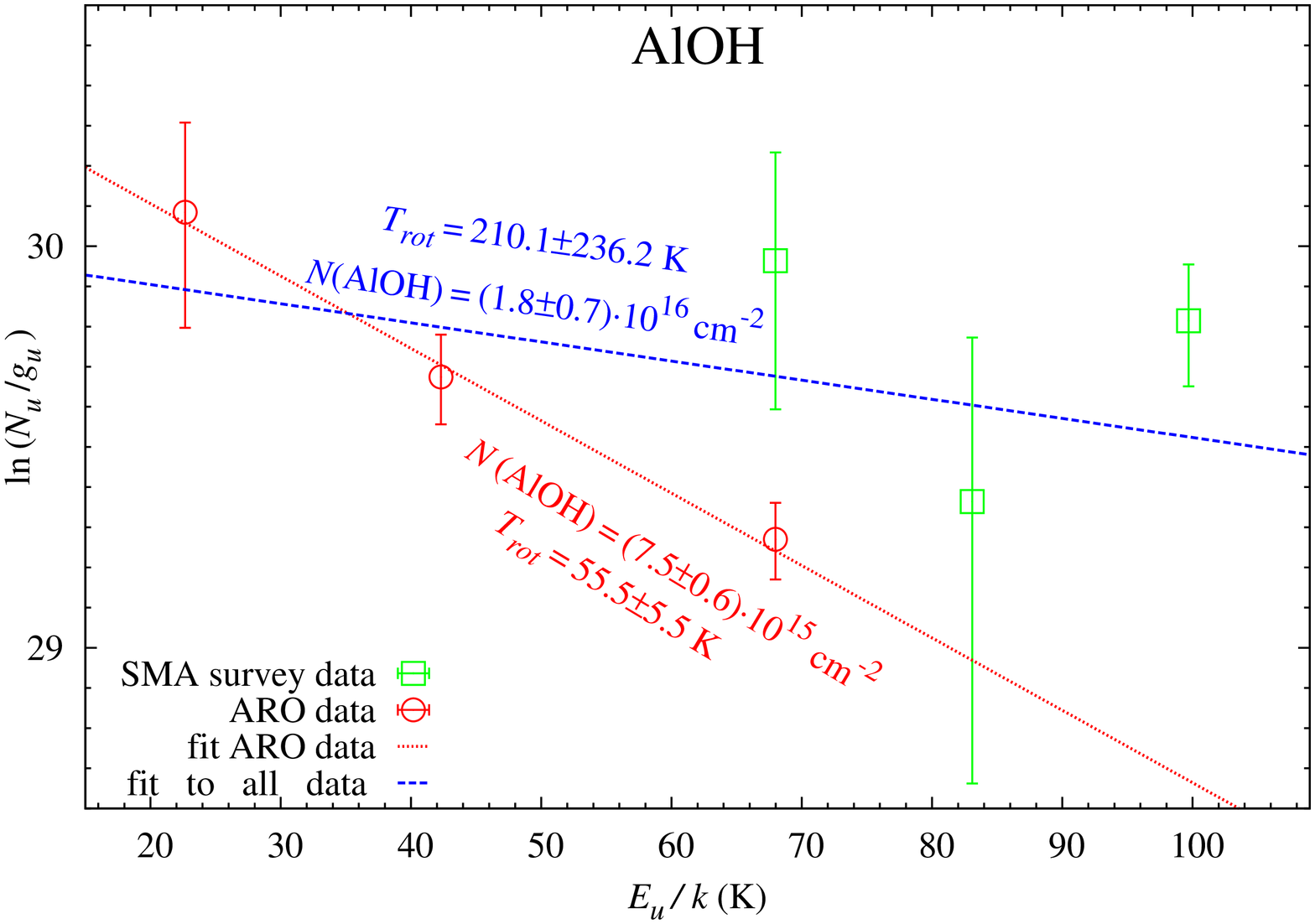}
\includegraphics[angle=270,width=0.45\textwidth]{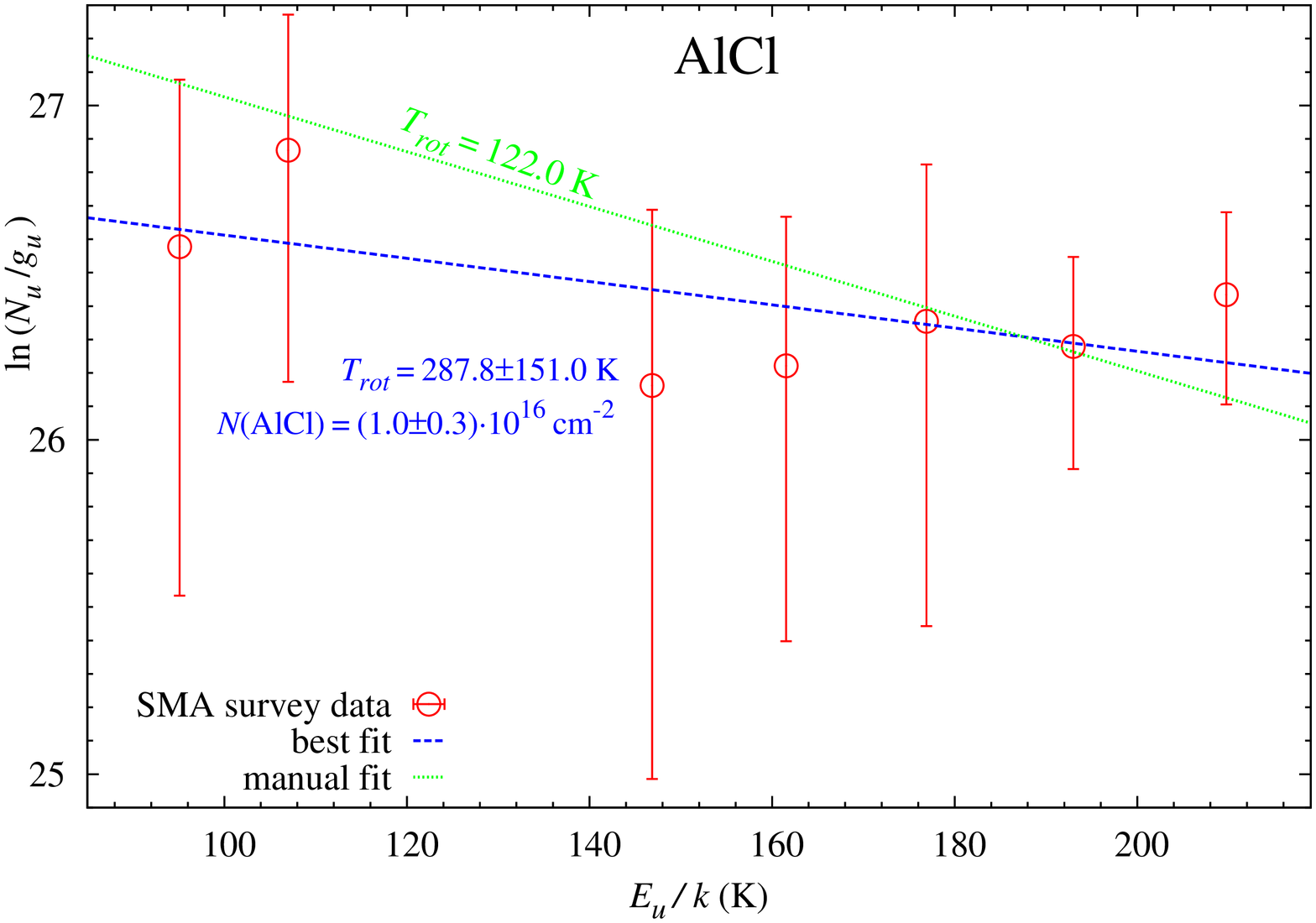}
\includegraphics[angle=270,width=0.45\textwidth]{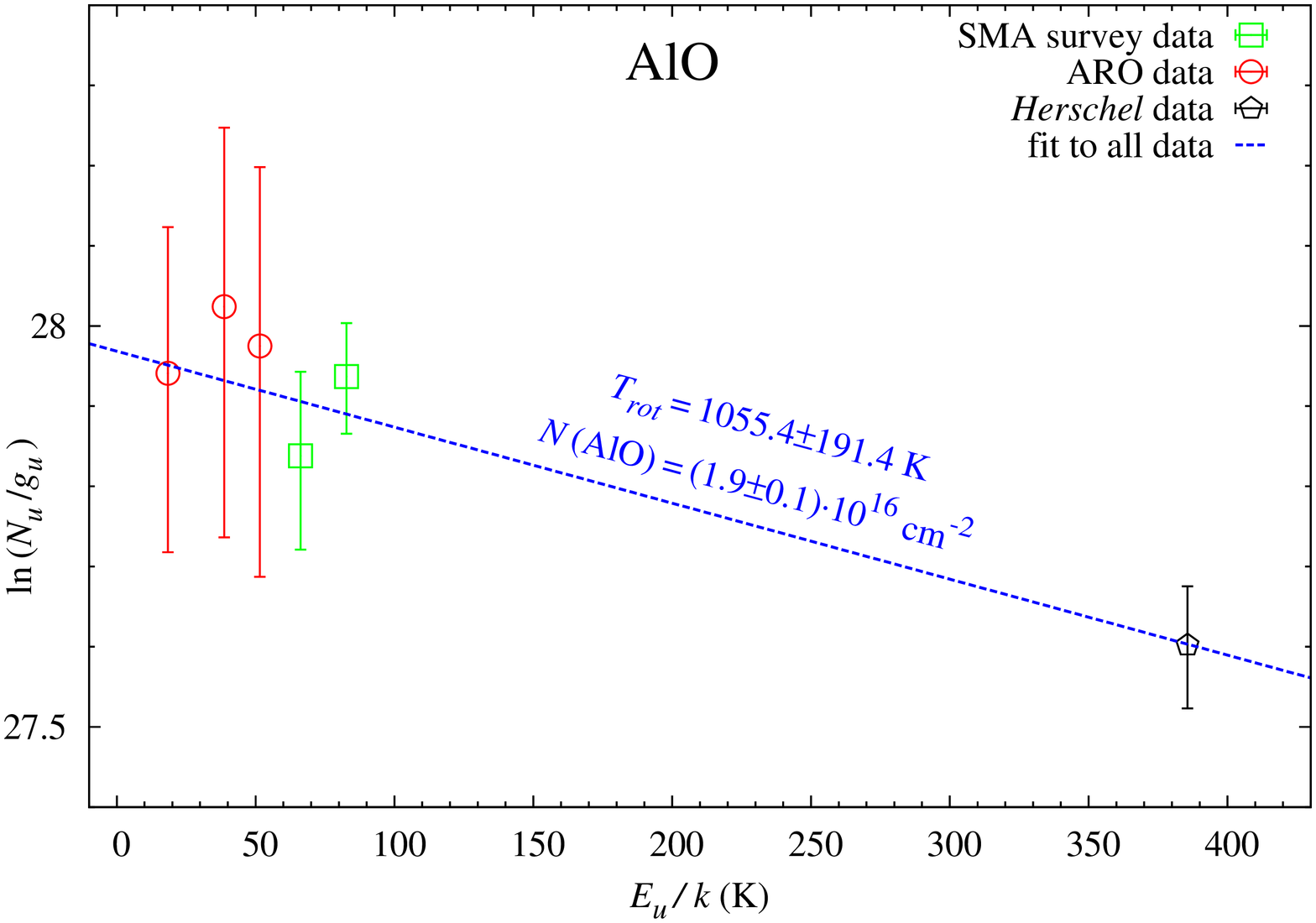}
\includegraphics[angle=270,width=0.45\textwidth]{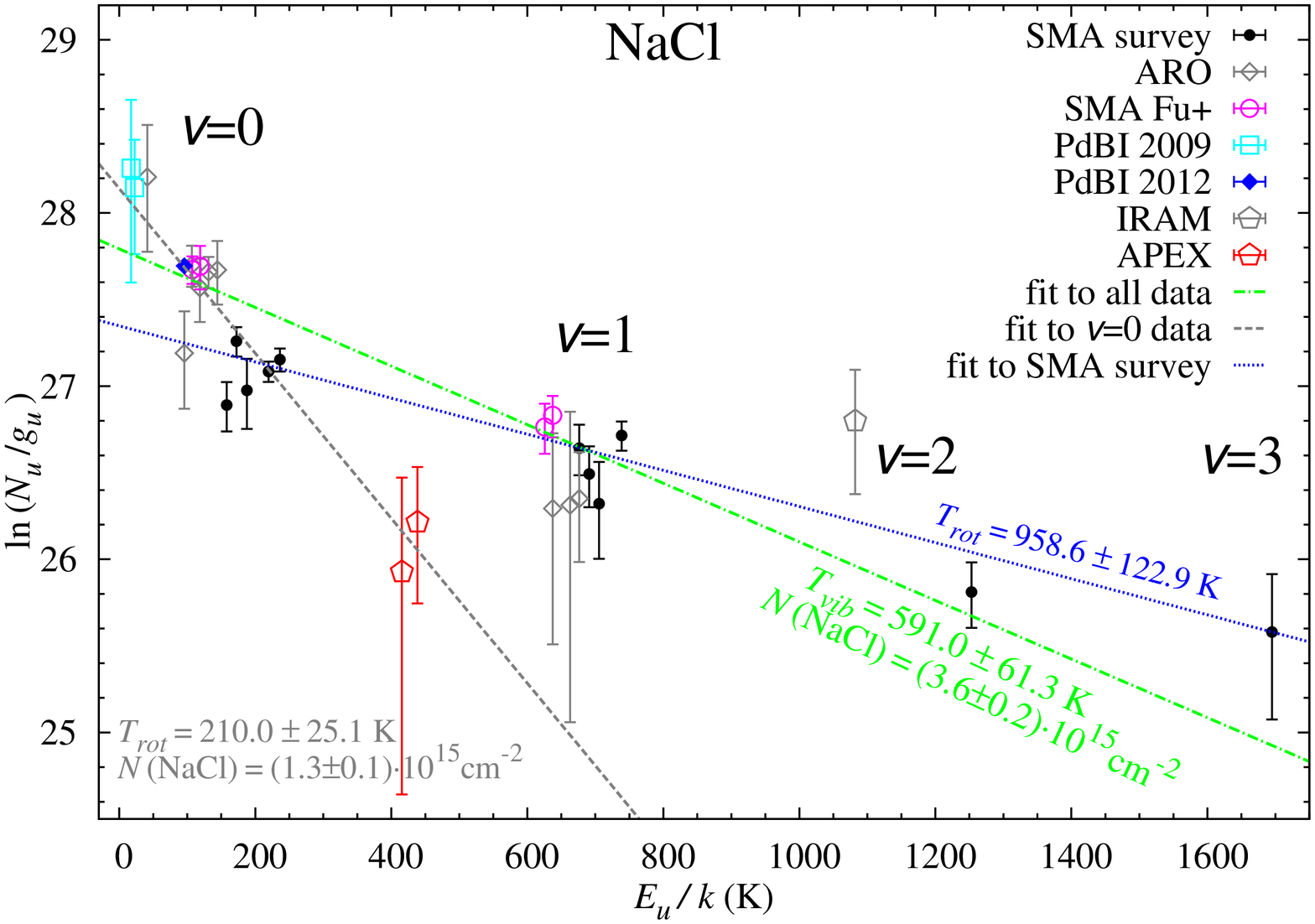}
\includegraphics[angle=270,width=0.45\textwidth]{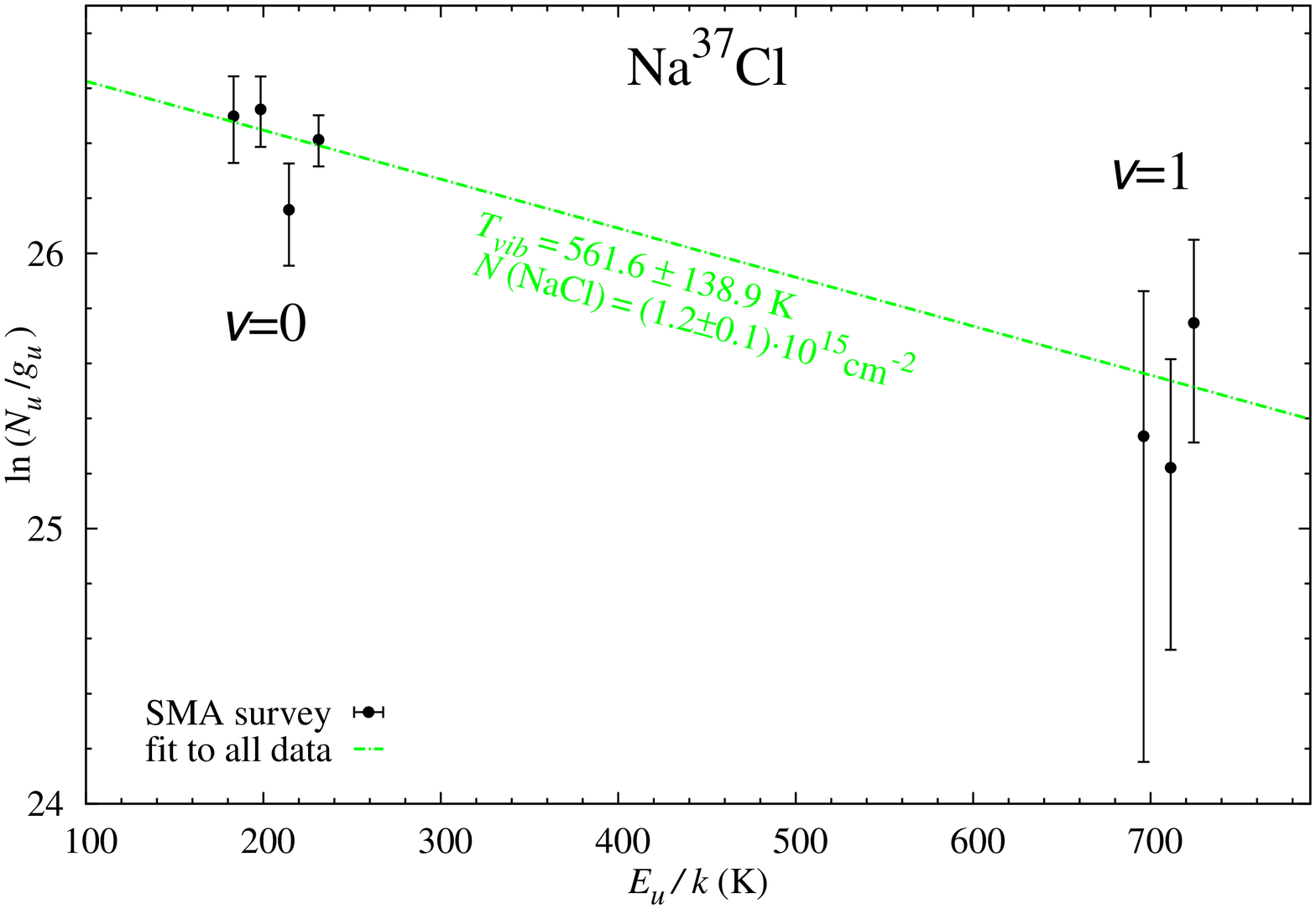}
\addtocounter{figure}{-1}
\caption{({\it Continued})}
\label{Fig-RDs2}
\end{figure*}

\acknowledgments
We thank Sandra Br{\"u}nken for making her IRAM spectra of VY\,CMa available to us.

\end{document}